\documentclass[traditionalabstract]{aa}  
\usepackage{graphicx}
\usepackage{txfonts}
\usepackage[colorlinks, citecolor=blue, linkcolor=magenta]{hyperref}
\usepackage[usenames, dvipsnames]{color}
\usepackage{overpic} 
\newcommand{\hls}{HLS\,J0918+5142}

\newcommand{\elev}{\rm{el}}
\newcommand{\az}{\rm{az}}

\newcommand{\aka}{a.\,k.\,a.}
\newcommand{\wrt}{w.\,r.\,t.}
\newcommand{\bm}{{\tt beammap}}
\newcommand{\bms}{{\tt beammaps}}

\newcommand{\cm}{\emph{common~mode}}
\newcommand{\cmoneb}{\emph{Most~Correlated~Pixels}}
\newcommand{\afternoon}{temperature-induced}

\newcommand{\trentemetre}{30-$\rm{m}$}
\newcommand{\baseline}{\emph{baseline}}
\newcommand{\taunu}{\tau_{\nu}}
\newcommand{\taumeter}{taumeter}
\newcommand{\airmass}{air mass}

\newcommand{\nico}[1]{#1}
\newcommand{\lp}[1]{#1}

\newcommand{\rev}[1]{#1}

\begin{document} 

   \title{Calibration and Performance of the NIKA2 camera at the IRAM 30-meter Telescope}

   \author{L.~Perotto \inst{\ref{LPSC}}
     \and  N.~Ponthieu \inst{\ref{IPAG}}
     \and  J.~F.~Mac\'ias-P\'erez \inst{\ref{LPSC}}
     \and  R.~Adam \inst{\ref{LLR}, \ref{CEFCA}} 
     \and  P.~Ade \inst{\ref{Cardiff}}
     \and  P.~Andr\'e \inst{\ref{CEA}}
     \and  A.~Andrianasolo \inst{\ref{IPAG}}
     \and  H.~Aussel \inst{\ref{CEA}}
     \and  A.~Beelen \inst{\ref{IAS}}
     \and  A.~Beno\^it \inst{\ref{Neel}}
     \and  S.~Berta \inst{\ref{IRAMF}}
     \and  A.~Bideaud \inst{\ref{Neel}}
     \and  O.~Bourrion \inst{\ref{LPSC}}
     \and  M.~Calvo \inst{\ref{Neel}}
     \and  A.~Catalano \inst{\ref{LPSC}}
     \and  B.~Comis \inst{\ref{LPSC}}
     \and  M.~De Petris \inst{\ref{Roma}}
     \and  F.-X.~D\'esert \inst{\ref{IPAG}}
     \and  S.~Doyle \inst{\ref{Cardiff}}
     \and  E.~F.~C.~Driessen \inst{\ref{IRAMF}}
     \and  P.~Garc\'ia \inst{\ref{Chine}, \ref{Chili}}
     \and  A.~Gomez \inst{\ref{CAB}} 
     \and  J.~Goupy \inst{\ref{Neel}}
     \and  D.~John \inst{\ref{IRAME}}
     \and  F.~K\'eruzor\'e \inst{\ref{LPSC}}
     \and  C.~Kramer \inst{\ref{IRAMF},\ref{IRAME}}
     \and  B.~Ladjelate \inst{\ref{IRAME}} 
     \and  G.~Lagache \inst{\ref{LAM}}
     \and  S.~Leclercq \inst{\ref{IRAMF}}
     \and  J.-F.~Lestrade \inst{\ref{LERMA}}
     \and  A.~Maury \inst{\ref{CEA}}
     \and  P.~Mauskopf \inst{\ref{Cardiff},\ref{Arizona}}
     \and  F.~Mayet \inst{\ref{LPSC}}
     \and  A.~Monfardini \inst{\ref{Neel}}
     \and  S.~Navarro \inst{\ref{IRAME}}
     \and  J.~Pe\~nalver \inst{\ref{IRAME}}
     \and  F.~Pierfederici \inst{\ref{IRAME}}
     \and  G.~Pisano \inst{\ref{Cardiff}}
     \and  V.~Rev\'eret \inst{\ref{CEA}}
     \and  A.~Ritacco \inst{\ref{LPSC},\ref{IRAME}}
     \and  C.~Romero \inst{\ref{IRAMF}}
     \and  H.~Roussel \inst{\ref{IAP}}
     \and  F.~Ruppin \inst{\ref{LPSC},\ref{MIT}}
     \and  K.~Schuster \inst{\ref{IRAMF}}
     \and  S.~Shu \inst{\ref{IRAMF}} 
     \and  A.~Sievers \inst{\ref{IRAME}}
     \and  C.~Tucker \inst{\ref{Cardiff}}
     \and  R.~Zylka \inst{\ref{IRAMF}}}
   \institute{
     Univ. Grenoble Alpes, CNRS, Grenoble INP, LPSC-IN2P3, 53, avenue des Martyrs, 38000 Grenoble, France
     \label{LPSC}
     \and
     Univ. Grenoble Alpes, CNRS, IPAG, 38000 Grenoble, France 
     \label{IPAG}
     \and
     LLR (Laboratoire Leprince-Ringuet), CNRS, École Polytechnique, Institut Polytechnique de Paris, Palaiseau, France
     \label{LLR}
     \and
     Centro de Estudios de F\'isica del Cosmos de Arag\'on (CEFCA), Plaza San Juan, 1, planta 2, E-44001, Teruel, Spain
     \label{CEFCA}
     \and
     Astronomy Instrumentation Group, University of Cardiff, UK
     \label{Cardiff}
     \and
     AIM, CEA, CNRS, Universit\'e Paris-Saclay, Universit\'e Paris Diderot, Sorbonne Paris Cit\'e, 91191 Gif-sur-Yvette, France
     \label{CEA}
     \and
     Institut d'Astrophysique Spatiale (IAS), CNRS and Universit\'e Paris Sud, Orsay, France
     \label{IAS}
     \and
     Institut N\'eel, CNRS and Universit\'e Grenoble Alpes, France
     \label{Neel}
     \and
     Institut de RadioAstronomie Millim\'etrique (IRAM), Grenoble, France
     \label{IRAMF}
     \and 
     Dipartimento di Fisica, Sapienza Universit\`a di Roma, Piazzale Aldo Moro 5, I-00185 Roma, Italy
     \label{Roma}
     \and
     Chinese Academy of Sciences South America Center for Astronomy, National Astronomical Observatories, CAS, Beijing 100101, China
     \label{Chine}
     \and
     Instituto de Astronom\'ia, Universidad Cat\'olica del Norte, Av. Angamos 0610, Antofagasta 1270709, Chile
     \label{Chili}
     \and
     Centro de Astrobiolog\'ia (CSIC-INTA), Torrej\'on de Ardoz, 28850 Madrid, Spain
     \label{CAB}
     \and  
     Instituto de Radioastronom\'ia Milim\'etrica (IRAM), Granada, Spain
     \label{IRAME}
     \and
     Aix Marseille Univ, CNRS, CNES, LAM (Laboratoire d'Astrophysique de Marseille), Marseille, France
     \label{LAM}
     \and 
     LERMA, Observatoire de Paris, PSL Research University, CNRS, Sorbonne Universit\'es, UPMC Univ. Paris 06, 75014 Paris, France  
     \label{LERMA}
     \and 
     Institut d'Astrophysique de Paris, CNRS (UMR7095), 98 bis boulevard Arago, 75014 Paris, France
     \label{IAP}
     \and
     Kavli Institute for Astrophysics and Space Research, Massachusetts Institute of Technology, Cambridge, MA 02139, USA
     \label{MIT}
     \and
     School of Earth and Space Exploration and Department of Physics, Arizona State University, Tempe, AZ 85287, USA
     \label{Arizona}
   }
   \date{Received July 2, 2019; Accepted January 15, 2020}
   \abstract
       {
         NIKA2 is a dual-band millimetre continuum
         camera of 2\,900 Kinetic Inductance Detectors (KID),
         operating at $150$ and $260\,\rm{GHz}$, installed at the IRAM 30-meter
         telescope 
          in Spain.
         Open to the scientific community since October 2017, NIKA2 
         will provide key observations for the next decade to 
         address a wide range of open questions in astrophysics and
         cosmology.}
       {We present {\lp the calibration method} and the performance assessment of NIKA2 after one year of observation.}
       {We use a large data set acquired between January 2017 and
         February 2018 including observations of primary and secondary
         calibrators and faint sources that span the whole range
         of observing elevations and atmospheric conditions encountered at the
         IRAM 30-m telescope. This allows us to test the stability of the
         performance parameters against time evolution and
         observing conditions. {\lp We describe a standard calibration
         method, referred to as the \emph{baseline} method, to go from
         raw data to flux density measurements. This includes the
         determination of the detector positions in the sky, the
         selection of the detectors, the measurement of
         the beam pattern, the estimation of the 
         atmospheric opacity, the calibration of absolute flux-density
         scale, the flat fielding, and the photometry. We
         assess the robustness of the performance results using the
         \emph{baseline} method against systematic effects by
         comparing to results using alternative methods.}  
       }
       {We report an instantaneous field of view (FOV) of 6.5'
         in diameter, filled with an average fraction of $84\%$ and 
         $90\%$ of 
         valid detectors
         at $150$ and $260\,\rm{GHz}$, respectively. The beam pattern
         is characterized by a FWHM of $17.6'' \pm 0.1''$
         and  $11.1''\pm 0.2''$, and a main beam efficiency of
         {\rev $47\% \pm 3\%$} and {\rev $64\% \pm 3\%$}
         at $150$ and $260\,\rm{GHz}$, respectively.
         The {\rev point-source} rms calibration uncertainties are about $3\%$ at $150\,\rm{GHz}$ 
         and $6\%$ at $260\,\rm{GHz}$. This demonstrates
         the accuracy of the methods that we have deployed to correct
         for atmospheric attenuation.
         {\lp The absolute
           calibration uncertainties are of $5\%$ and the systematic
           calibration uncertainties evaluated at the IRAM
           \trentemetre\ reference Winter observing conditions are
           below $1\%$ in both channels.}
         The noise equivalent
         flux density (NEFD) at $150$ and $260\,\rm{GHz}$ are of
         $9 \pm 1\, \rm{mJy}\cdot s^{1/2}$ and
         $30 \pm 3\, \rm{mJy}\cdot s^{1/2}$. 
         This state-of-the-art performance confers NIKA2 with
         mapping speeds of $1388 \pm 174$ and
         $111 \pm 11 \, \rm{arcmin}^2\cdot \rm{mJy}^{-2}\cdot
         \rm{h}^{-1}$
         at $150$ and $260\,\rm{GHz}$.}
       {With these unique capabilities of fast dual-band mapping at
         high (better that 18'') angular resolution, NIKA2 is providing an unprecedented view
         of the millimetre Universe.}      

   \nopagebreak    
   \maketitle
   \nopagebreak    
\section{Introduction}
\label{se:intro}
Sub-millimetre and millimetre domains offer a unique view of the
Universe from nearby astrophysical objects, including planets,
planetary systems, galactic sources, nearby galaxies, to high-redshift
cosmological probes, \emph{e.g.} distant dusty star-forming galaxies,
clusters of galaxies, Cosmic Infrared Background (CIB), Cosmic Microwave
Background (CMB).

Ground-based {\lp continuum} millimetre experiments have made spectacular progress in
the past-two decades thanks to the advent of large arrays of
high-sensitivity detectors~\citep{Wilson2008_AZTEC,
  Siringo2009_LABOCA, Staguhn2011_GISMO, Swetz2011_ACT, Monfardini2011_NIKA, Polarbear_2012, Bleem2012_SPT, Holland2013_SCUBA2,
  Dicker2014_MUSTANG2, BKS_2015, Adam2017, BICEP3_2018}. 
This fast growth will continue as experimental efforts are
driven by two challenges: improving the sensitivity to the
polarisation to detect the signature of
the end-of-inflation gravitational waves in the CMB, and {\rev
improving the angular resolution. The latter has several important
scientific implications: reaching arcminute angular resolution to
exploit the CMB secondary anisotropies as cosmological probes, and
sub-arcminute resolution to unveil the inner structure of faint or
complex astrophysical objects and to map the early universe down to
the confusion limit.}
Furthermore, new
generation of sub-arcminute millimetre experiments, like NIKA2, which
combines high angular resolution with a high mapping speed and a large
coverage in the frequency domain, will achieve a breakthrough in 
our detailed understanding of the formation and evolution of
structures throughout the Universe.

The New IRAM KID Array Two, NIKA2, is a sub-arcminute-resolution
high-mapping speed camera observing simultaneously a 6.5' diameter
field of view (FOV) in intensity in two
frequency channels centred at 150 and $260\,\rm{GHz}$ and in
polarisation at $260\, \rm{GHz}$~\citep{Adam2018}. NIKA2 was installed
at the IRAM \trentemetre\ telescope in October 2015 with a partial readout
electronics, and operated in the final instrumental configuration since
January 2017. After a successful
commissioning phase that ended in October 2017, NIKA2 is
open to the community for science-purpose observations for the next
decade. NIKA2 will provide key observations both at the galactic scale
and at high redshifts to address a plethora of open questions, including
the environment impact on dust properties, the star formation processes
at low and high redshifts, the evolution of the large-scale structures
and the exploitation of galaxy clusters for accurate cosmology.

At the galactic scale, progress in understanding the star formation
process relates to an accurate characterization of dust properties in
the interstellar medium (ISM). NIKA2 will provide the high-resolution
high-mapping speed dual-wavelength millimetre observations that are
required for the determination of the mass and emissivity of
statistically significant samples of dense, cold, star-forming
molecular clouds~\citep{Rigby2018}.
Deep multi-wavelengths surveys of nearby galaxies and of large areas
of the galactic plane also allows for setting constraints on
environmental-related variations of the dust properties.
Furthermore, NIKA2 observations are needed for a
detailed study of the inner molecular cloud filamentary structure that
hosts Solar-mass star progenitors~\citep{Bracco2017}, to
understand the evolution process that culminates in star
formation (see \emph{e.g.}~\citet{Andre2014} for a review). Ultimately, these
observations are also helpful to understand planet formation within
proto-planetary disks.

For cosmology, NIKA2 observations will have two major
implications. On the one hand, they represent a unique opportunity to
study the evolution of the galaxy cluster mass calibration with
redshift and {\lp dynamical state} for their accurate exploitation as cosmological probe. 
Galaxy clusters are efficiently detected via the thermal
Sunyaev-Zel'dovich (tSZ) effect~\citep{SZ1970} up to high redshifts {\lp ($z<1.5$)}, as was recently
proven by CMB
experiments~\citep{Hasselfield2013_ACT_SZ, Reichardt2013_SPT_SZ, Planck2016_SZcat}.
The exploitation of the vast SZ-selected galaxy cluster catalogues is
currently the most powerful approach for cosmology with galaxy
clusters~\citep{Planck_2016_SZ_cosmo}. However, the accuracy of the tSZ cluster
cosmology relies on the calibration of the relation between the tSZ
observable and the cluster mass and on the assessment of both its redshift
evolution and the impact of the complex cluster physics on its calibration. 
Previous arcminute resolution experiments only allowed detailed studies
of the intra cluster medium spatial distribution for low redshift clusters (z <
0.2). Sub-arcminute resolution high mapping speed experiments are
required to extend our understanding of galaxy cluster towards high
redshift~\citep{Tony2019}. The first high-resolution
tSZ mapping of a galaxy cluster with NIKA2 has been reported
in \citet{Ruppin2018}. Furthermore, NIKA2 capabilities for the
characterization of high-redshift {\lp ($0.5<z<0.9$)} galaxy clusters
have been verified using numerical simulation~\citep{Ruppin2019}, and
their implication for cosmology has been discussed
in \citet{Ruppin2019b}.

On the other hand, in-depth mapping of large extragalactic fields with
sub-arcminute resolution with NIKA2 will provide unprecedented insight
on the distant universe. 
{\lp Indeed, the high-angular resolution of NIKA2 is key to reduce the
confusion noise, which is the ultimate limit of single-dish cosmological
surveys~\citep{Bethermin2017_simu}, and the high mapping speed allows
to cover large area.}
{\rev This unique combination will result in detecting hundreds of
dust-obscured optically-faint galaxies up very high redshift
($z \sim 6$) during their major episodes of star formation.}
This will help quantifying the star formation up to $z \sim 6$.
{\rev Note that galaxy redshifts will have to be obtained with
spectroscopic follow-up observations (e.g. with NOEMA) or
multi-wavelength spectral energy distribution fittings
(e.g. in the GOODS-N field thanks to the tremendous amount of
ancillary data). Galaxy formation studies will also
benefit from the large instantaneous field-of-view, high-resolution
observations of NIKA2.}

Indeed, hundreds of dust-obscured optically-faint galaxies will be
detected up to very high redshift ($z \sim 6$) during their major
episodes of star formation. The high-angular resolution of NIKA2 is
key to reduce the confusion noise, which is the ultimate limit of
single-dish cosmological surveys~\citep{Bethermin2017_simu}, and the
high mapping speed allows to cover large area. This unique combination
will help quantifying the star formation history up to $z \sim
6$. Note that galaxy redshifts will have to be obtained with
spectroscopic follow-up observations (e.g. with NOEMA) or
multi-wavelength spectral energy distribution fittings (e.g. in the
GOODS-N field thanks to the tremendous amount of ancillary data).

The current generation of sub-arcminute resolution experiments also
include the Large APEX Bolometer Camera
(LABOCA~\citep{Siringo2009_LABOCA}) at the Atacama
Pathfinder Experiment (APEX) 12-meter telescope, which covers a
12' diameter FOV at $345\, \rm{GHz}$ {\lp at an angular resolution of about
19''}; AzTEC at the 50-meter Large Millimeter Telescope, which operates with a
single bandpass centred at either 143, 217 or
$270\,\rm{GHz}$~\citep{Wilson2008_AZTEC}, {\lp and which has a beam
FWHM of 5, 10 or $18''$, respectively}; the Submillimeter Common User Bolometer
Array Two (SCUBA-2~\citep{Holland2013_SCUBA2,Dempsey2013_SCUBA2}) on the
15-meter James Clerk Maxwell Telescope, which simultaneously
images a FOV of about 7' at $353\,\rm{GHz}$ and $666\,\rm{GHz}$ {\lp with a main beam FWHM of 13''
and 8'' in the two frequency channels, respectively}
; MUSTANG-2 at the 100-meter Green Bank telescope,
which maps a 4.35' FOV at 
$90\,\rm{GHz}$ {\lp with 9'' resolution}~\citep{Dicker2014_MUSTANG2, Stanchfield2016_MUSTANG2}.
Therefore, NIKA2 is unique
in combining an angular resolution better than 20'', an instantaneous FOV of a
diameter of 6.5' and multi-band observation capabilities at $150$ and
$260\,\rm{GHz}$.

Most of the other millimetre instruments consist of bolometric cameras. By contrast,
NIKA2 is based on the Kinetic Inductance Detectors (KID)
technology~\citep{Day2003, Doyle2008_LEKID, Shu2018_LEKID}. This concept has been
first demonstrated with a pathfinder instrument,
NIKA~\citep{Monfardini2010_NIKA, Monfardini2011_NIKA}.
Installed at the IRAM 30-m telescope until 2015, NIKA demonstrated
state-of-the-art performance~\citep{Catalano2014}, and obtained
breakthrough results
(see \emph{e.g.},~\citet{Adam2014, Adam2017_kSZ}.
NIKA has been crucial in optimizing the NIKA2 instrument and data
analysis. 

A thorough description of the NIKA2 instrument is presented in \citet{Adam2018},
along with the results of the commissioning in intensity based on the
data acquired during the two technical campaigns of 2017.
{\lp In the present paper, we propose a standard calibration method, which is
referred to as the \baseline\ calibration, to go from raw data to
stable and accurate flux density measurements.} 
amount of data acquired between January 2017 and February 2018.
Regarding the performance of the polarization capabilities, their
assessment will be addressed in a forthcoming paper.
To achieve a reliable and high-accuracy estimation
of the performance, we perform extensive testing of the
stability with respect to both the analysis methodological
choices and to the observing conditions.
{\lp First, the methodological choices and hypothesis may have an impact on
the performance results and the systematic errors. At each step of the
calibration procedure and for each performance metrics, we compare the
results obtained using the \baseline\ method to alternative approaches to
ensure the robustness against systematic effects.}
Second, we check the stability of the results using a large number of
independent data sets corresponding to various observing conditions.
Specifically, most of the performance assessment relies on data
acquired during the February 2017 technical campaign (N2R9) and the
October 2017 (N2R12) and January 2018 (N2R14) first and second
scientific-purpose observation campaigns. {\lp These observation
campaigns are referred to as the \emph{reference observation campaigns}.}
Each campaign consists of about 1300 observation scans lasting between
two and twenty minutes for a total observation time of about 150 hours. 

This paper constitutes a review of NIKA2 calibration and
performance assessment in intensity. It is intended to be a reference
for observations with NIKA2, which will last at least for ten years. 
The outline of the paper is as follows:
Sect.~\ref{se:instru}~to~\ref{se:dataproc} give short summaries of the
instrumental set up, the observational modes and the data analysis methods
that have been used for the calibration and the performance
assessment. Sect.~\ref{se:geometry}~to~\ref{se:sensitivity} detail the
dedicated calibration methods, extract the key characteristic results
and discuss their accuracy and robustness. The field-of-view
reconstruction and the KID selection for science purpose are discussed
in Sect.~\ref{se:geometry}. The beam pattern is characterized in
Sect.~\ref{se:beam}, along with the main beam
full-width at half maximum and the beam
efficiency. Sect.~\ref{se:opacity} is dedicated to the derivation of
the atmospheric opacity. The methods that we have proposed to
calibrate are gathered in Sect.~\ref{se:calibration}, while
Sect.~\ref{se:photometry} presents the validation of these methods and
the calibration accuracy and stability assessment. The noise
characteristics and the sensitivity are discussed in
Sect.~\ref{se:sensitivity}. Finally, Sect.~\ref{se:summary} summarizes
the main measured performance characteristics and {\lp briefly
describes next steps for future improvements on NIKA2.}

\section{General view of the instrument}
\label{se:instru}

NIKA2 simultaneously images a FOV of
6.5' in diameter at 150 and $260\,\rm{GHz}$. It also has polarimetry
capabilities at  $260\,\rm{GHz}$, which are not discussed here. 
To cover the 6.5'-diameter FOV without degrading the
telescope angular resolution, NIKA2 employs a total of around
2\,900\,KIDs split over three distinct arrays, one for the $150\,\rm{GHz}$
band and two for the $260\,\rm{GHz}$ band.

A detailed description of the instrument can be found in
\citet{Adam2018} and \citet{Calvo2016JLTP}. We briefly present here the main sub-systems
focusing on the elements that are specific to NIKA2
or that drive the development of dedicated procedures for the data
reduction or calibration.

\subsection{Cryogenics}

{\lp KIDs are superconducting detectors, which in the case of NIKA2 are made of
thin-aluminium films deposited on a silicon substrate~\citep{Roesch2012_LEKID}.
For an optimal sensitivity, they must operate at a temperature of
around 150\,mK, that is roughly one order of magnitude lower than the
aluminium superconducting transition temperature.}  
For this reason,
NIKA2 employs a custom-built dilution fridge to cool down the focal plane, and the
refractive elements of the optics. Overall, a total mass of around
$100\, \rm{kg}$ is kept deeply in the sub-Kelvin regime. Despite the complexity
and huge size of the system, the operation of NIKA2 does not require
external cryogenic liquids and can be fully operated remotely.

\subsection{Optics}
\label{se:instru_optics}
The NIKA2 camera optics include two cold mirrors and six lenses. The
filtering of unwanted (off-band) radiation is provided by a suitable stack of
multi-mesh filters {\lp as thermal blockers} placed at all temperature
stages between 150\,mK and room temperature. {\lp Two aperture stops,
at a temperature of 150\,mK, are conservatively designed to limit the
entrance pupil of the optical system to the inner 27.5\,m diameter of
the primary mirror. }
An air-gap dichroic plate splits the 150\,GHz (reflection)
from the 260\,GHz (transmission) beams. {\lp This element, which is
made of a series of thin micron-like membranes separated by calibrated
rings and mounted on a native ring in stainless steel, has been designed to
resist to low temperature-induced deformation.}
As discussed in Sect.~\ref{se:flat_field}, the air-gap technology has
been proven to be efficient
in preserving the planarity of the dichroic, but shows sub-optimal
performance in transmission. Moreover, a grid polariser ensures the
separation of the vertical and horizontal components of the linear
polarizations on the 260\,GHz channel. Band-defining filters,
custom-designed to optimally match the atmospheric windows while
ensuring robustness against average atmospheric condition at 260\,GHz,
are installed in front of each array. A half-wave polarization
modulator is added at room temperature when operating the instrument
in polarimetry mode.

Hereafter, the detector array illuminated by the 150\,GHz
($2\, \rm{mm}$) beam is named Array 2 (A2), 
while in the 260\,GHz ($1.15\, \rm{mm}$) channel, the array mapping the
horizontal component of the polarization is referred to as Array 1 (A1)
and the one mapping the vertical component is called Array 3 (A3). The
150\,GHz observing channel is referred to as the $2\, \rm{mm}$ band
and the 260\,GHz channel as the $1\, \rm{mm}$ band. 

\subsection{KIDs and electronics}
\label{se:array}

Array 2 consists of 616\,KIDs, arranged to cover a 78\,mm diameter
circle. Each pixel has a size of $2.8\times2.8\textrm{\,mm}^2$, which
is the maximum pixel size allowed not to degrade the theoretical
\trentemetre\ telescope angular resolution. In the
case of the 260\,GHz band detectors, the pixel size is $2\times
2\mathrm{\,mm}^2$, to ensure a comparable sampling of the focal
plane. {\rev This results in a sampling slightly above $\lambda/D$ in
this channel, where $D$ is the diameter aperture, as discussed in
Sect.~\ref{se:grid_distortion}.}
In order to ensure a full coverage of the 6.5' FOV, a total of
1,140 pixels is needed in each of the two 260\,GHz arrays A1 and A3. 

The key advantage of the KID technology is the simplicity of the cold
electronics and the multiplexing scheme. In NIKA2, each block of around 150
detectors is connected to single coaxial line linked to the readout
electronics~\citep{Bourrion2016}. {\lp Hence Array 2 is connected to four
different readout feed-lines, while Array 1$\&$3 are both equipped with eight
feed-lines.}
The warm electronics required to digitize
and process the pixels signals is composed of twenty custom-built readout
cards (one per feed-line).

\subsection{KID photometry and {\tt tuning}}
\label{se:tuning}

KIDs are superconducting resonators whose resonance
frequency shift linearly depends on the incoming optical power.
{\lp This was theoretically demonstrated in~\citet{Swenson2010} and
confirmed for NIKA KIDs, which have a similar design to the current
NIKA2 KIDs, using laboratory measurements, as discussed
in~\citet{Monfardini2014JLTP}.}
The measurement of the KID frequency shift $\Delta f$ is critical for the use of
KIDs as mm-wave detectors. 

For the KID readout, an excitation signal is sent into the cryostat on the
feed-line coupled to the KID.
The excitation tones produced by the electronics are amplified by a
cryogenic (4\,K) low noise amplifier after passing through the KIDs and
being analysed by the readout electronics. {\lp Each KID is thus
associated with an excitation tone at a frequency $f_{\rm{tone}}$, which
corresponds to an estimate of its resonance frequency for a reference
background optical load.}
The transmitted signal can be described by its
amplitude and phase, or, as is common practice for KID, by its $I$
(in-phase) and $Q$ (quadrature) components
with respect to the excitation signal.
The goal is now to relate the measured variations of the KID response
to the excitation signal $(\Delta I, \Delta Q)$, which are induced by incident light, to
$\Delta f$. For this, the electronics modulates the excitation tone
frequency $f_{\rm{tone}}$ at about 1\,kHz with a known frequency variation $\delta f$
and the read out gives the induced transmitted signal variations
$(dI, dQ)$. Projecting linearly $(\Delta I, \Delta Q)$ on $(dI, dQ)$ therefore
provides $\Delta f$. This quantity, in Hz, constitutes the raw KID
time-ordered data, which are sampled at a frequency of
$23.84\,\rm{Hz}$. For historical reasons, this way of deriving KID
signals has been nicknamed \emph{RfdIdQ}. More details on this process
are given in \citet{Calvo2013}.
Once ingested into the calibration pipeline, the raw data will be further converted
into astronomical units (Sect.~\ref{se:calibration}).\\

In addition to light of astronomical origin, any change in the
background optical load (due, for example, to changes in
the atmospheric emission with elevation) contributes as well to
the shift of the KID resonance frequencies. In
order to maximize the sensitivity of a KID, the excitation signal $f_{\rm{tone}}$
must always be near the KID resonance frequency. We therefore have
developed a tuning algorithm that performs this optimization. A {\tt tuning} is performed at the beginning
of each observation scan to adapt the KIDs $f_{\rm{tone}}$ to the working background
conditions.
This process takes only a few seconds.
These optimal conditions are further maintained by performing
continuous {\tt tunings} between two scans while NIKA2 is not observing, to
match regularly with the observing conditions.

\subsection{Bandpasses}
\label{se:instru_bandpass}

\begin{figure}[ht!] 
\begin{center}
\includegraphics[clip,trim={0, 1cm, 0, 2cm},width=0.5\textwidth]{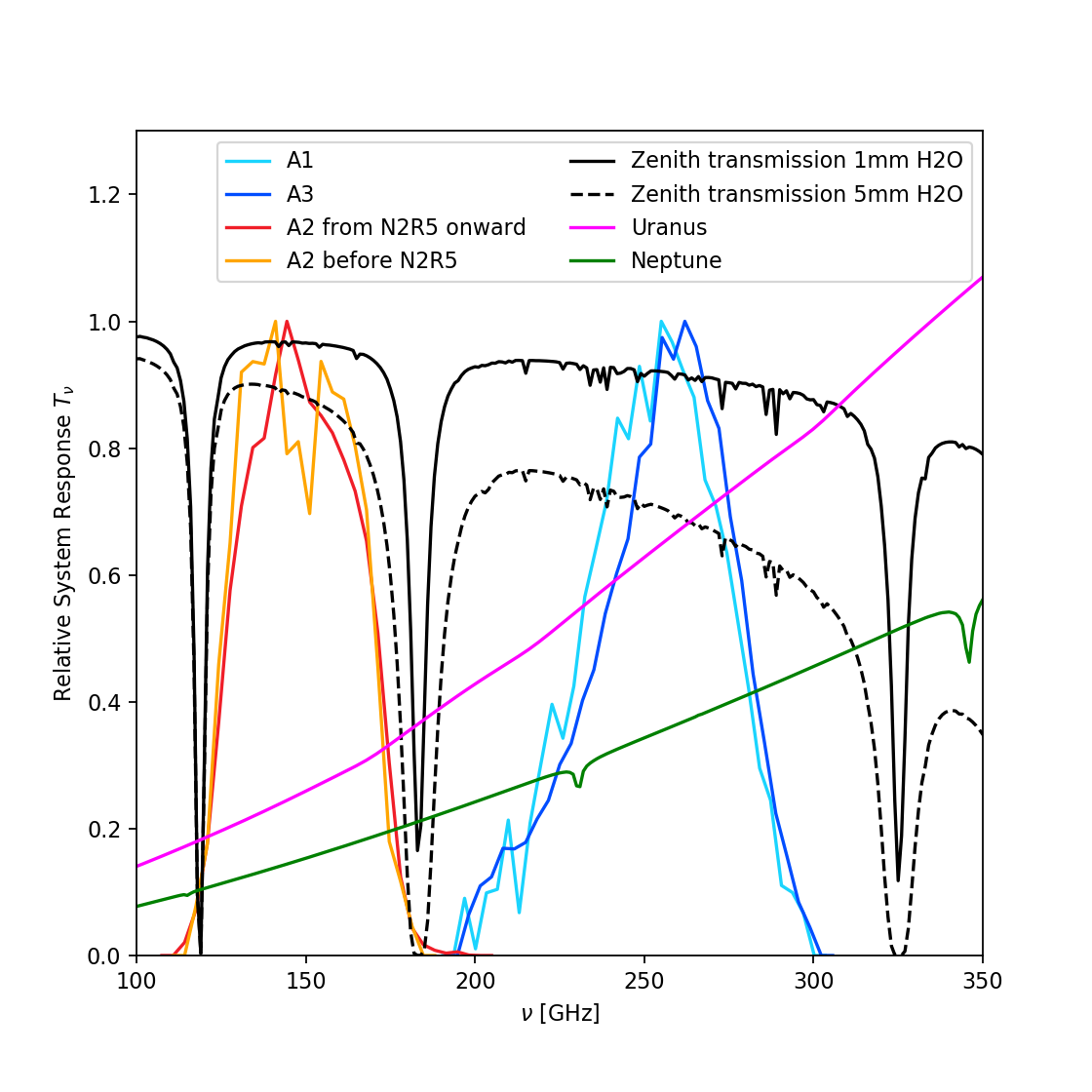}
\caption[NIKA2 transmission]{Relative system response of the three KID
  arrays as a
  function of frequency. For illustration we also show in black
  the atmospheric transmission obtained with the ATM model \citep{ATM,
    Pardo2001} for two values of precipitable water vapour (1 and
  $5\,\rm{mm}$).
  The spectra of the model of Uranus (pink) and the model of
  Neptune
  (green), as discussed in Appendix~\ref{ap:ref_flux_calibrator}, are
  also plotted for illustration with arbitrary
  normalization with respect to the NIKA2 transmission.} 
 \label{spectralband1}
\end{center}
\end{figure}

The NIKA2 spectral bands have been measured in laboratory using a
Martin-Puplett interferometer built in-house~\citep{Durand2007_these}.
The measurement relies on using the difference of two black
body radiations used as input signal for the interferometer. 
{\lp Both arrays and filter bands were considered in the
measurements, while the dichroic element was not included.}
Figure~\ref{spectralband1} shows the relative spectral response for
the three arrays. Notice that Array 2 was
upgraded in September 2016 (during the so-called N2R5 technical campaign) and that
the spectral transmissions are slightly different (red and orange lines in
Fig.~\ref{spectralband1}).

The two arrays operating at 260 GHz, mapping different linear polarisations,
exhibit a slightly different spectral behaviour as can be
seen on Fig.~\ref{spectralband1}. {\lp Besides the effect of the
optical elements in front of the two arrays, this may be explained by
a tiny difference in the silicon wafer, the difference of the
Aluminium film thickness of the KID arrays and/or the different
responses for two polarisations of the
detector~\citep{Adam2018, Shu2018_bandpass}.}

It is clear from Fig.~\ref{spectralband1} that the atmosphere will
modify the overall transmission of the system, especially at the
transmission tails for A2. {\lp To highlight this effect we compute an effective
frequency $\nu_{\rm{eff}}$ computed as the weighted integral of the
frequency considering the NIKA2 bandpass and the SED of Uranus, for
various atmospheric conditions.}
In Table~\ref{tab:frequencies}, we give $\nu_{\rm{eff}}$ and the
bandwidths $\Delta \nu$ computed both at zero atmospheric opacity and for the
reference IRAM \trentemetre\ winter-semester observing conditions
(defined by an atmospheric content of $2\,\rm{mm}$ of precipitable
water vapour and an observing elevation of $60^o$).

\begin{table}[!htbp]
  \caption[]{Effective frequencies $\nu_{\rm{eff}}$ and bandwidths
    $\Delta \nu$
    of the three arrays computed in two different observing conditions
    defined with the precipitable water vapour (pwv) contents in mm and
    the observing elevation in degrees).}
  \label{tab:frequencies}
  \centering    
  \begin{tabular}{lrrr}
    \hline\hline
    \noalign{\smallskip}
    & Array 1 & Array 3 & Array 2 \\
    \noalign{\smallskip}
    \hline
    \noalign{\smallskip}
    $\nu_{\rm{eff}}$ \small{(0mm, 90deg)} [GHz] & 254.7 & 257.4 &  150.9 \\
    $\Delta \nu$ \small{(0mm, 90deg)} [GHz] &  49.2 & 48.0  &   40.7 \\
    $\nu_{\rm{eff}}$ \small{(2mm, 60deg)} [GHz] & 254.2 & 257.1 &  150.6 \\
    $\Delta \nu$ \small{(2mm, 60deg)} [GHz] &  48.7 &  47.9 &    39.2 \\
    \noalign{\smallskip}
    \hline
  \end{tabular}
\end{table}

{\lp In laboratory spectral characterization allows the bandpass of the
three arrays to be measured with uncertainties better than one
percent. The bandpass characterization will be further improved using
\emph{in-situ} measurements with a new Martin-Puplett interferometer designed
to be placed in front of the cryostat window, which will allow to
account for the whole optical system, including the dichroic plate. As
it will be discussed in Sect.~\ref{se:calibration}, the \emph{baseline} calibration method
only resorts to the bandpass measurements for color correction
estimations in order to mitigate the bandpass uncertainty propagation
in the flux density measurement.}
In fact, for each array, we define reference frequencies 
to define NIKA2 photometric system. These are 260~GHz for the A1 and A3 and
150~GHz for A2.

\section{Observations}
\label{se:observations}

This section presents the different observation modes that are used at
the IRAM \trentemetre\ telescope for both commissioning and
scientific-purpose observations with NIKA2. {\lp Each observation
campaign is organized as observing pool allowing to optimize
observations of several science targets in a flexible way.}
Most of these observing scans are on-the-fly (OTF) raster scans,
which consist of a series of scans at constant azimuth or elevation
(right ascension or declination) and varying elevation or azimuth
(declination or right ascension).
Their characteristics have been tailored for NIKA2 performance.

\subsection{Focus}
\label{se:axial_focus}

Observation pools start with setting the telescope focus since NIKA2 large
FOV alleviates the need to adjust the pointing beforehand.  
We have designed a specific focus procedure that takes
advantage of the dense sampling of the FOV allowing to map a source
in a short integration time. We perform a series of five successive one-minute
raster scans of a bright (above a few Jy) point source at five
axial offsets of the secondary mirror (M2) along the optical
axis. As the scan size is $1'\times 5'$, the main contribution to each
map mainly comes from the KIDs located in the central part of the FOV.

Elliptical Gaussian fits on the five maps provide estimates of
the flux and FWHM along minor and major axes for each focus. 
The best axial focus in the central part of the array is then
estimated as the maximum of the flux or the minimum of the FWHM using
parabolic fits of the five measurements.

As presented in more details in Appendix~\ref{ap:focus_surfaces}, the focus
surface, that is defined as the locations of the best focus across the whole FOV
is not flat but rather slightly bowl-shaped.
To account for the curvature of the focus surfaces and optimize the
average focus across the FOV, we add -0.2\,mm to the best axial focus
in the central part of the array. This focus offset is measured on data using
a dedicated sequence of de-focused scans, as discussed in
Appendix~\ref{ap:focus_surfaces}. It is in agreement with expectations
derived with optical simulation using ZEMAX\footnote{Web site: \tt{www.zemax.com}}. 

{\lp Axial focus offsets are measured every other hour during daytime and
are systematically checked after sunrises and sunsets, while one or
two checks suffice during night. 
Lateral focus offsets can also be checked in a similar way, but are
found to stay constant over periods of time that cover several
observing campaigns.}

\subsection{Pointing}
\label{se:pointing}

Once the instrument is correctly focused, we can estimate pointing corrections
before scientific observations.
Based on general operating experience at the \trentemetre\ telescope, we use the so-called
{\tt pointing} or {\tt cross-type} scans to monitor the pointing during observations. The
telescope executes a back and forth scan in azimuth and a back and forth scan in
elevation, centred on the observed source. We fit Gaussian profiles
from the timelines of the reference detector, which is chosen as a
valid detector located close to the center of Array 2. {The choice of
a reference detector operated at $150\,\rm{GHz}$ is suitable since most
of our pointing sources are radio sources.} We use the
estimated position of the reference detector to derive the current pointing
offsets of the system in azimuth and elevation. This correction is
propagated to the following scans. The pointing is monitored in an
hourly basis.

{\lp In addition, we perform \emph{pointing sessions} in order to
refine NIKA2 pointing model. A pointing session consists in observing
about 30 sources on a wide range of elevations and azimuth angles while
monitoring the pointing offsets that are measured for each
observation. During N2R9 technical campaign, the rms of
the residual scatter after pointing offset correction was $1.62''$ in
azimuth and $1.37''$ in elevation. We conservatively report rms pointing errors $<3''$.}

\subsection{Skydip}
\label{se:skydip}

A skydip scan consists in a step-by-step sky scan along a large range
of elevations.
{\lp NIKA2 skydips are not used for the scan-to-scan atmospheric 
calibration. For this purpose, the KIDs are used as total power
detectors to estimate the emission of the atmosphere and hence, the
atmospheric opacity, as discussed in Sect.~\ref{se:opacity}. 
NIKA2 skydips therefore serve to calibrate the KID responses with
respect to the atmospheric background for atmospheric opacity
derivation.}

Unlike heterodyne receivers for which skydips can be
conducted continuously slewing the telescope in elevation, the NIKA2
camera cannot resort to such method, as the KIDs need to be retuned
for a given \airmass. A NIKA2 skydip, which is quoted {\tt skydip},
comprises eleven steps in the elevation range from 19 to 65 degrees,
regularly spaced in \airmass. For each step, we acquire about twenty
seconds of data to ensure a precise measurements. KIDs are tuned at
the beginning of each constant elevation sub-scan (hence once per
\airmass).

{\tt Skydips} {\lp are typically performed every eight hours for a wide spanning
of the atmospheric conditions through an observation campaign.}

\subsection{Beam map}
\label{se:beammaps}

A \bm\ scan is a raster scan in ($\az$, $\elev$) coordinates tailored to map a
bright compact source, often a planet, with steps of 4.8'', that are
small enough
{\rev to ensure a half-beam sampling, which gives around 90\% fidelity, for
each KID.} 
A scan of 
$13\times7.8$~arcmin$^2$ is acquired either with the telescope
performing a series of continuous slews at fixed elevation or at fixed azimuth. 
A continuous scanning slew defines a subscan. 
The fixed-elevation scanning has the advantage of suppressing the air-mass variation
across a subscan, while the fixed-azimuth scanning offers an
orthogonal scan direction to the former:
the combination of both gives a more accurate determination of the far side
lobes.
The scan size is optimized to enable maps to be made for all
KIDs, even those located at the edges of the array. Larger size in the scanning
direction allows for correlated noise subtraction.
During subscans, the telescope moves at
65\,arcsec/s.
{\rev The need to have high-fidelity sampling of 11'' beams 
along the scan direction translates into a maximum speed of
97\,arcsec/s, which ensures to have 2.7 samples per beam, for our nominal
acquisition rate of 23.8\,Hz,} and is thus met with margins. For the
sake of scanning efficiency with the \trentemetre\ telescope, the minimal duration for
subscans is 10\,s. For \bm\ scans,
subscans last 12\,s and the entire scan lasts about 25\,min.

{\tt Beammaps} are key observations for the calibration. {\lp Whereas
a single \bm\ acquired in stable observing conditions could suffice,
\bm\ scans are performed on a daily basis.}
More details on these observations are given in Sect.~\ref{se:geometry}
where we describe how to actually exploit them to derive individual KID
properties.

\section{Data Reduction: from raw data to flux density maps}
\label{se:dataproc}

The raw KID data ($I$, $Q$, $\Delta I$, $\Delta Q$) and the telescope
source-tracking data are synchronized by the NIKA2 acquisition system using a
clock that gives the absolute astronomical time, that is the telescope
pulses per second, to define the NIKA2 raw data. From the KID raw
data, we compute a quantity that is proportional to the KID
frequency shift using the \emph{RfdIdQ} method, as described in
Sect.~\ref{se:tuning}. This quantity, which is hence proportional to
the input signal, constitutes the KID time-ordered information (TOI).

We have developed a dedicated data reduction pipeline to
produce calibrated sky maps from NIKA2 raw data. This pipeline was first 
developed for the data analysis during the commissioning campaigns and
is currently used for science-purpose data reduction. The calibration
and performance assessment relies on this pipeline. 
A detailed description of this software will be presented in a companion
paper \citep{Ponthieu2019}, as well as an application to blind source
detection. Here we summarize the main steps of the data
reduction. {\lp Moreover, we focus the discussion on the treatment of
point-like or compact sources, which are used for the
performance assessment.}

\subsection{Low level processing}
\label{se:ll_proc}
We isolate the relevant fraction of the data for scientific
utilisation and {\lp we mask KIDs that do not meet the selection criteria, as
discussed in Sect.~\ref{se:avg_kidpar},} or
timeline accidents (glitches). Specifically, we flag out cosmic rays,
which impact only one data sample per hit due to the KID fast time
constant~\citep{Catalano2014},
and the KIDs for which the noise level exceed $3\sigma$ of the average
noise level of all other KIDs of the same array.  

\subsection{Pointing reconstruction}
\label{se:ptg}
We produce a timeline of the pointing positions of each KID with
respect to the targeted source position (usually located at the center of the
scan) using two sets of information. First, the control system of the
telescope provides us with the absolute pointing of a reference point
of the focal point, which is coincident with the reference KID after
the pointing correction are applied, as described in
Sect.~\ref{se:pointing}. Second, we estimate the offset positions of
each KID with respect to the reference KID using a dedicated
procedure that is referred to as the focal plane reconstruction, as
presented in Sect.~\ref{se:geometry}. After this step, we are able to
distinguish KIDs that are on-source from those that are off-source,
which is a key information for dealing with the correlated noise. 
  
\subsection{TOI calibration}
\label{se:flux_calib}
The KID TOI in units of Hz (frequency shifts) are converted to
Jy/beam in two steps. First, the KID data are inter-calibrated using the
calibration coefficients, \aka\ relative gains, 
as discussed in Sect.~\ref{se:flat_field} and the
absolute scale of the flux density is set using the absolute
calibration method discussed in
Sect.~\ref{se:calibration_method}. Second, the instantaneous line-of-sight
atmospheric attenuation $\rm{exp}[{-\taunu x(t)}]$ is corrected
using the zenith opacity at the observing frequency $\nu$ $\taunu$, which is estimated for a given scan
as discussed in Sect.~\ref{se:opacity}, and the instantaneous \airmass\
$x(t)$. {\lp The latter is estimated as $(\sin \elev_t)^{-1}$ using the observing elevation
$\elev_t$, as obtained using the pointing reconstruction.}     

\subsection{Correlated noise subtraction}
\label{se:toi_proc}
The TOI of each KID include a prominent low-frequency component of correlated noise of two
different origins: the atmospheric component, which is dominant
{\lp with some rare exceptions} and common to all KIDs, and the electronic noise, which is common to the KIDs connected
to a same electronic readout feed-line (see
Sect.~\ref{se:array}). The subtraction of this correlated noise is a key
step of the data processing, as correlated noise residuals are an
important limiting factor of the sensitivity. We have devised several
dedicated methods for this purpose. This will be thoroughly
discussed in \citet{Ponthieu2019}, whereas here we illustrate the
general principle and only discuss the method routinely used for
the calibration.

\begin{figure}[ht!]
\begin{center}
\includegraphics[clip, angle=0, scale=0.4]{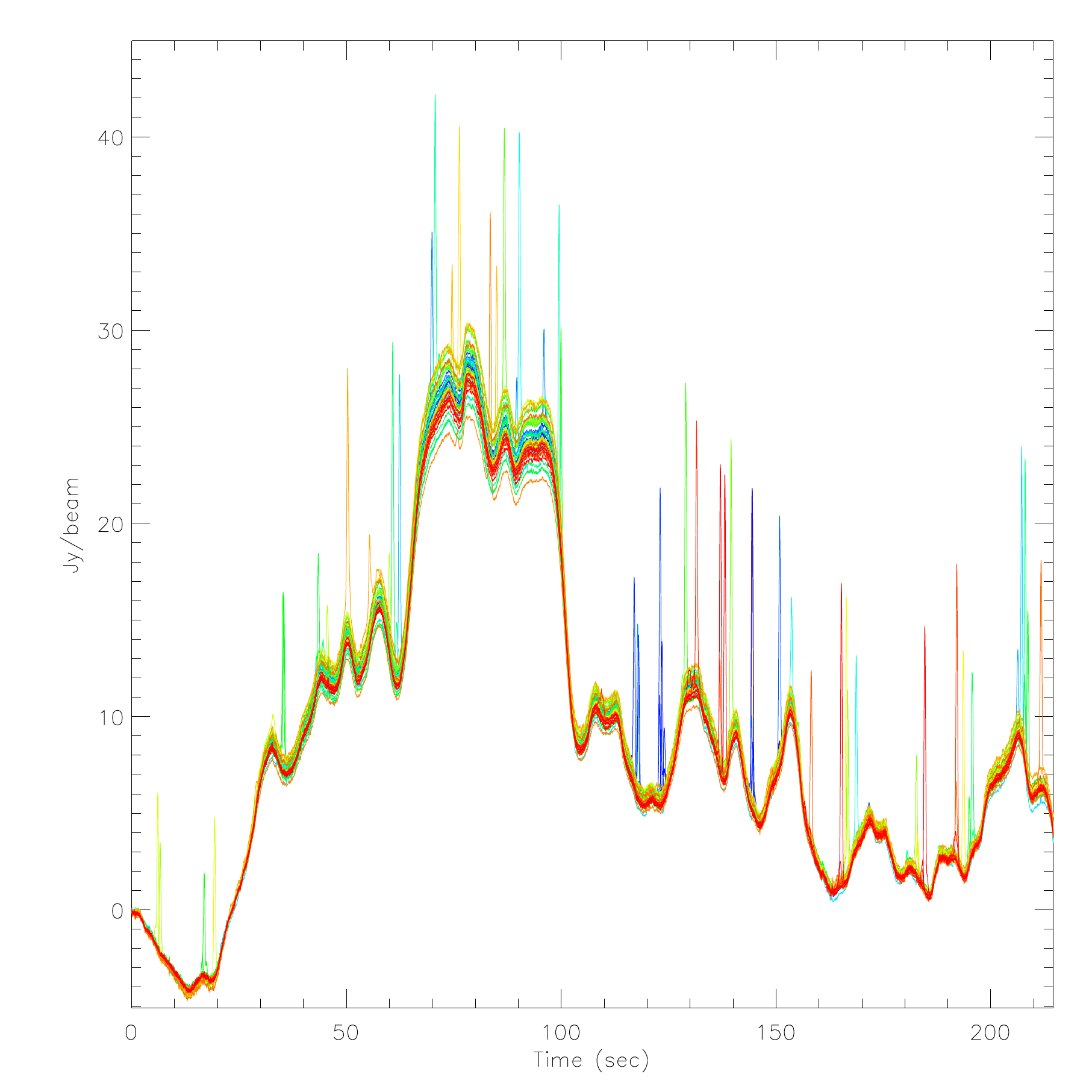}
\includegraphics[clip, angle=0, scale=0.4]{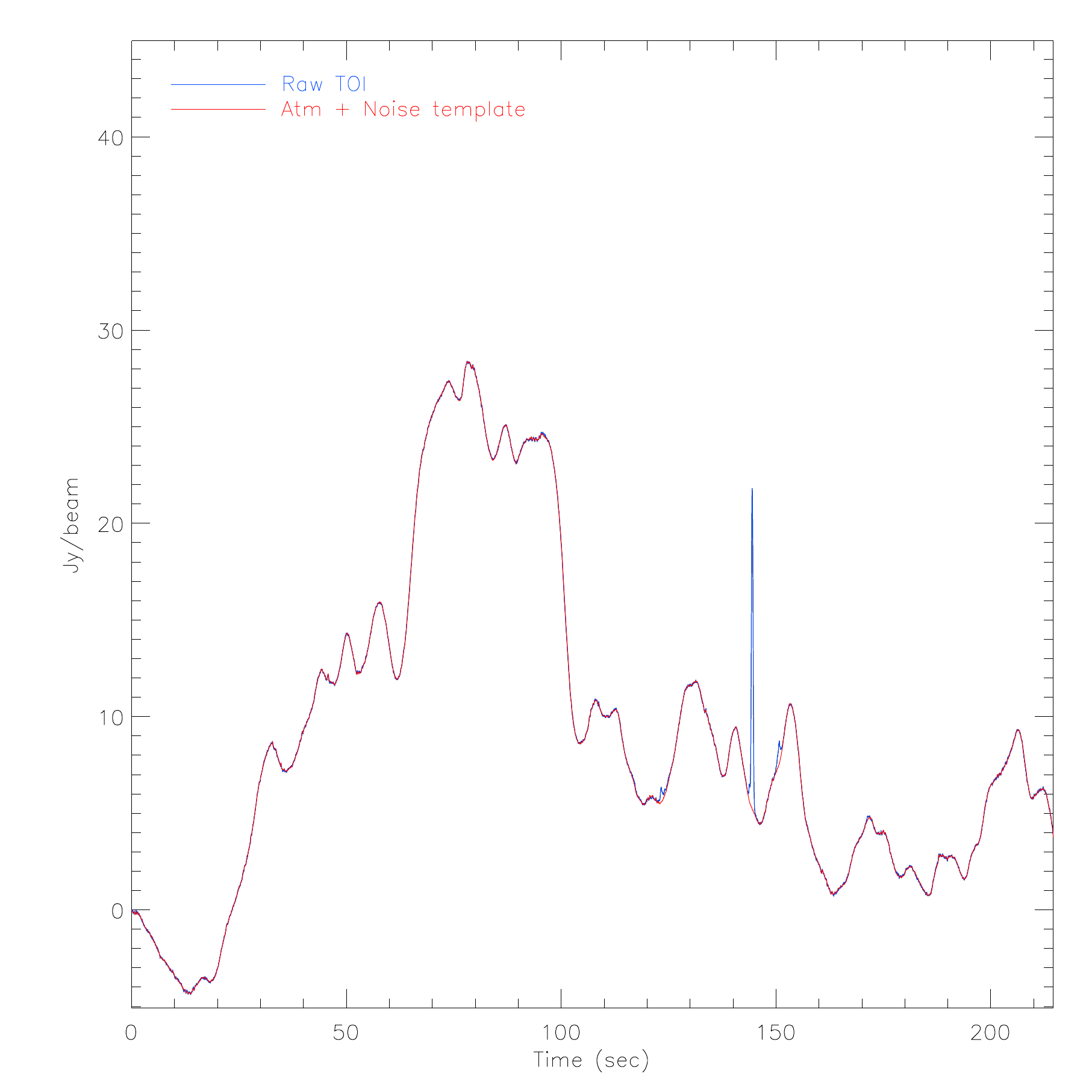}
\caption[Example of Time-Ordered-Information]{Example of variations of KID
  time-ordered information. \emph{Top:} Example of 40 KID {\lp calibrated} TOIs during an observation
  of Uranus. The low frequency correlated component (atmospheric and electronic
  noises) is clearly seen. \emph{Bottom:} One of these TOIs (in blue) and the
  \cm\ that is subtracted from it (in red). {\lp The zero level is arbitrary.}}
\label{fig:nika_toi}
\end{center}
\end{figure}

As illustrated on the upper panel of Fig.~\ref{fig:nika_toi}, the
low-frequency noise component is seen by all KIDs at the same time,
while the astrophysical signal {\lp (Uranus in this case)} is shifted
from one KID to another.
{\lp In this figure the KID TOIs have been rescaled to be null
at $t=0$.} 
A simple average of the KID TOIs provides an
estimate of the low-frequency noise component, that we referred to as
a \emph{common mode}, while the signal is averaged out. The common
mode shown as a red line on the lower panel of
Fig.~\ref{fig:nika_toi}, is then subtracted to each KID TOI.

\begin{figure}[ht!] 
  \begin{center}
    \begin{overpic}[clip=true, trim={0.5cm, 0, 0, 0.5cm},width=0.40\textwidth]{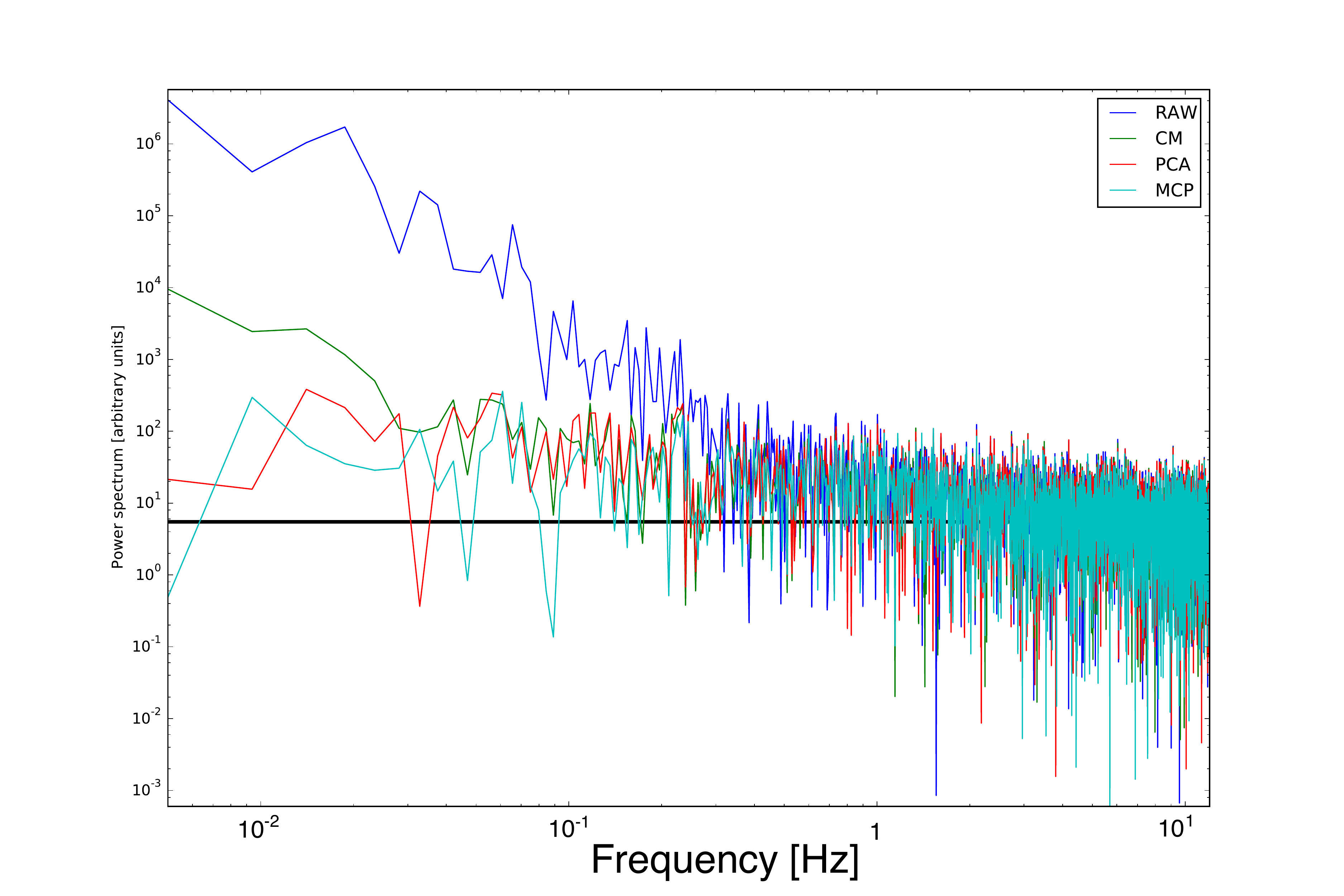}
      \put(2,15){\rotatebox{90}{\scriptsize P$_{\nu}$ [arbitrary units]}}
  \end{overpic}
    \begin{overpic}[clip=true, trim={0.5cm, 0, 0, 0.5cm},width=0.40\textwidth]{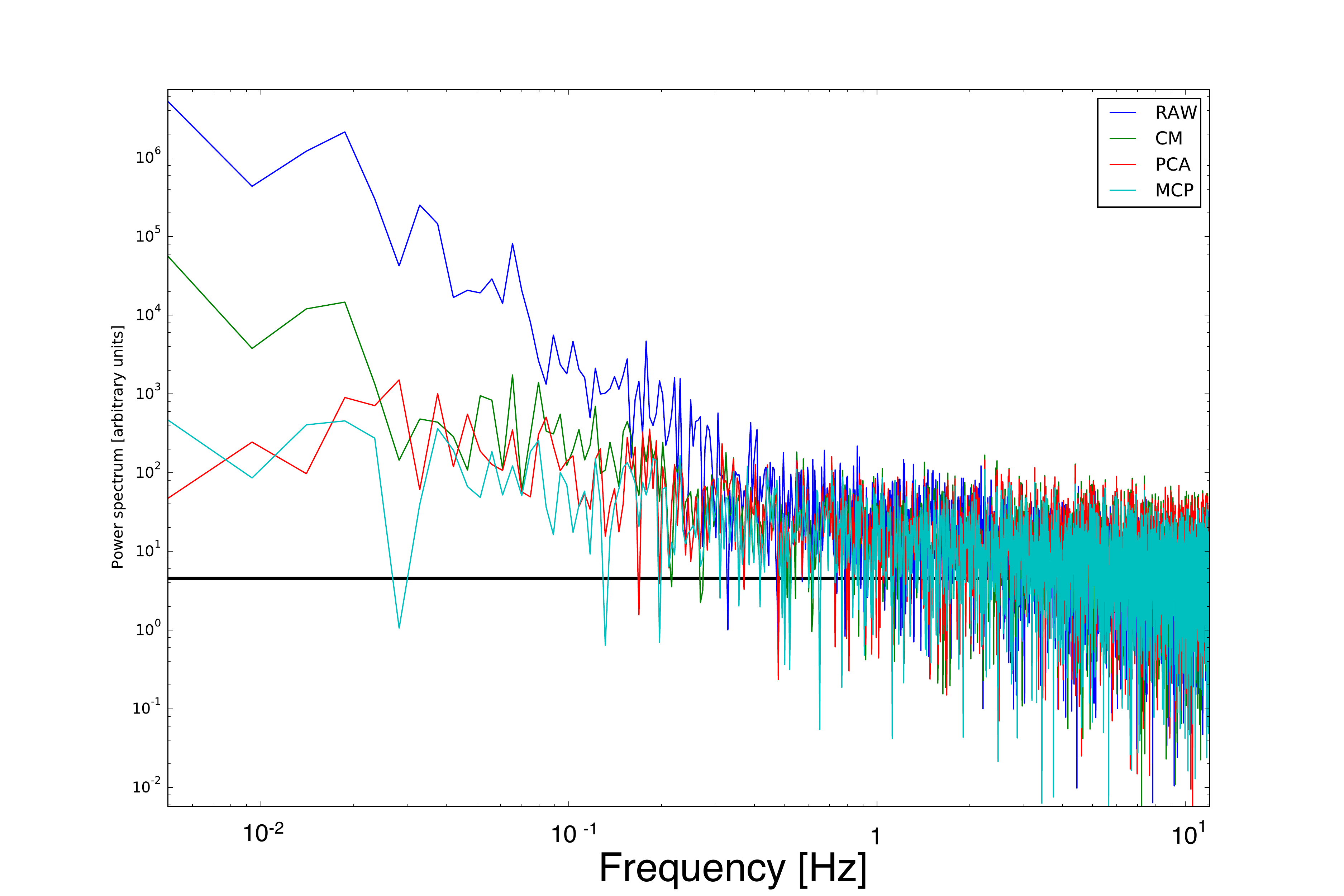}
      \put(2,15){\rotatebox{90}{\scriptsize P$_{\nu}$ [arbitrary units]}}
    \end{overpic}
    \begin{overpic}[clip=true, trim={0.5cm, 0, 0, 0.5cm},width=0.40\textwidth]{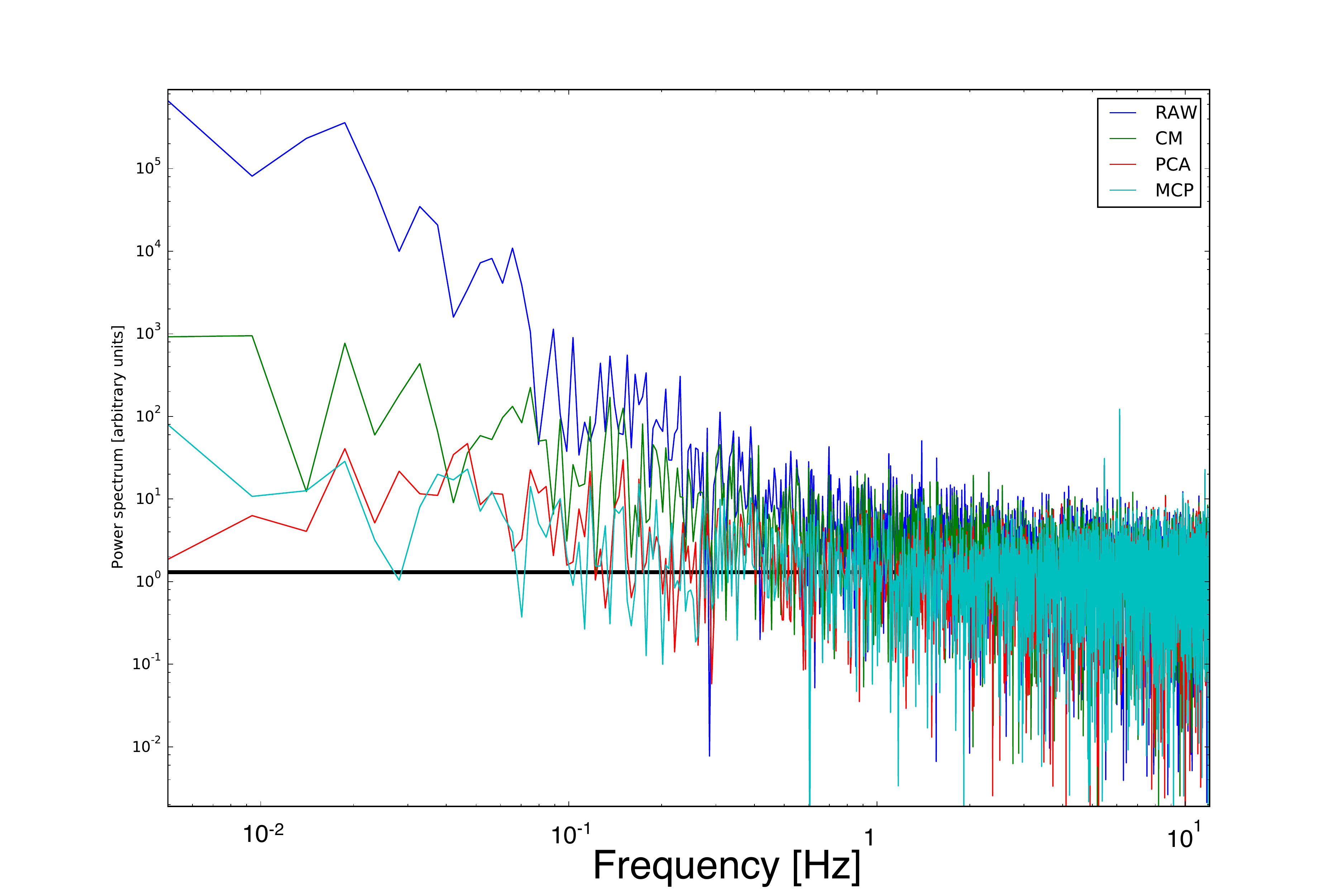}
      \put(2,15){\rotatebox{90}{\scriptsize P$_{\nu}$ [arbitrary units]}}
    \end{overpic}
  \end{center}
\caption[Noise power spectra]{
  The data noise power spectra are shown for the three NIKA2 arrays (A1, A3, and
  A2 from top to bottom). 
  The power spectra are given for the raw
  data (blue), and for noise decorrelated data using the common mode
  (labelled CM, green), the PCA (red) and the \cmoneb\ (labelled MCP,
  cyan) methods.
  \label{rmspws}}
\end{figure}

{\lp For the calibration and performance assessment, we use an
atmospheric and electronic noise
decorrelation method named \cmoneb\, which comprises two
additional technicalities with respect to the common mode
method.} First, the signal contamination of the common mode estimate
is mitigated by discarding on-source KID data samples before averaging
the rescaled TOI. {\rev This is achieved by deriving a mask per TOI
from the pointing information (Sect.~\ref{se:ptg}), which is zero if
the KID is close to the source, and equal to unity otherwise. In the case of a
point source, the mask consists in a disk of a minimum radius of 60''
centred on the source, whereas for diffuse emission, tailored masks
driven by the source morphology are built using iterative methods, as
for example in~\citet{Ruppin2017}.}
Second, instead of a single common mode subtraction to
all KIDs, we estimate an accurate common mode for each
KID. Calculating the KID-to-KID cross-correlation matrix, we
identify the most correlated KIDs. Then, we build an
inverse noise weighted co-addition of the timelines of the 
KIDs that are the most correlated with the KID under
concern. Furthermore, we have tested on simulations that this method
does preserves the flux of {\lp point-like or moderately extended
sources.}

{\rev Regarding diffuse emission, the noise decorrelation induces a
filtering effect at large angular scales that must be corrected
for to fully recover the large scale signal. A method to correct for the
spatial filtering, which relies on the evaluation of the data
processing transfer function using simulations, is
described in~\citet{Adam2015}. The data processing transfer function
depends on the morphological properties of the extended source under
concern. 
An example of the data processing transfer function for
NIKA2 observation towards a galaxy cluster is given
in~\citet{Ruppin2018}, evidencing a prominent filtering at
angular scales larger than the 6.5' diameter FOV.} 

In Fig.~\ref{rmspws} we present the noise power spectra of
a typical KID TOI both before any data reduction and using three noise
decorrelation methods, which are the simple common mode (CM) method
used in Fig.~\ref{fig:nika_toi}, a method based on a principal component
analysis (PCA) and the \cmoneb\ method (MCP). We observe that after decorrelation the
$1/f$-like noise in the power spectra (principally due to atmospheric
emission drifts)
is significantly reduced leading to nearly flat spectra down to {\lp
0.5~Hz}, with lower $1/f$-like residual noise for the PCA and
the \cmoneb\ methods than the common mode decorrelation at low
frequencies. {\rev The two former methods further subtract a
substantial fraction of the correlated noise that originates from the
electronic readouts.} {\lp Moreover, we have checked using simulations that
the \cmoneb\ method was more efficient than the PCA in preserving the
astrophysical signal. The former is thus preferred over the latter.} 

\subsection{Map projection}
\label{se:map_projection}
We use the pointing information to project the cleaned (low-frequency
noise subtracted) calibrated TOI of all the valid KIDs of an array
onto a flux density map (tangential projection). This map $M_p$ is produced using an inverse
variance noise weighting of all of the data samples that fall into a map
pixel as defined using a nearest grid point scheme. We also compute
the associated count map $H_p$ defined as the number of data samples
per map pixels. The map resolution
is chosen small enough (typically $2''$ per map pixel) to alleviate
the need for more refined interpolation scheme. The noise variance
$\sigma_k$ for each KID $k$ is evaluated by the standard deviation of the
KID TOI far from the source position. 
{\lp The variance map $\sigma_p^2$ is inhomogeneous and varies as the
inverse of $H_p$. Its normalisation is evaluated using the
homogeneous background map variance, that is the
variance of $M_p\sqrt{H_p}$ calculated far from the source.}

{\lp To account for the residual
correlated noise while evaluating the variance map, we resort to an
effective approach.
First, we compute the map of the signal-to-noise ratio (SNR) as the ratio of
$M_p$ and the noise map $\sigma_p$, that is the square root of the
variance map. We observe that the distribution of the SNR map over the pixels far from the source is
well-approximated with a Gaussian but has a width larger than the
expected unity. This is due to the remaining correlations between KID TOIs
before projection. Then, we multiply the noise map 
$\sigma_p$ by the required factor so
that the width of SNR distribution becomes normalized.
This normalizing factor ranges from 1.2 to 1.5 depending on the observing conditions. This
constitutes an effective approach to account for the pixel-to-pixel
correlation matrix off-diagonal terms alleviating the need of
accurately measure them.}

When several scans of the same source are averaged, we apply an inverse
variance weighting as well. 
The variance map of the sum of scans is also corrected to ensure unity-width
SNR distribution.

\subsection{Observation scan selection}
\label{se:data_selection}

For calibration and performance assessment, we select scans in average
observing conditions by performing mild selection cuts. These scan
cuts rely on zenith opacity estimates $\taunu$ in NIKA2 bands, as
described in Sect.~\ref{se:opacity}, on the elevation and on the
observation time of the day. We select the scans satisfying the
following criteria:
\begin{itemize}
\item[i)] $\tau_{\rm{A_3}} < 0.5$, where $\tau_{\rm{A_3}}$ is the $\taunu$ estimate for
  Array 3; 
\item[ii)] $x\, \tau_{\rm{A_3}} < 0.7$ and $\elev > 20^{o}$, where
$\elev$ is the observing elevation and $x$, the
  \airmass, which depends on the elevation as $x=(\sin{\elev})^{-1}$. This
  threshold corresponds to a decrease of the astrophysical signal by a
  factor of two;
\item[iii)] observation time from 22:00 to 9:00 UT and from 10:00 to
  15:00 UT, that excludes the sunrise period and the late afternoon.
\end{itemize}
{\lp In the following sections, these selection cuts are referred to as the 
'\emph{baseline} scan selection'.}  
As discussed in Sect.~\ref{se:beam_variation}, the late afternoon
observations are often affected by time-variable broadening of the
telescope beams caused by (partial) solar irradiation of the primary
mirror and/or anomalous atmospheric refraction.
Around sunrise, the focus shifts continuously due to the ambient temperature
change until the temperature stabilizes, so that the scans taken from
9:00 to 10:00 UT are likely not to be optimally focused.
After the focus stabilisation, the middle of the day period ranging
from 10:00 to 15:00 UT offers stable observing conditions
provided that the telescope is not pointed too
close to the Sun.
Otherwise, further scan selection based on
the exact sequence of observations and on beam monitoring might be
needed before using these observations for performance assessment.
{\lp In summary, the \emph{baseline} scan selection retains 16 hours of
observations a day and discards observations affected by an
atmospheric absorption exceeding 100\%.}

\section{Focal Plane Reconstruction}
\label{se:geometry}

{\lp From the operational point-of-view the KIDs are defined by the
resonance frequency and not by their physical position in the focal
plane. Therefore, to find the position on the sky we need to deploy
a dedicated method that we refer to as the FOV reconstruction.}
The FOV reconstruction thus consists in matching the KID frequency tones
to positions in the sky and in performing a KID selection.
Although all the 2,900 KID are responsive, some of them are affected by
cross-talk or are noisy due to an inaccurate tuning of their
frequency, and must be discarded for further analysis. We use \bms,
which enable an individual map per KID to be constructed 
to measure the KID positions and relative gains, as
discussed in Sect.~\ref{se:fov_geometry}. The measured KID positions
are further
checked by matching with the design positions, as presented in
Sect.~\ref{se:grid_distortion}. In Sect.~\ref{se:avg_kidpar}, we
present the final KID selection and FOV geometry, as obtained by
repeating the procedure on a series of \bms.

\subsection{Reconstruction of the FOV positions and KID properties}
\label{se:fov_geometry}

In order to be able to produce a map, one needs to associate a pointing
direction to any data sample of the system. The telescope provides
pointing information for a reference position in the focal
plane. These information consist of the
absolute azimuth and elevation $(\az_t^{\rm{ref}},\elev_t^{\rm{ref}})$
of the source, together with offsets
$(\Delta\az_t^{\rm{ref}}, \Delta\elev_t^{\rm{ref}})$ \wrt~these{\lp, which 
depends on the scanning strategy.}
We then need to know the relative pointing offsets of each detector
with respect to this reference position. We use
\bms\ for this purpose (see Sect.~\ref{se:beammaps}).

We apply a median filter per KID timeline whose width is set to 31
samples, that is equivalent to about 5~FWHM at 65\,arcsec/s and for the
sampling frequency of
23.84\,Hz. Then, we project one map per KID in Nasmyth
coordinates. The median filter removes
efficiently most of the low frequency atmospheric and electronic
noise, albeit with a slight ringing and flux loss on the
source. However, at this stage, we are only interested in the location
of the observed source.
To derive the Nasmyth coordinates from the
provided $(\az_t,\elev_t)$ and $(\Delta\az_t,\Delta\elev_t)$
coordinates, we build the following quantities at time~$t$:

\begin{equation}
\left( {\begin{array}{c}
\Delta x_t\\
\Delta y_t \\
\end{array}} \right) = 
\left( {\begin{array}{cc}
\cos\elev_t & - \sin\elev_t  \\
\sin \elev_t & \cos \elev_t \\
\end{array}} \right)
\left( {\begin{array}{c}
\Delta \az_t^{\rm{ref}}\\
\Delta \elev_t^{\rm{ref}} \\
\end{array}} \right)
\end{equation}

Note that $\Delta\az_t^{\rm{ref}}$ is already corrected by the $\cos\elev_t^{\rm{ref}}$ factor to
have orthonormal coordinates in the tangent plane of the sky and be immune to
the geodesic convergence at the poles.
The data timelines are then projected onto $(x,\, y)$ maps. 
We fit a 2D elliptical Gaussian on each KID Nasmyth map. {\lp The centroid
position of this Gaussian provides us with an estimate of the KID
pointing offsets \wrt\ the telescope reference position
($\az_t+\Delta\az_t$, $\elev_t+\Delta\elev_t$) in Nasmyth
$(x,y)$ coordinates (independent of time).

To convert from Nasmyth offsets to $(\az,\elev)$ offsets, we apply the
following rotation:
\begin{equation}
\left( {\begin{array}{c}
\Delta \az^k_t\\
\Delta \elev^k_t\\
\end{array}} \right) = 
\left( {\begin{array}{cc}
\cos\elev_t & \sin\elev_t  \\
-\sin \elev_t & \cos \elev_t \\
\end{array}} \right)
\left( {\begin{array}{c}
\Delta x^k\\
\Delta y^k\\
\end{array}} \right)
\end{equation}
where $k$ is a KID index. Adding these offsets to
$(\Delta \az_t^{\rm{ref}}, \Delta \elev_t^{\rm{ref}})$ gives the
absolute pointing of each KID in these coordinates.}

{\lp Furthermore, the fitted Gaussian per KID further provides us with
a first estimate of the KID FWHM, ellipticity and sensitivity. 
We apply a first KID selection by removing outliers to the statistics
on these parameters. We also discard manually KIDs that show a
cross-talk counterpart on their map. We repeat this
procedure using the \emph{baseline} TOI decorrelation method instead of the
median filter. Specifically, we apply the \cmoneb\ noise subtraction
presented in Sect.~\ref{se:dataproc} to the KID timelines, which are
then used to produce maps per KID. Therefore this alleviates the flux loss
induced by the median filter.}
{\rev This also ensures that the \bms\ are treated in the same way as
the scientific observation scans will.} Finally, a second iteration of
the KID selection is performed.

This analysis is repeated on all \bms\ to obtain statistics and
precision on each KID parameter, together with estimates on KID
performance stability, as discussed in the next sections.

\begin{figure*}[!thbp]
\begin{center}
\includegraphics[trim=3cm 14cm 6cm 4cm, clip=true, width=0.32\linewidth]{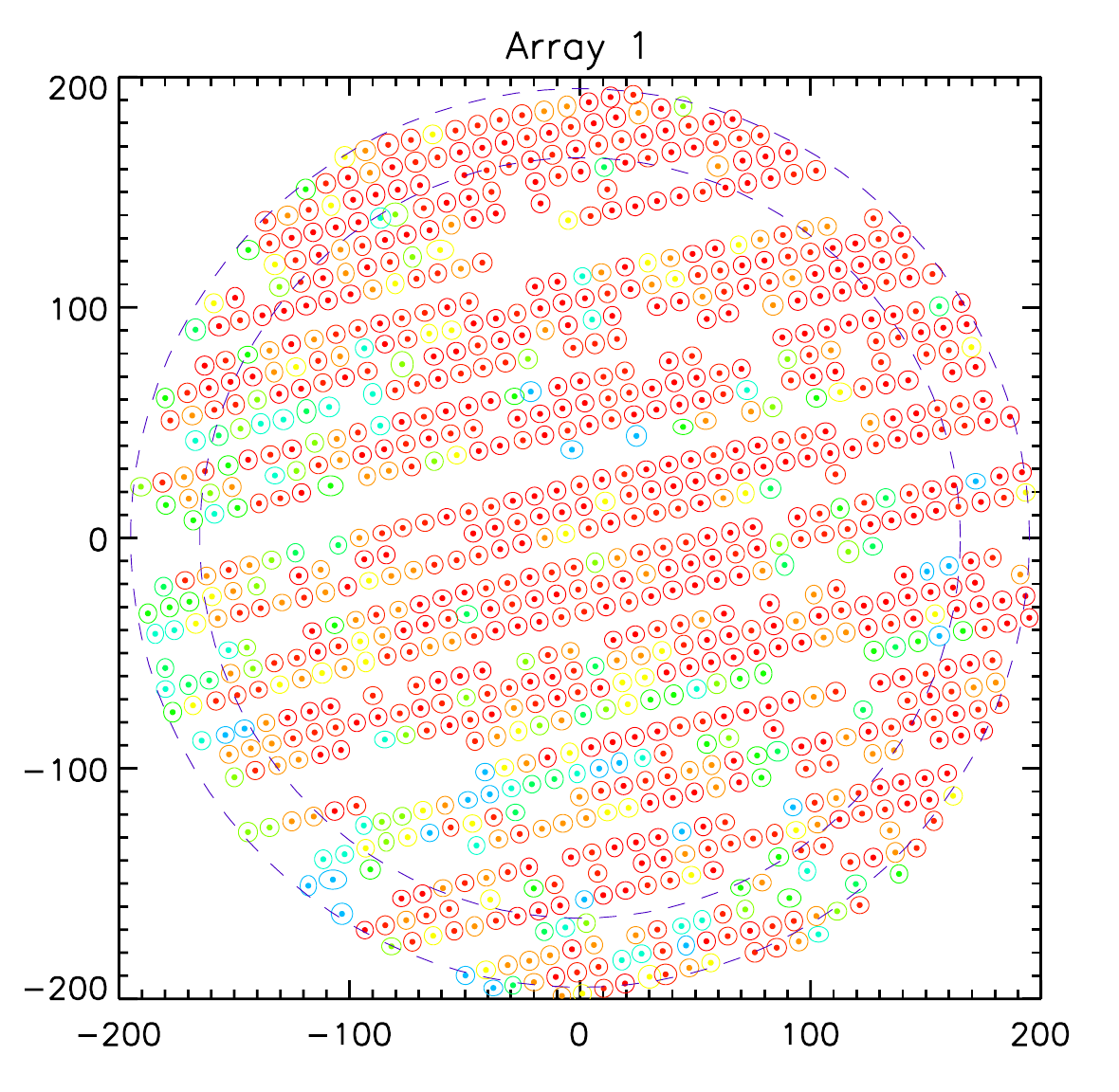}
\includegraphics[trim=3cm 14cm 6cm 4cm, clip=true, width=0.32\linewidth]{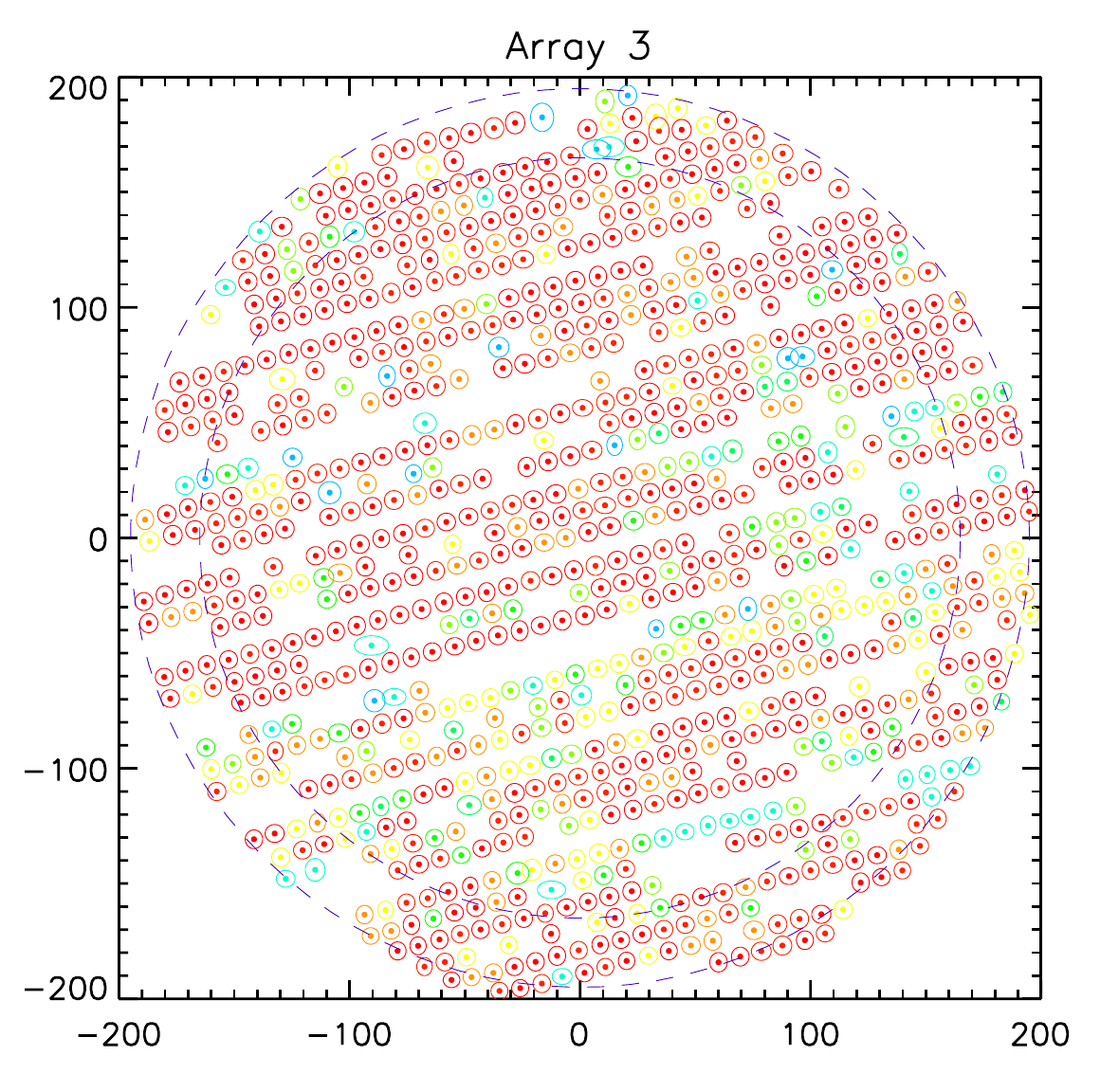}
\includegraphics[trim=3cm 14cm 6cm 4cm, clip=true, width=0.32\linewidth]{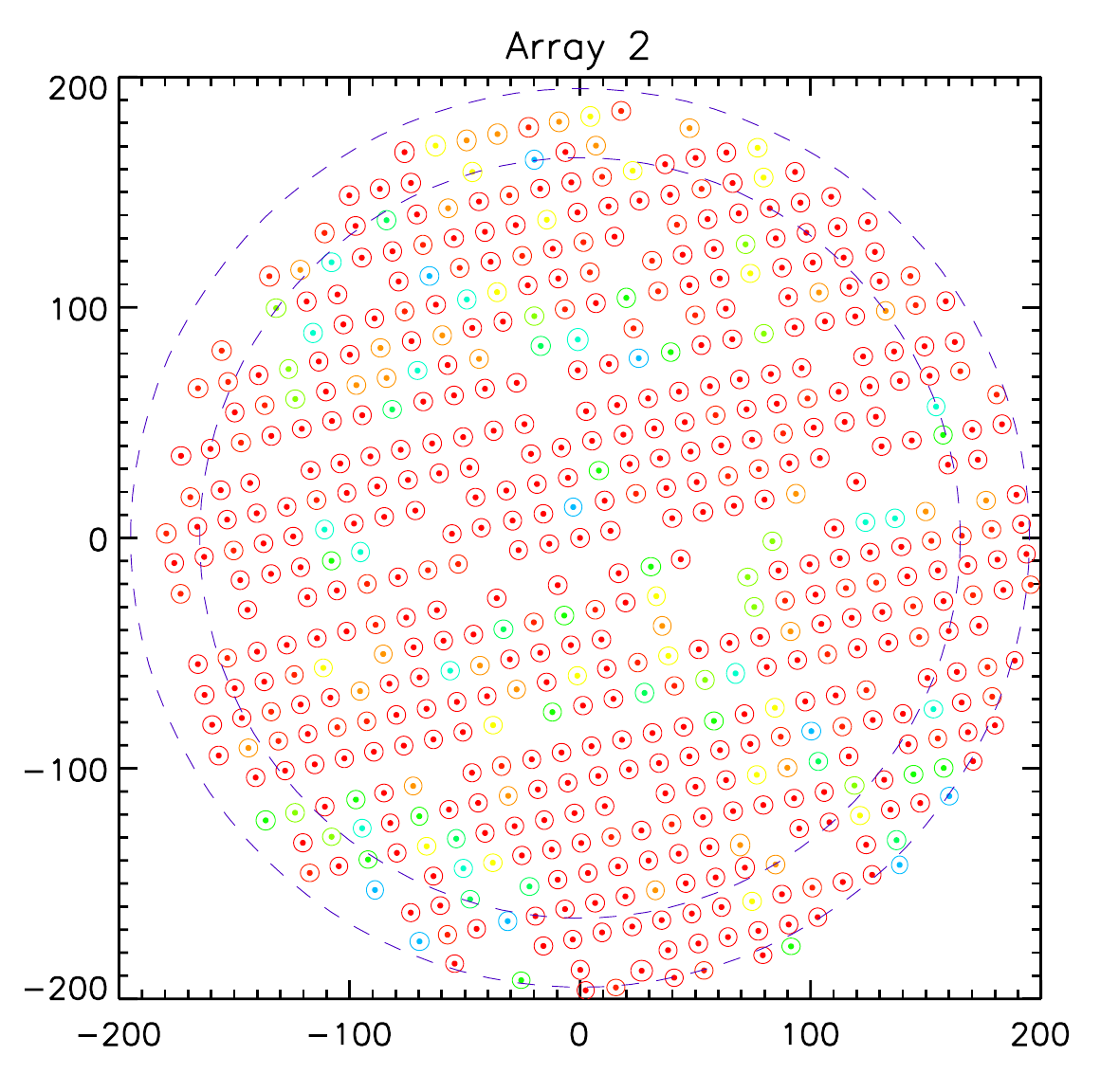}
\caption[KID selection in the FOV]{Average detector positions
  for arrays A1, A3, and A2. The three plots show the detectors that
  met the selection quality criteria for at least two \bms\ during two technical
  campaigns. These consist of 952, 961, and 553 detectors for A1, A3
  and A2, respectively. The color
  indicates how many times a KID was identified as valid on a \bm,
  {\lp ranging from blue for the KIDs valid in at least two \bms, to red for the KIDs
  valid in all (ten) \bms.} 
  The inner and outer
  dash-line circles correspond to circular regions of 5.5\,arcmin and 6.5\,arcmin,
  respectively. Units are arcseconds.}
\label{fig:avg_fov_color}
\end{center}
\end{figure*}

\subsection{FOV grid distortion}
\label{se:grid_distortion}

We compare the reconstructed KID positions in the FOV to their design
positions in the array. We fit the 2D field translation and rotation that allow
matching the measured KID positions with the design positions using a 2D
polynomial mapping function. {\lp We find that a matching can be
obtained using a 2D polynomial function of degree one, which corresponds to a linear
transformation and a rotation only.} We call distortion cross-terms
between the two spatial coordinates in the polynomial fit.

The aim is twofold. First we obtain a detailed
characterization of the real geometry of NIKA2 focal plane. Secondly,
{\lp this analysis is also used for KID
selection}. The most deviant KIDs, whose measured position deviates
by more than $4''$ from the design position are discarded. 

We present the global results of the grid distortion
analysis using the KID positions given
by the focal plane geometry procedure, as described in
Sect.~\ref{se:fov_geometry}, applied to a \bm\ scan acquired
during the first scientific campaign (\aka\ N2R12). 
The initial
number of KID considered in this analysis results from
a first KID selection, which consists in discarding the KIDs that are the most
impacted by the cross-talk effect or the {\tt tuning} failures,
applied in the FOV geometry obtained from the \bm\ scan. More details
on the KID selection are given in the next section. 
The results are gathered in Table~\ref{ta:gridmatch}.

\begin{table}[!htbp]
  \caption[Field-of-view deformations]{Field-of-view
    deformation. Example of mapping of the observed KID positions in the
    sky to their mechanically designed positions. The initial table of
    selected KIDs is given by the focal plane geometry procedure, as
    described in Sect.~\ref{se:fov_geometry}, applied to a \bm\ scan
    acquired during the N2R12 campaign. 
    More than 90\% of the detectors are within less than 5 arcseconds
    of their expected position.}
  \label{ta:gridmatch}
  \centering
  \begin{tabular}{r|c|c|c}
    \hline
    \hline
    Characteristic &  Array 1  &	Array 3   &	Array 2  \\
    \hline
    \small{$\lambda$ [mm]}  &  1.15     &      1.15      & 2.0  \\ 
    \small{Design detectors} & 1140  &  1140 & 616  \\
    \small{Selected KID}\tablefootmark{a}    &  866  &  808  & 488  \\
    \small{Well-placed KID}\tablefootmark{b}          &  864  &  808  & 488  \\
    \small{Median deviation\tablefootmark{c}  [arcsec]}    & 1.01    &     0.95   &    0.75  \\
    \small{Mean distortion\tablefootmark{d} [arcsec]}                       & 1.09    &     1.01   &    0.84  \\
    \small{Array center\tablefootmark{e} [arcsec]}  & (1.9, -5.1) & (2.3, -6.2) &  (9.6, -7.8) \\
    \small{Scaling\tablefootmark{f} [arcsec/mm]}   &  4.9     &	4.9      &    4.9 \\
    \small{Rotation angle\tablefootmark{g} [degree]} & 77.3     &	76.3      &    78.2  \\
    \small{Grid step\tablefootmark{h} [arcsec, mm] }     & 9.8, 2.00 &	9.7, 2.00  &    13.3, 2.75 \\
    \small{Grid step\tablefootmark{i} [$\lambda$/D] } & 1.11     &	1.10      &	0.87  \\
    \small{Modelled grid step\tablefootmark{j} [$\lambda$/D]}  & 1.09     &      1.09      & 0.93    \\
    \hline
  \end{tabular}
  \tablefoot{ \\
    \tablefoottext{a}{Initial number of KIDs selected in a FOV geometry using a \bm\ scan of the N2R12 campaign;}
    \tablefoottext{b}{{\lp Number of KIDs for which the best-fit sky
  position is less than 4'' away from the expected position;}}           
    \tablefoottext{c}{{\lp Median angular offset [arcsec] between the expected and measured sky position of the KIDs;}}
    \tablefoottext{d}{{\lp Average best-fit cross term of the polynomial model across the FOV [arcsec];}}
    \tablefoottext{e}{Array center in Nasmyth coordinates;}
    \tablefoottext{f}{Averaged scaling between measured {\lp KID position
  grid} and the designed one;}
    \tablefoottext{g}{Rotation from the design to Nasmyth coordinates;}
    \tablefoottext{h}{Center-to-center distance between neighbour detectors;}
    \tablefoottext{i}{Center-to-center distance between neighbour detectors using the
  reference frequencies ($260\,\rm{GHz}$ and $150\,\rm{GHz}$) and a
  27\,m {\lp entrance pupil diameter} (see Sect.~\ref{se:instru_optics});}
    \tablefoottext{j}{Center-to-center distance between neighbour detectors modelled using ZEMAX simulation.}
  }
\end{table}

Most of the selected KID are also well-placed, that is at less than 4'' from
their expected position. Moreover, on average the position of each
detector is known to about one arcsecond. We find
that Array 1 has some of the most deviant detectors (above 4''
from their expected position). These detectors are excluded from
further analysis. The two 1\,mm arrays have almost
the same center but this center differs by $7''$ and $2''$ in the two Nasmyth
coordinates, respectively, from the 2\,mm array center.
This has no significant impact on the pointing and the focus settings
at the precision of which they are measured.
{\lp The center-to-center distance between
contiguous detectors, referred to as grid step, has been estimated in
$mm$ and arcseconds. The ratio of the grid step in mm to
the grid step in arcseconds gives compatible effective focal lengths
of about $42.4\pm0.3\,\rm{m}$ at both observing wavelengths.} The sampling is above $\lambda/D$ at
1\,mm, assuming a 27\,m effective diameter aperture. Note that
the rotation angle between the array and the Nasmyth
coordinates was designed as $76.2^{\rm{o}}$, less than two degrees
away from what is observed.

These results have been compared to expectations obtained using ZEMAX
simulation. 
We generated a grid diagram for the NIKA2 optical system and found a maximum
grid distortion of $2.7\%$ in the $6.5'$ FOV. We notice that the
strongest distortion appear in the upper right corner of the Nasmyth plane, which is
also the area of the largest defocus w.r.t. the center (see Appendix~\ref{ap:focus_surfaces}).
An expected distortion of $2.7\%$ is at most a 5'' shift from the
center to the outside of the array. The quoted distortions between the
measured and design positions are well within the expected
maximum distortions from the NIKA2 optics.

\subsection{KID selection and average geometry}
\label{se:avg_kidpar}

In order to identify the most stable KIDs, we compare the KID parameters
obtained using the FOV reconstruction procedure, as described in
Sect.~\ref{se:fov_geometry}, with several \bms. In the following, we
show results as obtained using {\lp ten} \bms\ acquired during two
technical observation campaigns in 2017.
For each KID we compute the average position on the focal plane and
the average FWHM. 
As discussed in Sect.~\ref{se:fov_geometry}, we perform a KID
selection while analysis a \bms. A few KIDs have very close resonance
frequencies and can be misidentified on some scans. A few others must
also be discarded because they appear identical
numerically due to the fact that a same (noisy) KID can sometimes be
associated to two different frequency tones in the acquisition system.
These KIDs are flagged out (less than 1\% of the design KIDs).
{\lp We count how many times a KID has been kept in
the KID selection per \bm\ and has been found at a position agreeing
with its median position within 4''.}
Using this, we define the {\emph valid} KIDs as the KIDs that met the selection
criteria in about {\lp $20\%$} of the FOV geometries (in two \bm\
analysis out of ten).

In Fig.~\ref{fig:avg_fov_color} we show the average focal plane
reconstruction. The colors, from blue to red,
represent the number of times that the KID has been retained after
KID selection. The eight feed lines for each of the two
$1\,\rm{mm}$ arrays {\lp can also be traced out in several ways} in this
figure. First, slightly larger spaces are seen between KID rows
connected to different feed lines than between KID rows of the same
feed line. Second, KIDs at the end of a feed line are less often valid
than the others
(see e.g. the FOV of Array 3). As the tone frequencies
increase with the position of the KID in the feed-line, some KIDs are
sometimes missing because their frequency lays above the maximum tone
frequency authorized by the readout electronics. This explains the
linear holes in the middle of the $1\,\rm{mm}$ arrays. For A1, this
end-of-feed-line effect is mixed with the effect of the KID gain
variation across the FOV, which mainly affects the lower left third of
the array, as discussed in Sect.~\ref{se:flat_field}.

For A1, A3 and A2, respectively, we found 952, 961, and 553 valid KIDs
 (selected at least twice). From this, we deduce the fraction of
valid detectors over the design ones, as given in Table~\ref{tab:number_of_kids}.

\begin{table}[!htbp]
  \begin{center}
    \caption[Number of detectors]{Summary of the number of valid detectors per array.}
    \label{tab:number_of_kids}  
    \begin{tabular}{lrrr}
      \hline
      \hline
      \noalign{\smallskip}
      Characteristics & Array 1 & Array 3  & Array 2  \\
      \noalign{\smallskip}
      \hline
      \noalign{\smallskip}
      Design detectors ($N_k$)  & 1140  & 1140   & 616  \\
      Valid detectors           & 952   & 961    & 553  \\ 
      Ratio $\equiv \eta$       & 0.84  & 0.84   & 0.90   \\
      \noalign{\smallskip}
      \hline
    \end{tabular}
  \end{center}    
\end{table}

The valid KIDs represent the KIDs
usable to produce a scientifically exploitable map of the flux
density using observations for which no sizeable {\tt tuning}
issues are experienced. 
{\lp To give an idea of the dependence of the KID selection on the
observing conditions,} we also
evaluate a number of \emph{very stable} KIDs as the KIDs that met the
selection criteria at least in {\lp $50\%$} of the FOV geometries (at least
five times out of {\lp ten}). We
found 840, 868 and 508 KIDs using this definition for A1, A3 and A2,
respectively. The fraction of very stable KIDs over the designed ones
is reported in Table.~\ref{tab:eta_used}.

\begin{table}[!htbp]
  \centering
  \caption[]{Fraction of \emph{very stable} KIDs in percent of the design
  KIDs. The row 'very stable KIDs' gives the fraction of KIDs that have been selected
  in {\lp $50\%$} of the analysed {\tt beammap} scans, while the rows 'Used KIDs' gather
  the median fraction of used KIDs in the data reduction processing
  after the conservative KID selection has been performed (see Sect.~\ref{se:dataproc})}
  \label{tab:eta_used}
  \begin{tabular}{llrrrr}
    \hline\hline
    \noalign{\smallskip}
    &  Data set   & A1      &   A3    &     A2 \\
    \noalign{\smallskip}
    \hline
    \noalign{\smallskip}
    Very stable KIDs [\%] & {\tt beammaps} & 74  &  76  &  82  \\
    \hline
    \noalign{\smallskip}
    Used KIDs [\%]  & N2R9     & 58 &  64  & 71 \\
               & N2R12    & 73 &  69  & 77 \\
               & N2R14    & 69 &  68  & 79 \\
               & Combined & 70 &  69  & 78 \\
    \hline
  \end{tabular}
\end{table}

\begin{figure*}[!thbp] 
\begin{center}
\includegraphics[clip=true, trim={0.1cm, 0cm, 3cm, 2cm},width=0.235\textwidth]{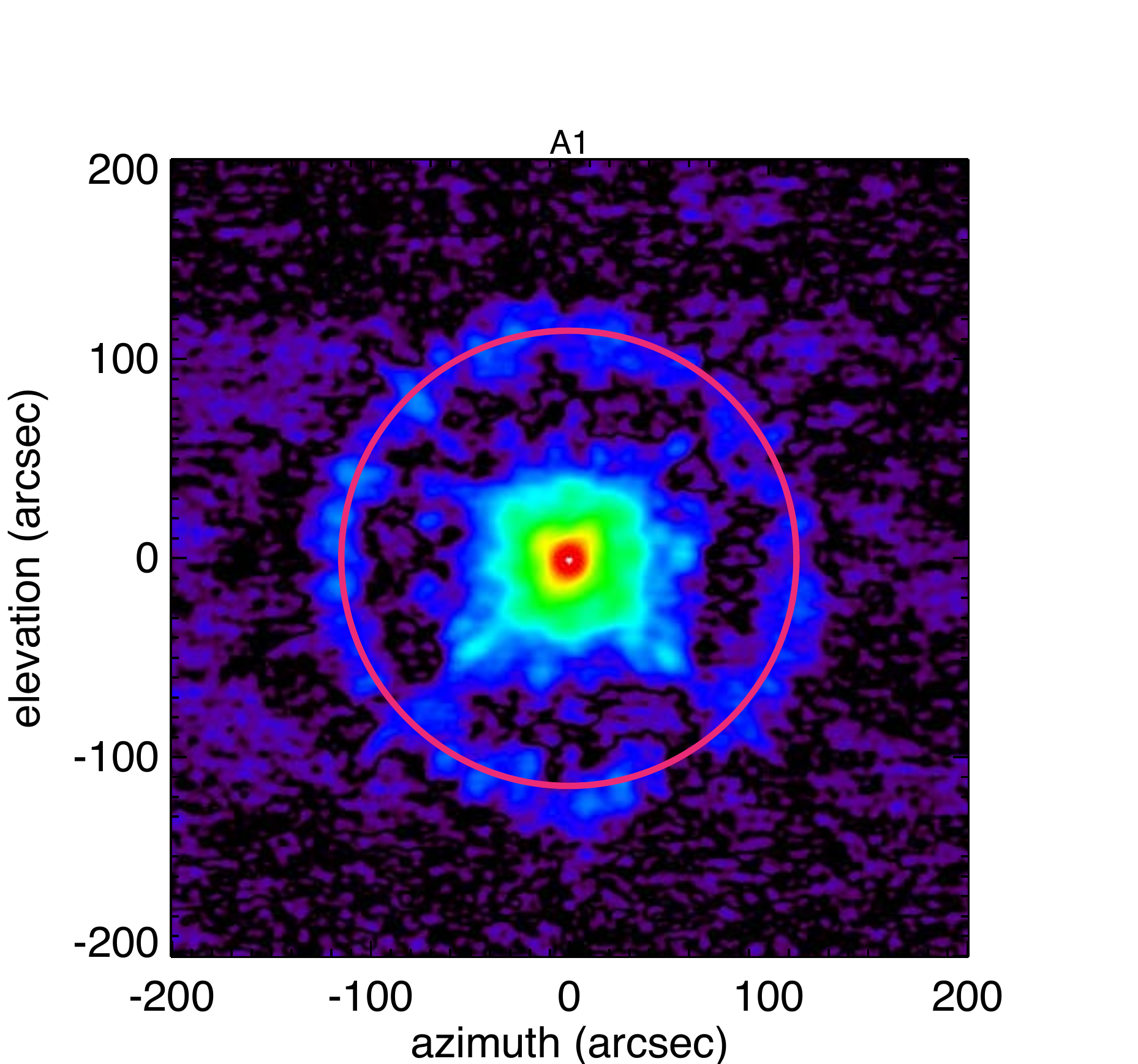}
\includegraphics[clip=true, trim={1cm, 0cm, 3cm, 2cm},width=0.225\textwidth]{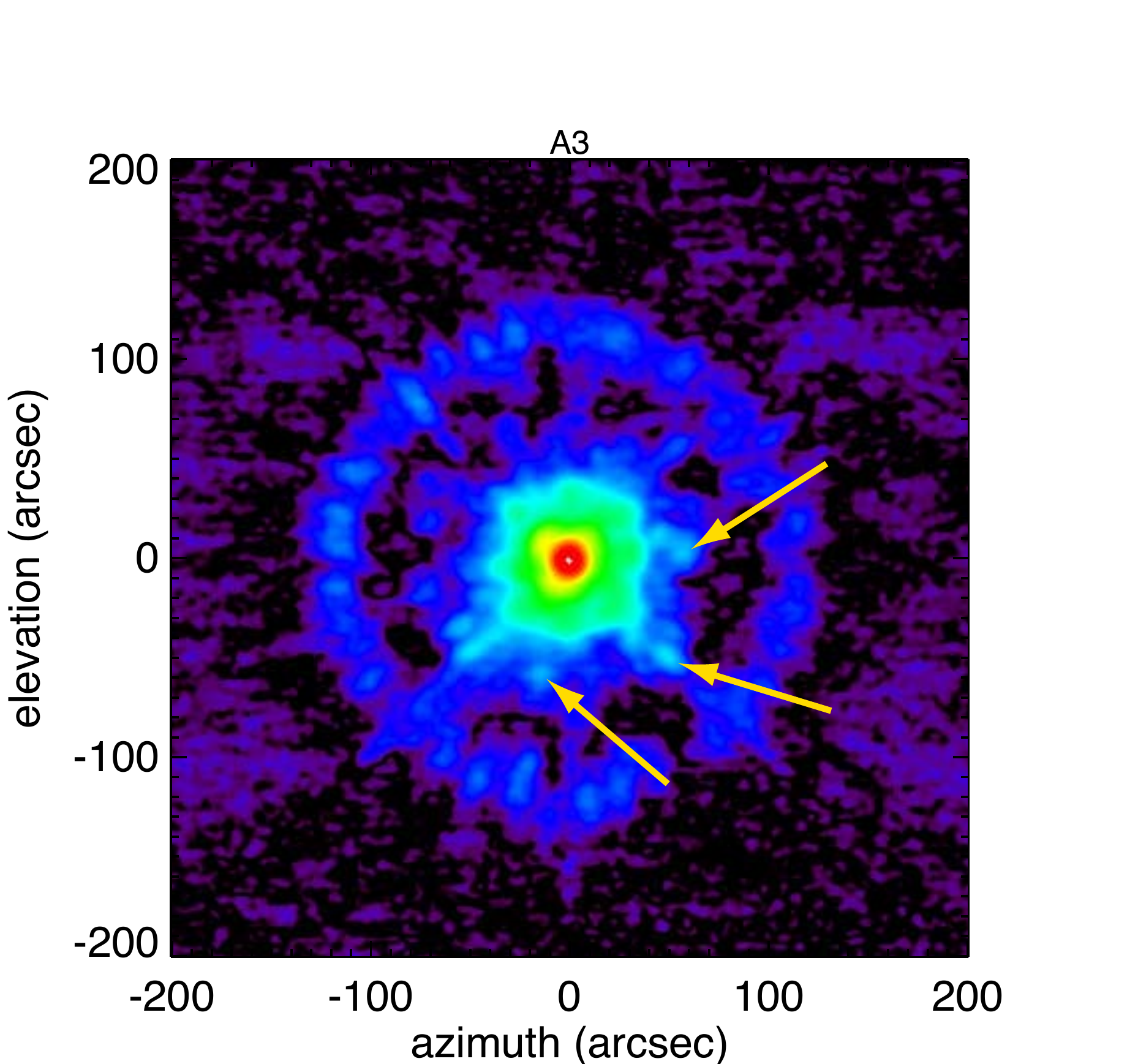}
\includegraphics[clip=true, trim={1cm, 0cm, 3cm, 2cm},width=0.225\textwidth]{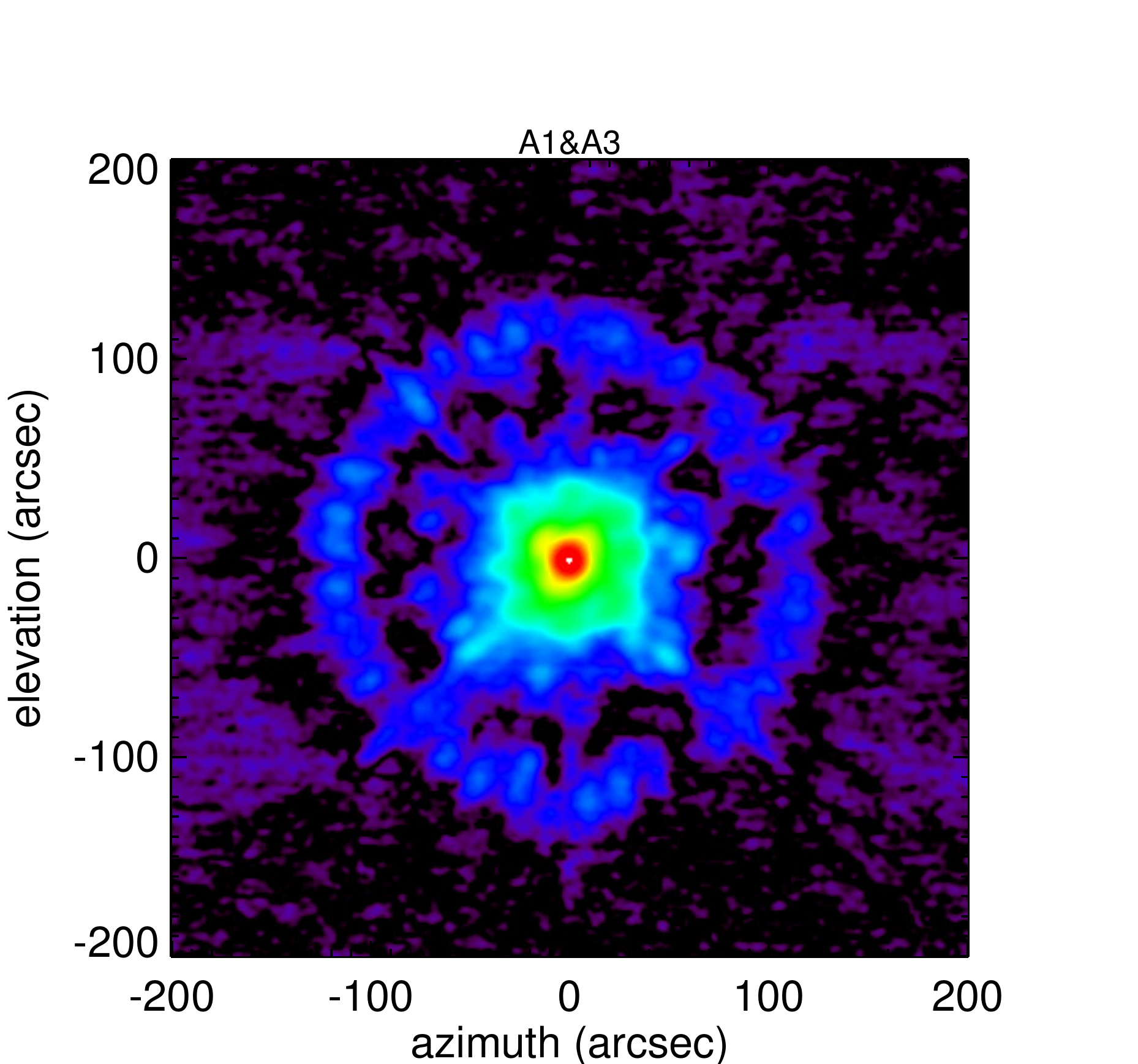}
\includegraphics[clip=true, trim={1cm, 0cm, 0cm, 1cm},width=0.27\textwidth]{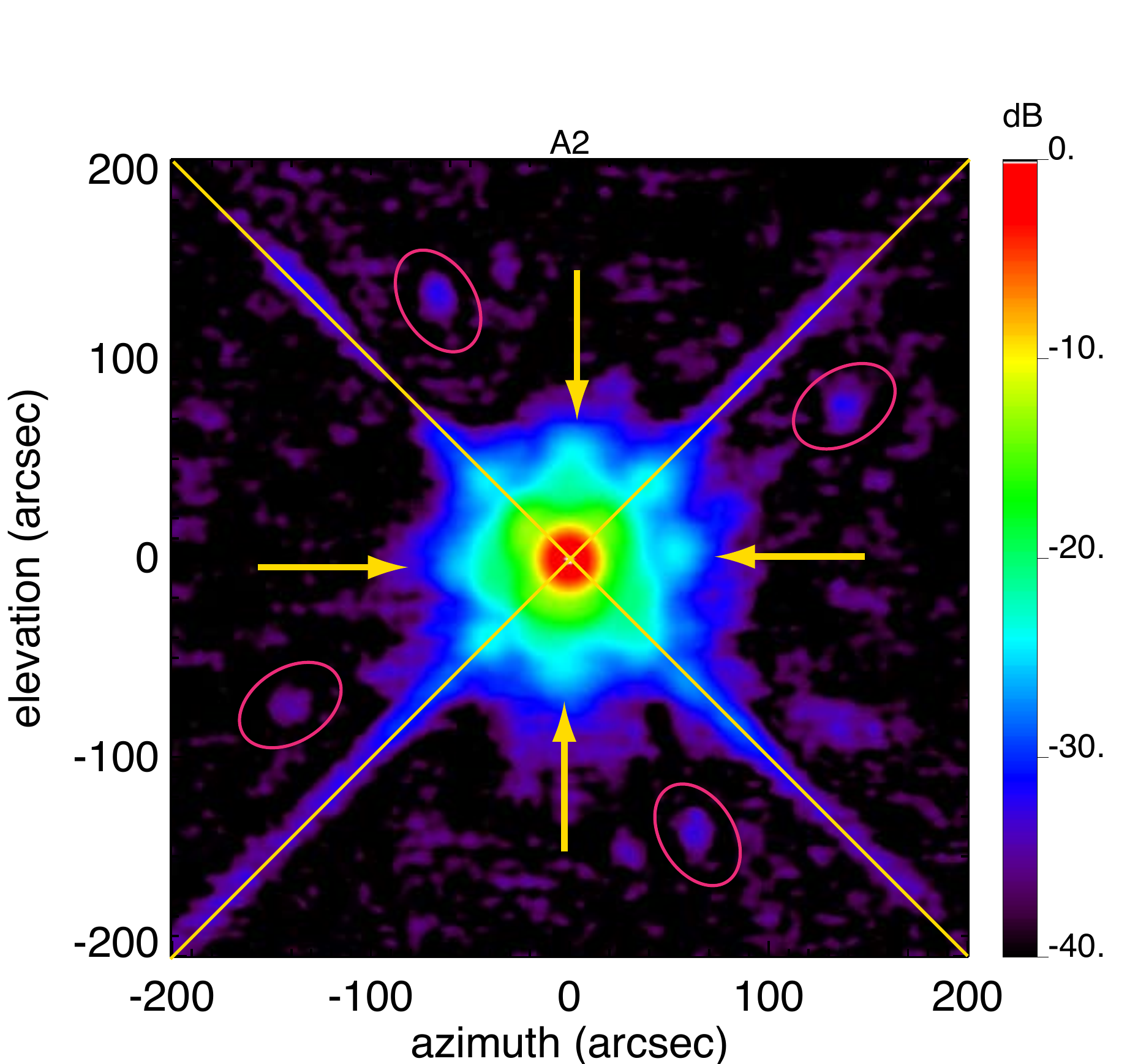}
\caption[Noticeable features of NIKA2 beam pattern.]{From left
to right, beam maps of A1, A3, the combination of the $1\,\rm{mm}$ arrays (A1$\&$3) and the
$2\,\rm{mm}$ array (A2) are shown in decibel. These maps, which
consist of the normalized combination of four {\tt beammap} scans of
bright point-like sources, are in horizontal coordinates. {\rev They
represent a zoom in the inner part of larger maps that cover a
sky area which extends over $10'$.} 
The solid lines and arrows highlight some noticeable features.
Red circle in the A1 map (first panel): diffraction ring seen in 1-mm maps
(the spokes are presumably caused by radial and azimuthal panel
buckling \citep{Greve2010}; Orthogonal yellow lines in the A2 map
(rightmost panel): diffraction pattern caused by the quadrupod
secondary support structure (prominently
seen in A2 map); Yellow arrows in the A3 map (third panel):
pattern of 3 spikes seen in $1\,\rm{mm}$ maps of unknown origin; Yellow
arrows in A2 map (fourth panel): four symmetrical spikes of the
first sidelobes; Pink ellipses: four spikes seen in A2 maps.}
\label{fig:features}
\end{center}
\end{figure*}

Practically, for the production of flux density maps, we perform
further selection of the valid KIDs using a conservative noise level
threshold of the KIDs at the \emph{low level processing}, as discussed in
Sect.~\ref{se:dataproc}. The number of \emph{used} KIDs for producing
science-purpose maps using the data reduction pipeline, as described
in Sect.~\ref{se:dataproc}, is thus significantly lower than the number of
\emph{valid} KIDs. We evaluate the median number of used KIDs using all scans
for each of the observation campaigns. The median fractions of used
KIDs with respect to the design ones for each campaigns and for the
combinations of all scans are given in Table.~\ref{tab:eta_used}. We
find median fractions of used KIDs of about $70\%$ for the
$1\, \rm{mm}$ arrays and of about $80\%$ for the $2\, \rm{mm}$ array,
with a notable exception at the N2R9 technical campaign, for which
these fraction are lower due to severe {\lp atmospheric} temperature-induced
unstable observing conditions (see
Sect.~\ref{se:beam_variation}). Moreover the median fractions of used
KIDs are close to the fractions of
\emph{very stable} KIDs from the FOV geometries. We stress that these numbers
depend on the choices made in the data reduction pipeline for data
sample cuts, and are thus subject to improvement. By contrast, the
fractions of valid KIDs $\eta$ constitute a conservative estimates of the
stable KIDs usable for science exploitation over all the functioning
KIDs. These are thus the relevant estimates for the instrument
performance assessment.  
 

\section{Beam Pattern}
\label{se:beam}

{\lp The NIKA2 full beam pattern originates from the KIDs illuminating the internal
and external optics, out to the IRAM \trentemetre\ telescope primary mirror.}
To characterize the full beam pattern, we use \bm\ observations. First,
deep integration maps of bright sources are produced to provide a
qualitative description of the complex beam structure in
Sect.~\ref{se:fullbeam}. Then we model the beam using three
complementary methods to estimate the main beam angular resolution
(Sect.~\ref{se:mainbeam}) and the beam efficiency
(Sect.~\ref{se:beam_efficiency}).

\subsection{Full beam pattern}
\label{se:fullbeam}

To study the two-dimensional pattern of the beam, we 
primarily use a map obtained from a combination of \bm\ observations
of strong point sources acquired during the N2R9 commissioning campaign.
Namely, we use \bm\ scans of Uranus
,  Neptune 
and the bright quasar 3C84. 
Furthermore, we checked the stability of our results on single scan
maps, combinations of scans for a single source, and combinations of
shallower scans but spanning a large range of scanning direction. {\lp
The \bm\ scans are reduced using the method discussed in
Sect.~\ref{se:dataproc} to produce maps.}
{\lp Figure~\ref{fig:features} shows the two-dimensional beam pattern
as measured with NIKA2 using the former \bm\ combination, for each of
the arrays and for the $1\,\rm{mm}$ array combination.}
{\lp The beam pattern is shown over a large dynamic range down
to about $-40\,\rm{dB}$ and out to radii of about 5’.
The telescope beam pattern further extends well beyond this radius, as for
example shown by lunar edge observations
at the IRAM \trentemetre\ telescope~\citep{Greve1998,
Kramer2013}. However,
this extended pattern is at present difficult to detect using the data
reduction pipeline discussed in Sect.~\ref{se:dataproc}, as
the extended error beams are both filtered and mixed with 
atmospheric and electronics large-scale correlated noise
residuals. {\rev We expect the filtering effect due to the data
processing to become non negligible for angular scales larger than
90'', which corresponds to the radial size of the mask used in the
correlated noise subtraction process (see Sect.~\ref{se:toi_proc}).}
The contributions to the beam pattern that stem from larger angular scales are
further discussed in Sect.~\ref{se:beam_efficiency}.} 

The NIKA2 beam maps reveal some noticeable features, which are
shown in Fig.~\ref{fig:features}. Ranging from strong and/or extended to
weak and/or spiky, they include:
\begin{enumerate}
\item the main beam and the underlying first error
  beam, which is due to large-scale deformations of the primary
  mirror, and the first side lobes, {\lp which correspond to various diffraction
  patterns. In particular, the 60'' and 85'' diameter (square-like shaped) side
  lobes at $1$ and $2\,\rm{mm}$, respectively, at a level lower than
  $-20\,\rm{dB}$, are due to the convolution of the primary mirror and the
  quadrupod diffraction pattern with the pixel (KID) transfer function;}
\item at a much lower level of about $-30\,\rm{dB}$, a diffraction ring shows    
up, which is presumably caused by panel buckling of the primary 
  mirror~\citep{Greve2010}, as shown with a red circle in the A1 panel;
\item also at a level of about $-30\,\rm{dB}$, the side lobes shown with green
  diagonal lines in the A2 panel are due to diffraction on the
  quadrupod holding the secondary mirror of the telescope, as expected
  from ZEMAX simulations;  
\item spikes of not fully understood origin marked by yellow
  arrows. The ones that are along the vertical and
  horizontal axes are reproduced by ZEMAX simulations but at a 
  shallower level, whereas the ones shown in the A3 panel in the
  diagonal directions may be due to the small cylindrical
  instrumentation box on the side of the M2 cabin. The origin of the
  asymmetry on the 1~mm arrays is unknown but most probably due to
  internal optics aberrations;
\item shallow spikes of unknown origin at a level less than $-30\,\rm{dB}$, which are circled by pink
  ellipses. The multiple images on the combined deep beam map indicate
  a rotation of these spikes with the observing elevation, which in
  turn point to diffraction related issue or a ghost image that are
  formed inside the cryostat. These shallow features are expected to
  have no significant impact on NIKA2 science results.
\end{enumerate}

We further quantify the respective level of the axi-symmetrical
features of the beam pattern by evaluating the beam radial profile
$B_\nu(r)$, which is the normalized radial brightness profile for the
array $\nu$, where $r$ is the radius from the beam center.
Although the profile cannot represent the sub-dominant non-axisymmetrical
features, which are seen in Fig.~\ref{fig:features} (quadrupod
diffraction pattern, spikes), it provides a useful
representation of the internal and central parts of the beam {\lp up to
 about $180''$.} We determine a beam profile from a beam map in centring to
the fitted value of the main beam center and in computing the
weighted average of the map pixels in annular rings.

\begin{figure}[!thbp]
  \centering
   \includegraphics[clip, width=\linewidth]{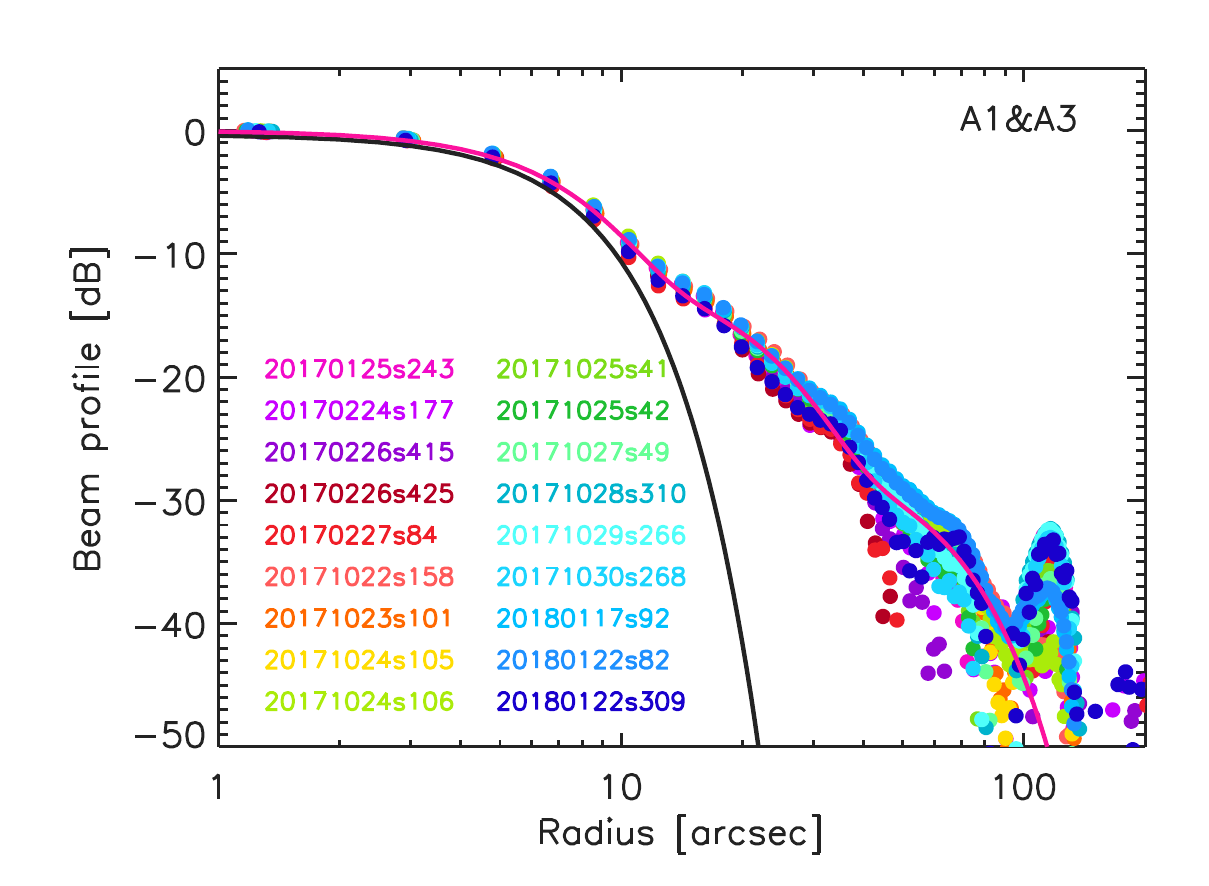}
   \includegraphics[clip, width=\linewidth]{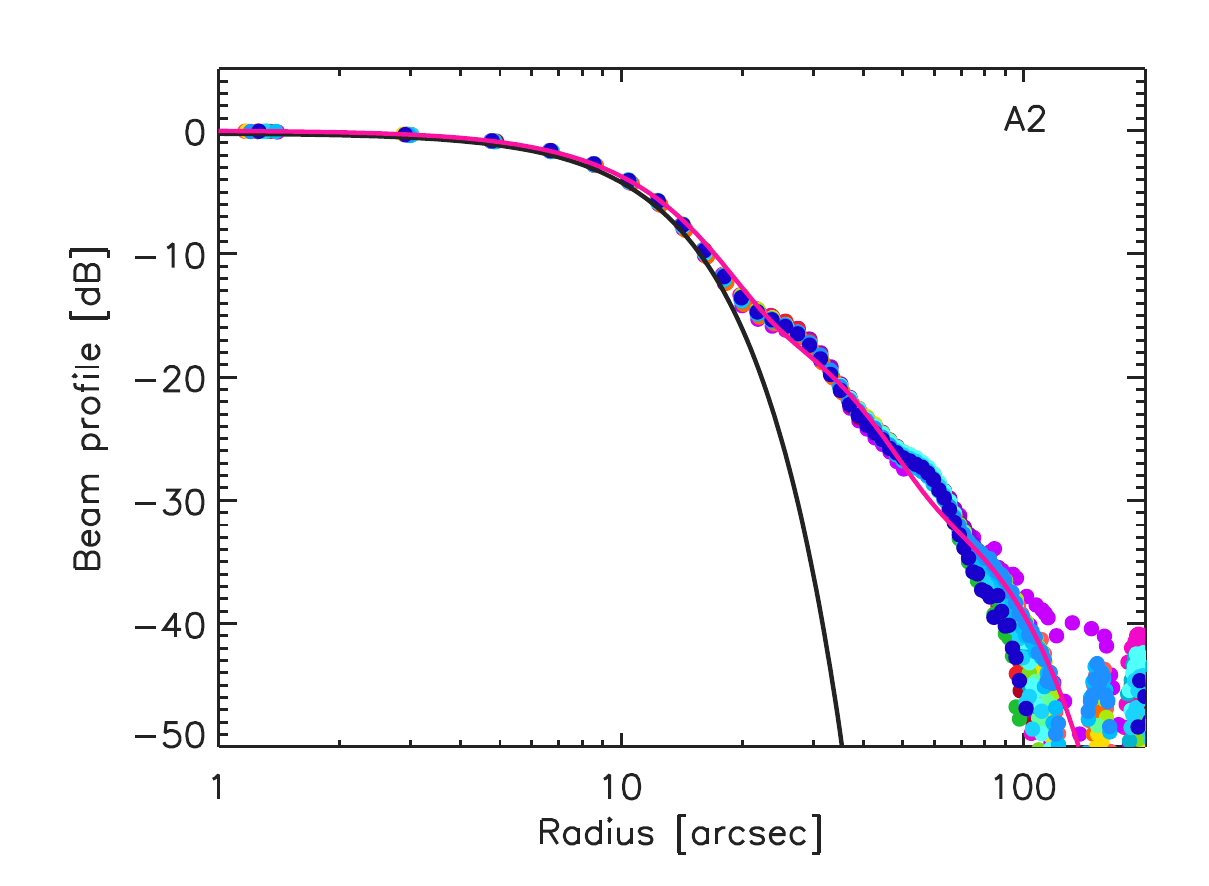}
  \caption[Stability of the beam profile]{{\lp Beam radial
    profiles given in decibel. 
    The data points are the beam profiles for a series of 18
    \bm\ scans acquired during the reference observational campaigns, labelled from the scan
    ID. The black line shows the main beam profile using the 'combined'
    FWHM, as given in Table~\ref{tab:fwhm}, while the pink
    line shows the median best-fit three-Gaussian profile, as defined
    in Eq.~\ref{eq:3gauss}.}}
  \label{fig:beam_prof}
\end{figure}

We checked the stability of the beam against various
observing conditions (source intensity, weather condition, focus
optimisation) by comparing the beam profiles of a series of 18 \bm\
observations.
This set of \bms, which is referred to as {\tt BM18}, has been
selected from all the available \bm\ scans at optimal focus using the
baseline scan selection criteria (Sect.~\ref{se:data_selection}).
The measured beam profiles using the {\tt BM18} data set are shown in
Fig.~\ref{fig:beam_prof}. {\lp The profiles consist of the main beam and
the first error beams and side lobes, which significantly contribute
at levels of less than $-10\,\rm{dB}$ at both observing
wavelengths. Moreover, at $1\,\rm{mm}$ the contribution of the
diffraction ring, which was marked with a red circle in Fig.~\ref{fig:features},
is seen as a peak at a level up to $-33\,\rm{dB}$ located at a radius
of about $115''$.} Calculating the {\lp rms of the relative}
difference of the beam profiles to the median beam profile, we find a
dispersion below $5\%$ at $1\,\rm{mm}$ and below $2\%$ at
$2\,\rm{mm}$.

{\lp To measure the relative level of the axi-symmetrical beam pattern
features, we further model the beam profiles using an empirical function,
which accounts for the main beam and for a significant fraction of the
error beams and the side lobes. We define this function $B_{3G}(r)$ as:
\begin{equation}
  B_{3G}(r) = \sum_{i=1}^{3} \mathcal{A}_i G_i(r) + \mathcal{B}_0,
  \label{eq:3gauss}
\end{equation}
where $\mathcal{A}_i$ is the amplitude of the Gaussian $G_i$ for
$i \in {1, 2, 3}$ and $\mathcal{B}_0$ a pedestal level accounting for
the residual background level in the map. The measured beam profiles
are fitted using Eq.~\ref{eq:3gauss} and the median best-fit
parameters are given in Table~\ref{tab:mean_3gauss_fit}. The errors
are evaluated as the standard deviation of the best-fitting parameter
values of the 18 \bm\ scans of {\tt BM18}.
These values are given to gain insight of beam profile, but are not
used for the calibration procedure. We find the level of the first error
beam with respect to the total beam at about $-11\,\rm{dB}$ and
$-13\,\rm{dB}$ at 1 and $2\,\rm{mm}$, respectively.}
%
\begin{table}[!th]
   \caption{{\lp Median best-fitting values of the parameters of the
  3-Gaussian beam profile, as defined in Eq.~\ref{eq:3gauss}, using
  the {\tt BM18} data set. For $i \in {1, 2, 3}$ $\bar{\mathcal{A}_i}
  =\mathcal{A}_i / \sum{\mathcal{A}_i}$.
  The FWHM for each of the Gaussian are given in arcseconds.}}
  \label{tab:mean_3gauss_fit}
  \begin{center}
    \begin{tabular}{rll}
      \hline\hline
      \noalign{\smallskip}
       parameters  &  $1\,\rm{mm}$  & $2\,\rm{mm}$ \\
       \noalign{\smallskip} 
      \hline
      \noalign{\smallskip} 
      $\bar{\mathcal{A}_1}$ [dB] &   $-0.33 \pm 0.09$   &  $-0.24 \pm 0.03$ \\
      $\bar{\mathcal{A}_2}$ [dB] &   $-11.4 \pm 1.0$     &  $-12.8 \pm 0.5$   \\
      $\bar{\mathcal{A}_3}$ [dB] &   $-26 \pm 7$       &  $-27 \pm 3$    \\
      FWHM$_1$  ['']             &   $10.8 \pm 0.2$    &  $17.4 \pm 0.6 $ \\
      FWHM$_2$  ['']             &   $30 \pm 2$        &  $42 \pm 3 $ \\
      FWHM$_3$  ['']             &   $81 \pm 10$       &  $99 \pm 7 $ \\     
       \noalign{\smallskip}   
      \hline
    \end{tabular}    
  \end{center}
\end{table}

{\lp For illustration, we show in
Fig.~\ref{fig:beam_prof} the median 
three-Gaussian profiles at $1$ and $2\,\rm{mm}$ (pink lines), and
the main beam profiles (black lines). }

\subsection{Main beam}
\label{se:mainbeam}

NIKA2 angular resolution {\lp is characterized using the FWHM of a
Gaussian fitted to the main beam. This principal Gaussian encloses
most of the measured flux density of a point-like source.} 

\subsubsection{Main beam characterization methods}
\label{se:mainbeam_methods}
To characterize the main beam and to derive an estimate of its FWHM, we
have developed three methods. The two first methods, quoted
{\tt Prof-3G} and {\tt Prof-1G}, rely on a fit of the beam profile to
benefit from the signal-to-noise ratio increase after azimuthally
averaging the signal. The last one by contrast,
consists in an elliptical Gaussian fit of the beam map for a better
2D modelling, and is labelled {\tt Map-1G}. They are presented in more
detail below. \\

\noindent {\tt Prof-3G} consists in fitting the beam profile using the
three-Gaussian function defined in Eq.~\ref{eq:3gauss}. The main beam
FWHM estimate is given by the best-fitting value of the FWHM for the
first Gaussian function. {\lp This main beam FWHM estimate is expected
  to be immune to the first error beams
and side lobes, which are well accounted for. For consistency checks,
we also relies on two other methods that rely on simpler beam models.}\\

\noindent {\tt Prof-1G} relies on fitting a single Gaussian to the beam
profile after masking the portion of the profile where the
contributions of the side lobes and error beams are the
largest. Specifically, the side lobe mask is designed to cut out {\lp the
radius ranging from an inner radius
$r_{\rm{in}} = 0.65\, \mathrm{FWHM}_0,$, where FWHM$_0$ is the
reference Gaussian beam FWHM (see Sect.~\ref{se:photometric_system})
to an outer radius $r_{\rm{out}} = 80''$,} centred on the beam
maximum.
The profile is estimated up to a radius of
$180''$, that is in the inner part of the beam map where the noise
variance is uniform.\\

\noindent {\tt Map-1G} consists in modelling the two-dimensional distribution of
the main beam using an 2D elliptical Gaussian of size $\sigma_x$ and
$\sigma_y$. We characterize the NIKA2 main beam using
\begin{equation}
  FWHM = 2 \sqrt{2\ln {2}\, \sigma_x\sigma_y}.
  \label{eq:fwhm_geom}
\end{equation}
As in {\tt Prof-1G}, we use masked versions of the
beam map to avoid side lobe and error beam contaminations. 
Whereas $r_{\rm{out}}$ is conservatively set to be $100''$,
$r_{\rm{in}}$ is let free to vary around a central value of about $8''$
for A1 and A3 and of about $12''$ for A2 to provide the best 2D Gaussian
fit.

\subsubsection{Data sets for the main beam determination}
\label{se:mainbeam_dataset}

We select a sub-set of the selected \bm\ scans described in
Sect.~\ref{se:fullbeam} by discarding scans of Mars. {\lp Indeed, \bms\
towards Mars unveil the complex full beam pattern, which extends beyond
radii of $100''$, so that the annulus sidelobe mask used in {\tt
Prof-1G} and {\tt Map-1G} is not sufficient to mitigate the error
beams and sidelobes effects.}
The 12 remaining \bm\ scans are analysed using the data reduction
pipeline of Sect.~\ref{se:dataproc} and projected onto maps
with a resolution of $1''$ and an angular size of $10'$. This data set
is referred to as {\tt BM12}.

We also consider series of shorter integration scans. We select
$5' \times 8'$ raster scans of moderately bright to very bright point
sources by thresholding the flux density estimates at $1~\rm{Jy}$ at both
wavelengths.
After the baseline scan selection, as described in
Sect.~\ref{se:data_selection}, the data set comprises 154 
scans towards the giant planets Uranus and Neptune, the secondary calibrator
MWC349 and the quasars 3C84, 3C273, 3C345 and 3C454 (aka
2251+158). For these short scans, which are referred to as {\tt
R154}, the data are reduced and projected onto $2''$
resolution maps. 

{\lp Finally, we use a series of 75 observation scans of Uranus and
Neptune, which includes both \bm\ and $5' \times 8'$ raster scans. 
This data set, which is referred to as {\tt UN75}, consists of all the
scans of Uranus and Neptune acquired during the \emph{reference}
observation campaigns (N2R9, N2R12 and N2R14).}

\subsubsection{Results}
\label{se:mainbeam_results}

We have derived the main beam FWHM for the three arrays and the
$1\,\rm{mm}$ arrays combination using the three methods presented in
Sect.~\ref{se:mainbeam_methods} and the data
sets of Sect.~\ref{se:mainbeam_dataset}.
Namely, our main beam FWHM estimates
consist of i) the median FWHM estimate using {\tt Prof-3G} on the
{\tt BM12} dataset, ii) the average FWHM estimate using {\tt Prof-1G}
on the {\tt UN75} data set and the {\tt Map-1G} average FWHM using
either {\tt BM12} or {\tt R154}. 
By comparing these results, we test the stability of the FWHM
estimates against the choices of the data set and of the estimation
method. 

In the case of Uranus, the FWHM estimates are further corrected for
the {\lp average beam broadening induced by the extension of the
apparent disc of the planet, which is $0.19''$ and $0.12''$ at 1 and
$2\,\rm{mm}$, respectively.}
{\lp During the observation period, Uranus disc diameter has varied
from $3.3''$ to $3.7''$. This diameter variation translates into beam
broadening variations of an amplitude of a few tenth of arcseconds,
which are neglected.}  

The results of this analysis are
gathered in Table~\ref{tab:fwhm}, including uncertainties evaluated as
the rms dispersion of single-scan based FWHM estimates.
{\tt Prof-1G} and {\tt Map-1G} results are in agreement within
uncertainties, whereas {\tt Prof-3G} yields slightly smaller FWHM.
{\lp The latter is more robust against the error beams and large radii
beam features than the formers, as expected.}
Combined results are obtained from an error-weighted
average of the four FWHM estimates for each array.
Because the rms errors estimated using the 12 \bm\ scans may be
optimistic considering the small statistic, they are conservatively
increased to match the uncertainty estimates based on the {\tt R154}
data set before performing the weighted average.  
The combined results, as given in
Table~\ref{tab:fwhm}, provide a robust evaluation of the
FWHM. Hence, we report main beam FWHMs of $11.1'' \pm 0.2''$ at
$1\, \rm{mm}$ and $17.6''\pm 0.1''$ at $2\, \rm{mm}$.  

\begin{table*}[!thbp]
  \caption[]{Estimates of the main beam FWHM in arcsec, using three estimation methods (see
    Sect.~\ref{se:mainbeam_methods}) and three data sets
    (see Sect.~\ref{se:mainbeam_dataset}), and their combination.}
  \label{tab:fwhm}
  \centering
  \begin{tabular}{llrrrr}
    \hline\hline
    \noalign{\smallskip}
    Method   &    Dataset   &  \multicolumn{4}{c}{FWHM ['']} \\
    \noalign{\smallskip}\cline{3-6}\noalign{\smallskip}
        &    &   A1 &  A3 & A1 $\&$ A3 &  A2  \\
    \noalign{\smallskip}
    \hline
    \noalign{\smallskip}
    {\tt Prof-3G}  &  {\tt BM12}    & $10.8 \pm 0.1$  &  $10.8 \pm 0.1$  & $10.8 \pm 0.1$  &  $17.4 \pm 0.1$  \\
    {\tt Prof-1G}  &  {\tt UN75}    & $11.3 \pm 0.4$  &  $11.2 \pm 0.4$  & $11.2 \pm 0.3$   & $17.4 \pm 0.2$  \\ 
    {\tt Map-1G}   &  {\tt R154}    & $11.3 \pm 0.2$  &  $11.1 \pm 0.2$  & $11.2 \pm 0.2$  &  $17.8 \pm 0.1$  \\ 
                   &  {\tt BM12}    & $11.2 \pm 0.1$  &  $11.1 \pm 0.1$  & $11.2 \pm 0.1$  &  $17.6 \pm 0.1$  \\
    \noalign{\smallskip}
    \hline
    \noalign{\smallskip}
    \multicolumn{2}{c}{Combined}               & $11.1 \pm 0.2$  & $11.0 \pm 0.2$  & $11.1 \pm 0.2$  &  $17.6 \pm 0.1$  \\
    \noalign{\smallskip}
    \hline
  \end{tabular}
\end{table*}

\subsubsection{Stability checks}
\label{se:mainbeam_stability}

\begin{figure}[!thbp]
\begin{center}
  \includegraphics[clip, width=0.4\textwidth]{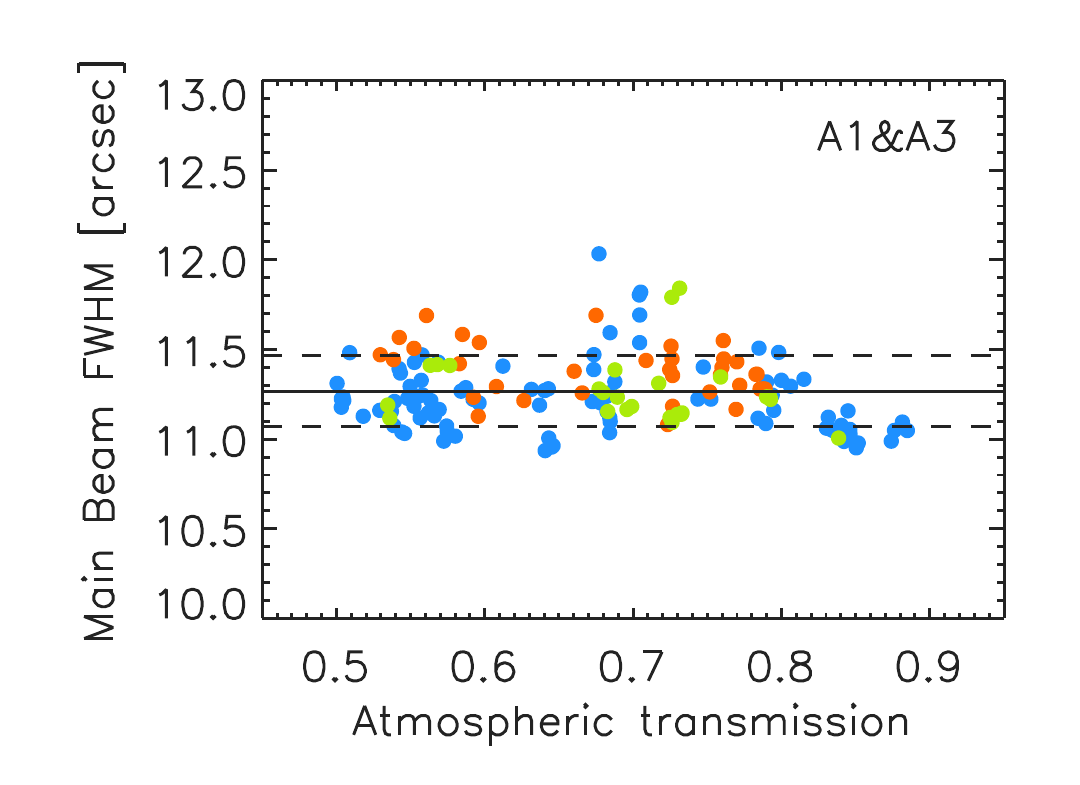}
  \includegraphics[clip, width=0.4\textwidth]{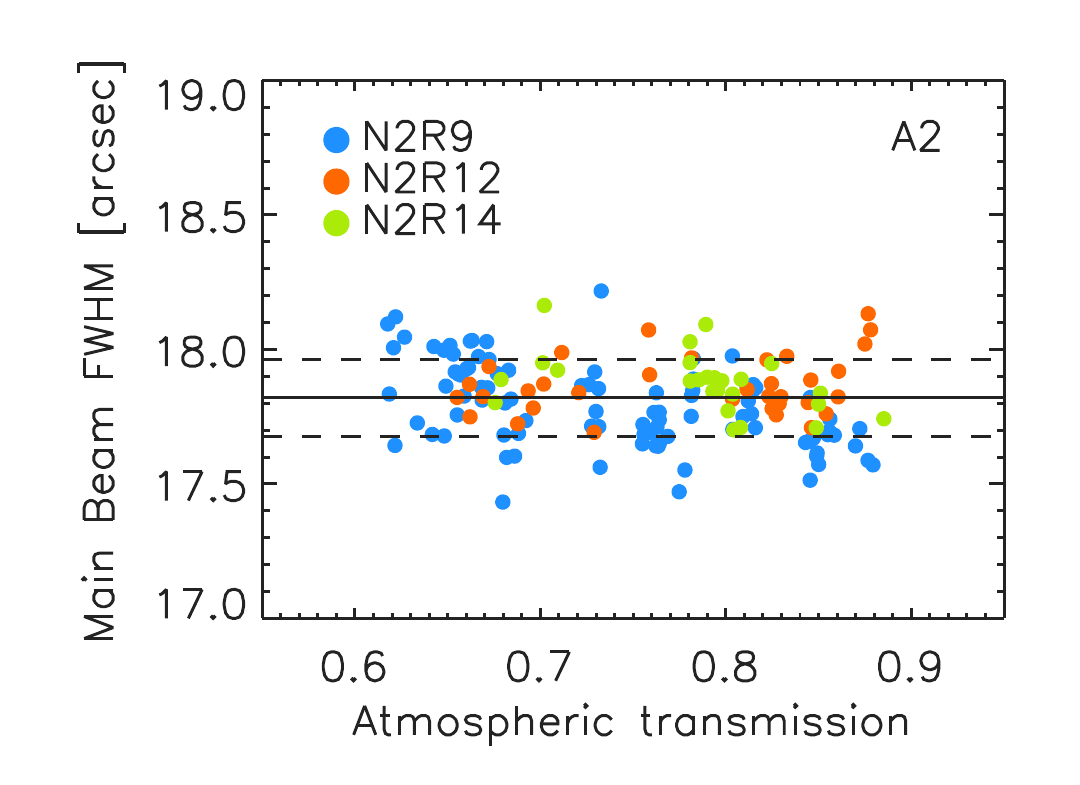}
  \caption[Main Beam FWHM]{Main beam FWHM estimates for the
    $1\,\rm{mm}$ (top) and $2\,\rm{mm}$ (bottom) channels are shown as
    a function of the atmospheric transmission estimated at the
    corresponding wavelengths using bright point source observations
  acquired during the \emph{reference} observation campaigns (N2R9, N2R12, N2R14).}
\label{fig:fwhm_map_atmtrans}
\end{center}
\end{figure}

Figure~\ref{fig:fwhm_map_atmtrans} shows the main beam FWHM estimates
using {\tt Map-1G} as a function of atmospheric transmission,
which is modelled as $\exp{\left(-\taunu \, x\right)}$. 
The main beam FWHM estimates using data of the three campaigns are in
agreement within rms errors. Moreover, the main beam FWHM is stable
against atmospheric conditions at both wavelengths. Slightly lower
values than average (about $11''$) are observed in the best
atmospheric conditions at $1\,\rm{mm}$ providing us with a lower limit
in the absence of correlated atmospheric noise residuals. We note
three scans acquired during the N2R12 campaign with larger FWHM than average at
$2\,\rm{mm}$ although the atmospheric transmission was excellent: this
is likely an effect of atmospheric instabilities
, which affected a large number of observation scans during N2R12. 


\subsection{Main beam efficiency}
\label{se:beam_efficiency}

We derive the main beam efficiency for each array, which is defined as the
ratio of the solid angle sustained by the main beam to the total beam
solid angle.

{\rev To estimate the total beam solid angle, we primarily use the maps
of the beam pattern presented in Sect.~\ref{se:fullbeam}, which
provide us with accurate representation of the main beam and of the
first error beams. However, at angular scales larger than the radius
of the noise decorrelation mask, 90'', the
filtering effect induced by the data processing becomes
non negligible (see also Sect.~\ref{se:dataproc}). Furthermore,
heterodyne observations at the IRAM \trentemetre\ telescope
towards the lunar edge and estimates of the forward beam efficiency using
skydips show that a significant fraction of the full beam stems
from angular scales much larger than $90''$~\citep{Greve1998,
Kramer2013}. Thus, the accurate assessment of the total beam solid
angle requires to account for the filtering effect and the large
angular scales contributions. To that aim, we resort to an hybrid
approach. We utilize both the full beam pattern measurements with
NIKA2 as presented in Sect.~\ref{se:fullbeam}, and the results of the
IRAM \trentemetre\ telescope beam pattern characterization using EMIR
front-end observations~\citep{Carter2012}, as reported
in \citet{Kramer2013}. These results came in two flavours.}

{\rev First, the main beam and error beams are measured using
observations towards the limbs of the Moon. As EMIR detectors were coupled to
the \trentemetre\ telescope entrance pupil via corrugated horns, the
contribution of the first error beam is significantly attenuated
compared to the NIKA2 case. In fact, the first error beam measured by
NIKA2 is not detected with EMIR, while the second error beam is
measured with both NIKA2 and EMIR at compatible levels. The third and
fourth error beams, which originate from the IRAM \trentemetre\
telescope frame misalignment and panel deformations, respectively, are
expected to be measured by EMIR without
attenuation~\citep{Kramer2013}. We assume these latter are
characteristics of the \trentemetre\ telescope without any significant
dependencies on the receiver instrument.}

{\rev Second, the IRAM \trentemetre\ telescope forward efficiencies
$F_{\rm{eff}}$ are derived using skydip scans with EMIR. Their
measurements allow us to estimate the far side lobes efficiencies,
which receive two different contributions. The \emph{forward}
spillover and scattering efficiencies are estimated as $F_{\rm{eff}}$
subtracted from the main beam and error beams efficiencies, and
the \emph{rearward} spillover and scattering efficiencies equal
$1-F_{\rm{eff}}$~\citep{Kramer2013}.}

{\rev To sum up, we estimate the solid angle of the total beam as
\begin{equation}
\begin{split}
  \Omega_{\rm{tot}}(\nu) =& \int_0^{2\pi} \left( \tilde{B}_{\nu}(r)
  + \mathcal{A}_{\nu}^{(3)} G_{\nu}^{(3)}(r) + \mathcal{A}_{\nu}^{(4)}
  G_{\nu}^{(4)}(r) \right)  2 \pi r dr  \\
  &+ \,  \Omega_{\rm{FSL}}
  (\nu),
  \end{split}
  \label{eq:omega_tot}
\end{equation}
where $\tilde{B}_{\nu}(r)$ is the normalised beam profile $B_{\nu}(r)$
for the array $\nu$, as discussed in Sect.~\ref{se:fullbeam}, after
rescaling with a factor of
$(1-\mathcal{A}_{\nu}^{(3)}-\mathcal{A}_{\nu}^{(4)})$; 
$\mathcal{A}_{\nu}^{(3)}$ and $\mathcal{A}_{\nu}^{(4)}$ are the
amplitudes of the third and fourth Gaussian error beams
$G_{\nu}^{(3)}$ and $G_{\nu}^{(4)}$, respectively, which are measured
with EMIR and extrapolated for the array $\nu$ following the
prescription given in \citet{Kramer2013}; $\Omega_{\rm{FSL}} (\nu)$ is
the contribution of the far side lobes to the total beam solid angle
of the array $\nu$, as derived from $F_{\rm{eff}}$ measurements.}

{\rev In addition, for cross-checks and stability tests, we also compute the
total beam solid angle up to a maximum radius
\begin{equation}
 \Omega_{r_{\rm{max}}} (\nu) = \int_0^{r_{\rm max}} B_{\nu}(r)  \,  2 \pi r dr, 
\label{eq:omega_rmax}
\end{equation}
where $r_{\rm{max}}$ is taken as the largest radius for which the
filtering effect of the data processing has no significant impact.
We take $r_{\rm{max}}$ as the radius of the noise decorrelation mask,
which is 90'' and evaluate {\rev $\Omega_{90}$} from the measured beam profiles
obtained using the {\tt UN75} data set (see Sect.~\ref{se:mainbeam_dataset}).
Results per observational campaigns, with uncertainties evaluated
as the rms scatter on the average, are given in
Table~\ref{tab:solid}. The $\Omega_{90}$ estimates are statistically
compatible from one campaign to another with some variations of the
average value due to the dispersion in the focus settings and
atmospheric conditions that prevail during each campaign. The combined
results, based on the whole {\tt UN75} data set, are the
inverse-variance weighted average of the $\Omega_{90}$ for the three
campaigns, while the error bars are conservatively estimated as the
maximum half-difference between the $\Omega_{90}$ estimates.}
  
\begin{table}[!h]
\caption{Estimates of the solid angle of the beam up to a radius of
{\rev $90''$, $\Omega_{90}$}, and rms uncertainties given in
arcsec$^{2}$ using Neptune and Uranus scans acquired during three
observation campaigns, and the combined result. For each case, the
number of scans is given in the column labelled 'nbs'.}
\label{tab:solid}
\centering
\begin{tabular}{l ccccc}
\hline\hline
\noalign{\smallskip}
Campaign  & nbs & A1 & A3 & A1\&A3 & A2 \\
\noalign{\smallskip}
\hline
\noalign{\smallskip}
N2R9      & 27  &  245$\pm$20    &  233$\pm$18  & 239$\pm$15  & 452$\pm$16 \\
N2R12     & 20  &  209$\pm$9    &  203$\pm$8 &  206$\pm$7   & 422$\pm$9 \\
N2R14     & 28  &  232$\pm$14    &  228$\pm$18 & 230$\pm$14 & 441$\pm$14 \\
Combined  & 75  &  219$\pm$18    &  211$\pm$15   & 215$\pm$17    &  432$\pm$15 \\
\noalign{\smallskip}
\hline
\end{tabular}
\end{table}

%


{\rev The main beam solid angle is evaluated from the main beam (mb) FWHM, as
$\Omega_{\rm{mb}} = 2 \pi\,  \sigma_{\rm{mb}}^2$, where
FWHM$ =2\sqrt{2\ln{2}}\, \sigma_{\rm{mb}}$. The main beam efficiency
\begin{equation}
\rm{BE} = \frac{\Omega_{\rm{mb}}}{\Omega_{\rm{tot}}}
\end{equation}
is evaluated using both the {\tt UN75} and the {\tt BM12} data sets,
as presented in Sect.~\ref{se:mainbeam_dataset}.} For cross-checks, we
compare the results based on three estimates of the total beam and
main beam solid angles:
\begin{itemize}
  \item{{\tt BE1} relies on the best-fitting parameters of the
    three-Gaussian model of the full beam, as given in
    Eq.~\ref{eq:3gauss}, to derive both the main beam solid angle and
    the two first errors beam contributions to the total beam solid
    angle. The main beam solid angle thus corresponds to the volume
    enclosed by the first Gaussian, as obtained using {\tt Prof-3G},
    while the normalised beam profile in Eq.~\ref{eq:omega_tot} is the
    normalised best-fitting $B_{3G}(r)$;}
  \item{{\tt BE2} consists in using the normalised beam profile
    measurements from the {\tt UN75} data set to estimate
    $\Omega_{\rm{tot}}$, while $\Omega_{\rm{mb}}$ is
    derived with the FWHM obtained using {\tt Prof-1G} (see
    Sect.~\ref{se:mainbeam_methods});}
  \item{{\tt BE3} is similar to {\tt BE2} but relies on the {\tt BM12}
    data set, while the main beam FWHM is derived using {\tt Map-1G}.}  
\end{itemize}
For all methods, the contributions of the third and fourth error beams
and of the far side lobes, which enters in Eq.~\ref{eq:omega_tot}, are
the same. 


The main beam efficiency estimates using the three methods are gathered
in Table~\ref{tab:beam_efficiency}: central values and error
bars are evaluated as the median and the rms error of the
estimates on individual observation scans, respectively. 
{\rev We combined the results of the three methods, which are in
agreement with each others, using an inverse variance-weighted
average and a quadratic mean of the rms uncertainties. 
The main beam efficiency uncertainties, as given in
Table~\ref{tab:beam_efficiency}, also include uncertainties on the
third and fourth error beam contributions and on the \trentemetre\
telescope forward efficiencies, as reported
in \citet{Greve1998, Kramer2013}.}
Using the combined results, we report main beam efficiencies of
{\rev $47 \pm 3 \%$ at $1\,\rm{mm}$ and  $64 \pm 3 \%$ at $2\,\rm{mm}$.}

\begin{table}[!h]
  \caption[]{Main beam efficiency estimated for each array or array
  combination, using three different methods, and the combined
  results, given in percent.}
  \label{tab:beam_efficiency}
  \centering
  \begin{tabular}{l cccc}
    \hline\hline
    \noalign{\smallskip}
    %
    Method & A1 &  A3 & A1 $\&$ A3 &  A2  \\
    \noalign{\smallskip}
    \hline
    \noalign{\smallskip}
    {\tt BE1}  &  $46 \pm 2$  & $47 \pm 2$  &  $47 \pm 3$  &  $62 \pm 3$  \\
    {\tt BE2}  &  $45 \pm 4$  & $46 \pm 4$  &  $45 \pm 4$  &  $63 \pm 3$  \\
    {\tt BE3}  &  $49 \pm 3$  & $50 \pm 3$  &  $50 \pm 3$  &  $64 \pm 2$  \\
    combined   &  $47 \pm 3$  & $48 \pm 3$  &  $47 \pm 3$  &  $64 \pm 3$  \\
    \noalign{\smallskip}
    \hline
  \end{tabular}
\end{table}
\begin{table}[!h]
\caption{Estimates of the total beam solid angle and main beam
efficiency including in turn further contributions to the full beam ranked by
their angular scales (see text). All solid angles are given in arcsec$^{2}$,
while the main beam efficiencies are in percent.}
\label{tab:solid_corr}
\centering
\begin{tabular}{lcc}
\hline\hline
\noalign{\smallskip}
&  A1\&A3 & A2 \\
\noalign{\smallskip}
\hline
\noalign{\smallskip}
$\Omega_{90}$        &    211 $\pm$  12 & 434 $\pm$ 13 \\\noalign{\smallskip}
$\Omega_{\rm{hyb}}$   &    264 $\pm$  13 & 504 $\pm$ 13  \\\noalign{\smallskip}
$\Omega_{\rm{tot}}$   &    290 $\pm$  14 & 541 $\pm$ 18  \\\noalign{\smallskip}
\noalign{\smallskip}
\hline
\noalign{\smallskip}
$\rm{BE}_{90}$       &   64 $\pm$ 5  &  80 $\pm$ 3  \\\noalign{\smallskip}
$\rm{BE}_{\rm{hyb}}$  &   52 $\pm$ 3  &  69 $\pm$ 2  \\\noalign{\smallskip}
$\rm{BE}$            &  47  $\pm$ 3  &  64 $\pm$ 3  \\\noalign{\smallskip}
\noalign{\smallskip}
\hline
\label{tab:solid_corr}
\end{tabular}
\end{table}

{\rev Finally, we compute the total beam solid angle and main beam
efficiency using the combination of the three previously described
methods, in three cases that successively include larger angular scale
contributions to the full beam: $\Omega_{90}$ and $\rm{BE}_{90}$ are
the total beam solid angle and main beam efficiency integrated up to
$90''$, which include the main beam and the two lower angular scales
error beams, as measured in NIKA2 beam maps; $\Omega_{\rm{hyb}}$ and
$\rm{BE}_{\rm{hyb}}$ further account for the third and fourth error
beams, as derived from EMIR front-end measurements; $\Omega_{\rm{tot}}$
and $\rm{BE}$ are the final estimates that comprise all beam
contributions. The results are given in Table~\ref{tab:solid_corr} for
the 1\,mm and 2\,mm channels. These give insight on the relative
importance of the contributions at various angular scales. For
example, 18$\%$ and 13$\%$ of the total beam solid angle stem mainly
from the two largest error beams integrated at angular
scales larger than 90'', and 9$\%$ and 7$\%$ originate from the far
side lobes at 1 and 2\,mm respectively.}

\section{Atmospheric opacity}
\label{se:opacity}

The atmospheric opacity constitutes the ultimate limitation of
ground-based experiments. Only a fraction of the source 
signal is transmitted by the atmosphere and reaches NIKA2 detectors. 
The relation between the observed flux density
$\tilde{S}_{\nu}$ and the top-of-the-atmosphere flux density $S_{\nu}$
is parametrized by the zenith opacity $\taunu$
and the \airmass\ $x$ as 
\begin{equation}
\tilde{S}_{\nu} = S_{\nu} \, e^{-\taunu  x}.
\label{eq:uncorr_flux}
\end{equation}

An accurate derivation of the atmospheric opacity for each scan is
of the utmost importance to retrieve the source signal and thus, to
ensure small calibration uncertainties.
We developed three atmospheric opacity derivation methods, which are described in
Sect.~\ref{se:opacity_methods}. In Sect.~\ref{se:opacity_tests}, we
present robustness tests.

\subsection{Atmospheric opacity estimation}
\label{se:opacity_methods}

We have developed three procedures for the atmospheric opacity
derivation: i) {\tt taumeter} relies on measurements
provided by the resident IRAM \taumeter\ operated at $225\,\rm{GHz}$;
ii) {\tt skydip} consists in using NIKA2 as a total-power \taumeter\ 
by resorting to a series of {\tt skydip} scans;
iii) {\tt corrected skydip} is a modified
version of the {\tt skydip} method that minimizes the dependence of the
estimated flux density on the opacity.

All methods i) do not rely on an ATM model nor on any
hypothesis on the atmospheric contents for the sake of robustness, and
ii) do not use the laboratory measurements of the bandpass (see 
Sect.~\ref{se:instru_bandpass}) for more accuracy.  

Sect~\ref{se:taumeter-method} presents the {\tt taumeter} method. The 
{\tt skydip} method is described in Sect~\ref{se:skydip-method} and
the selection of the used {\tt skydip} scans is discussed in
Sect.~\ref{se:skydip-selection}.
{\tt corrected skydip} is presented in Sect.~\ref{se:corrected-skydip}.

\subsubsection{The {\tt taumeter} method}
\label{se:taumeter-method}

The IRAM \trentemetre\ telescope facility is equipped with a
resident \taumeter\ operated at 225\,GHz. Every four
minutes, it performs elevation scans at fixed azimuth
to monitor the atmospheric opacity.
The IRAM science support
team provides us with time-stamped zenith opacities at $225\,\rm{GHz}$
$\tau_{225}$, as derived from the \taumeter\ measurements. The
$\tau_{225}$ estimates come in two different flavours: one relying
on a linear model and the other on an exponential fitting model. They
are then filtered by removing outliers and by using a threshold on
goodness-of-fit criteria.
Based on IRAM experience, we use the linear fit and filtered $\tau_{225}$
data for the NIKA2 analysis. The time-stamped $\tau_{225}$ estimates,
which are sampled about every 4 minutes, are interpolated to the time
of the NIKA2 scans (we consider the time of the middle of the
scan). For cross-check we also produce a smooth version of time-stamped
$\tau_{225}$ data by
filtering with a running median of seven samples, which is then
interpolated to the NIKA2 scan times.

We fit the relations between the IRAM
$225\,\rm{GHz}$ \taumeter\ opacities and NIKA2 band pass opacities using
observation of calibration sources which spans a large range of air
masses. This method has the advantages of not relying on atmospheric
model nor on the bandpass measurements in laboratory.
We use a series of 64 scans of MWC349, which consists of the
\baseline\ selected subset of scans from the 68 available scans for
this source during N2R9.
It constitutes an homogeneous data set in flux density but
heterogeneous in atmospheric conditions: zenith opacities at
$225\,\rm{GHz}$ range from 0.08 to 0.32 and elevations from $23$ to $73$
degrees, spanning a large range of \airmass\ as required.
NIKA2 opacities $\taunu$, for $\nu$ corresponding to the observing
frequency of Array 1, 2, 3 and the combination of Arrays 1 and 3, are estimated
from the $225\,\rm{GHz}$ \taumeter\ median-filtered linear-based opacity
estimates $\tau_{225}$ as
\begin{equation}  
  \tau_\nu =  a_\nu^{225}\tau_{225} + b_\nu^{225}.
  \label{eq:taumeter_model}
\end{equation}
The parameters $a_\nu^{225}$ and $b_\nu^{225}$ are fitted
to the data set so that the source flux density 
\begin{equation}  
  S_\nu = \tilde{S}_\nu\,\, e^{(a_\nu^{225}\tau_{225} + b_\nu^{225}) \, x}, 
  \label{eq:opacorr_taumeter}
\end{equation}
is constant within scans.

\begin{table}[!htbp]
  \begin{center}
    \caption[IRAM \taumeter\ to NIKA2 opacity model]{{\lp Best-fit
    parameters and rms uncertainties to infer NIKA2 opacities from the IRAM \taumeter\ measurements.}}
    \label{tab:tau225-to-taunika}  
    \begin{tabular}{lrrrr}
      \hline
      \hline
      \noalign{\smallskip}
      Parameters & Array 1 & Array 3  & Array 1$\&$3 & Array 2  \\
      \noalign{\smallskip}
      \hline
      \noalign{\smallskip}
      $a_\nu^{225}$         & $1.94$   &  $1.90$ &  $1.92$ & $0.94$ \\
      $b_\nu^{225}$         & $-0.04$  & $-0.07$ & $-0.06$ & $0.00$ \\
      $\Delta a_\nu^{225}$  & $0.15$  & $0.08$  &  $0.09$ & $0.10$ \\
      $\Delta b_\nu^{225}$  & $0.05$  & $0.03$  & $0.04$ & $0.03$ \\
      \noalign{\smallskip}
      \hline
    \end{tabular}
  \end{center}    
\end{table}

We tested two estimators of the flux stability. The first one relies
on minimizing the standard deviation of the measured-to-median flux
densities ratio after atmospheric opacity correction using
Eq.~\ref{eq:opacorr_taumeter}. The second one is obtained by minimizing
\begin{equation}
\chi^2 = \sum_{i=1}^{N} \frac{1}{\sigma^2} \, \left( \frac{S_\nu}{\rm{Med}(S_\nu)} -1 \right)^2,  
\end{equation}
where $\sigma$ is the rms uncertainty of the flux density estimates. Note
that these estimators do not depend on
the absolute scale of the flux density of the source. Both estimators
yield consistent results that are combined and gathered in
Table.~\ref{tab:tau225-to-taunika}. The quoted errors
$\Delta a_\nu^{225}$ and $\Delta b_\nu^{225}$ are 1-$\sigma$ errors of
the fit.

{\rev We note that the atmospheric opacity correction using {\tt
taumeter} opacities yields stable flux density measurements with
respect to the atmospheric transmission by construction, as we will
verify in Sect.~\ref{se:photometry}. This means that the best-fitting
parameters of the model in Eq.~\ref{eq:taumeter_model} also correct
for any line-of-sight opacity dependent systematic effects on the flux
densities, as it will be further discussed in
Sect.~\ref{se:corrected-skydip} and Sect.~\ref{se:opacity_tests}.}

Because the IRAM \taumeter\ observes at a fixed azimuth, the
{\tt taumeter} opacities are not the line-of-sight opacities
for the observation scans. As this will be checked in
Sect.~\ref{se:photometry}, this induces larger rms errors of
the top-of-the-atmosphere flux density estimates compared to
opacity correction methods that relies on NIKA2 skydip-based
measurements. The {\tt taumeter} method will thus be used
as an alternative method in case of failure of the NIKA2 skydip-based
methods and to perform consistency checks.

\subsubsection{NIKA2 skydip-based method}
\label{se:skydip-method}

The NIKA2 {\tt skydip} method for the opacity derivation consists in
using the NIKA2 instrument as an in-band total-power \taumeter. {\lp The opacity
integrated in the NIKA2 bandpasses and in the line-of-sight of the
observing scan is thereby directly obtained.} 
This idea,
which was successfully tested with NIKA~\citep{Catalano2014}, relies
on the fact that the resonance frequency of each KID varies linearly
with the total power, as discussed in Sect.~\ref{se:tuning}.
First, we have to calibrate the relationship between total
power and opacity. Then, we can use this calibration to measure the
opacity during a given scan. 

First, we detail the opacity calibration. For each KID $k$, the
absolute value of its resonance frequency $f_{\rm{reso}}^k$ varies with the
atmospheric load according to
\begin{equation}
f_{\rm{reso}}^k  = c_0^k - c_1^k T_{\rm{atm}}[1-e^{-\taunu x}]
\label{eq:skydip}
\end{equation}
where $c_0^k$ is a constant equal to the KID resonance frequency at zero
opacity, $c_1^k$ is a calibration conversion factor in Hz$/$K,
$T_{\rm{atm}}$ is the temperature of the atmosphere. 
By assuming a homogeneous plane-parallel atmosphere, the
\airmass\ $x$ is defined from the elevation as
$x = \left(\sin\elev\right)^{-1}$. {\lp The Earth sphericity can be safely
neglected at the elevations under discussion here.}
We take $T_{\rm{atm}}$ as a constant equal to 270\,K. However, the opacity is
expected to slightly depend on the atmospheric temperature. For
example, in poor weather conditions (6\,mm of water vapour contents),
the zenith opacity in {\lp both observing bands} can vary by
about 10\% for temperature variations of 10\,K. The
effect of the temperature variations on the final
calibration of NIKA2 response to the sky load is mitigated by using
several dedicated observation scans regularly distributed all along an
observation campaign.

\begin{figure}[!htbp]
\begin{center}
\includegraphics[trim={9cm 0cm 0cm 6.5cm}, clip=true, width=0.9\linewidth]{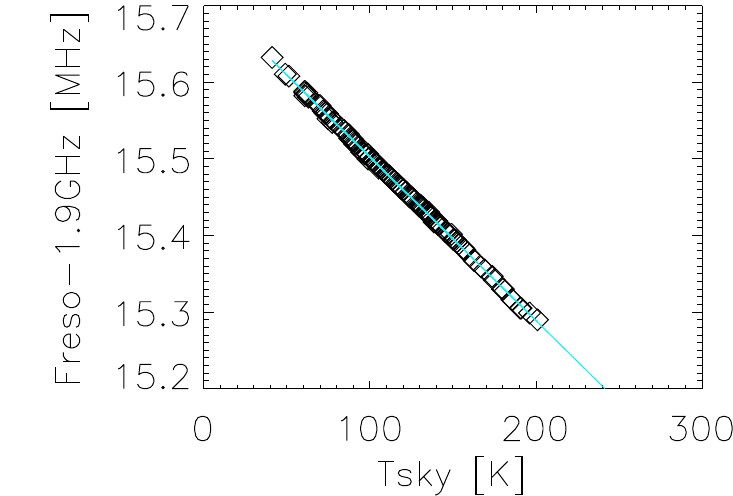}
\caption[]{Example of the global skydip fit for a KID.
Each square point represents one step in a skydip (made of eleven
elevation steps). A series of 12 {\tt skydip} scans are jointly used
spanning zenith opacities from 0.15 to 0.50 {\lp in the $1\,\rm{mm}$ band}. The horizontal axis gives the sky
effective temperature $T_{\rm{sky}} = T_{\rm{atm}}[1-e^{-\taunu\, x}]$ in Kelvin, where $\taunu$ is the
skydip zenith opacity found in the fit. The vertical axis shows the relative
resonance frequency of the KID with respect to 1.9\,GHz, given in MHz. The blue line is the linear
model using the best-fit $c_0^k$ and $c_1^k$ coefficients (see
Eq.~\ref{eq:skydip}).}
\label{fig:skydipfitexample}
\end{center}
\end{figure}

The $c_0^k$ and $c_1^k$ are determined using {\tt skydip} scans, which
consist in moving the telescope in elevation step by step, as
defined in Sect.~\ref{se:skydip}. For each KID $k$, the evolution
of $f_{\rm{reso}}^k$ is monitored as a function of the \airmass\ in each
elevation step to perform a joint fit of the zenith opacity $\taunu$ and
the $c_0^k$ and $c_1^k$ coefficients.
All skydips, obtained under various opacity conditions, are analysed
together to break the degeneracy between the opacity and the
Hertz-to-Kelvin conversion factor $c_1^k$. The degeneracy occurs mostly for low opacity
conditions for which we can only determine the combination
$c_1^k \taunu x$. The procedure has two steps. First, all the {\tt
skydip} scans are analysed individually to extract $f_{\rm{reso}}^k$ for each
stable elevation and for each KID. Secondly, a simultaneous fit is done
for all parameters (one $\taunu$ per skydip, and a set of $c_0^k$ and
$c_1^k$ for all KIDs). Figure~\ref{fig:skydipfitexample} illustrates the
fitting procedure.
{\lp This fit is performed on block of 40 KIDs. We check that the
resulting $\taunu$ from the different blocks are consistent within rms
errors, which are equal to about $4\times 10^{-3}$ at 1\,mm and
$1\times 10^{-3}$ at 2\,mm.}
Once the $\taunu$ values are estimated for each {\tt skydip} (as the average over the
blocks), we compute, while fixing the $\taunu$, the $c_0^k$ and $c_1^k$
final values for each KID $k$ with a linear fit. We thus retrieve
the coefficients of all the KIDs even though some of them could not
contribute to the $\taunu$ determination. 

{\lp We have observed that the  $c_0^k$ and $c_1^k$ coefficients vary
between observational campaigns due to a change in the KID properties
from one cool-down to another.}
However, by comparing the results of different skydips, we have verified that the
coefficients $c_0^k$ and $c_1^k$ are stable, within the fitting errors, on very
long time scales within a cool-down cycle. The coefficients can thus be
applied to the whole observing campaign for the opacity derivation. 
Specifically, the opacity 
is retrieved for each observation scan by
inverting Eq.~\ref{eq:skydip} as:
\begin{equation}
\taunu =   \rm{Med}\left( -\frac{1}{x} \log\left( \frac{f_{\rm{reso}}^k - c_0^k}{c_1^kT_{\rm{atm}}} +1 \right)\right), 
\end{equation}
where the median is evaluated using all the valid
KIDs of the arrays under concern. Hence, we are able to derive an opacity
integrated in the NIKA2 bandpasses and in the line-of-sight of the
source in the considered observation scan.

\subsubsection{{\tt Skydip} scan selection}
\label{se:skydip-selection}

The {\tt skydip} opacity derivation requires to have on hands a
sizeable amount of {\tt skydip} scans --
typically ten to twenty -- that i) span the whole opacity range and
ii) avoid highly perturbed atmosphere to meet the plane-parallel
atmosphere assumption. To that aim, we perform a {\tt skydip}
scan twice a day during a scientific campaign. Then, the ($c_0^k$, $c_1^k$)
determination process relies on a selection of the {\tt skydip} scans.

For each {\tt skydip} scan and 
for each KID, 
we compute the difference between the measured KID resonance frequency and the model
given in Eq.~\ref{eq:skydip} taken at the best-fit values of the
($c_0^k$, $c_1^k$) parameters{\lp, which is named $df_{\rm{reso}}^k$.} Then, we
determine two indicators of the fit quality per {\tt skydip}. {\lp First,
for each block of 40 KIDs, the standard deviation of $df_{\rm{reso}}^k$ is calculated
over all the KIDs of the block. This standard deviation per KID block
is called $\sigma_{40}$. For each {\tt skydip}, we evaluate the
median rms, which is the median $\sigma_{40}$ over all the KID blocks.} 
%
\begin{figure}[!htbp]
\begin{center}
\includegraphics[clip=true,width=\linewidth]{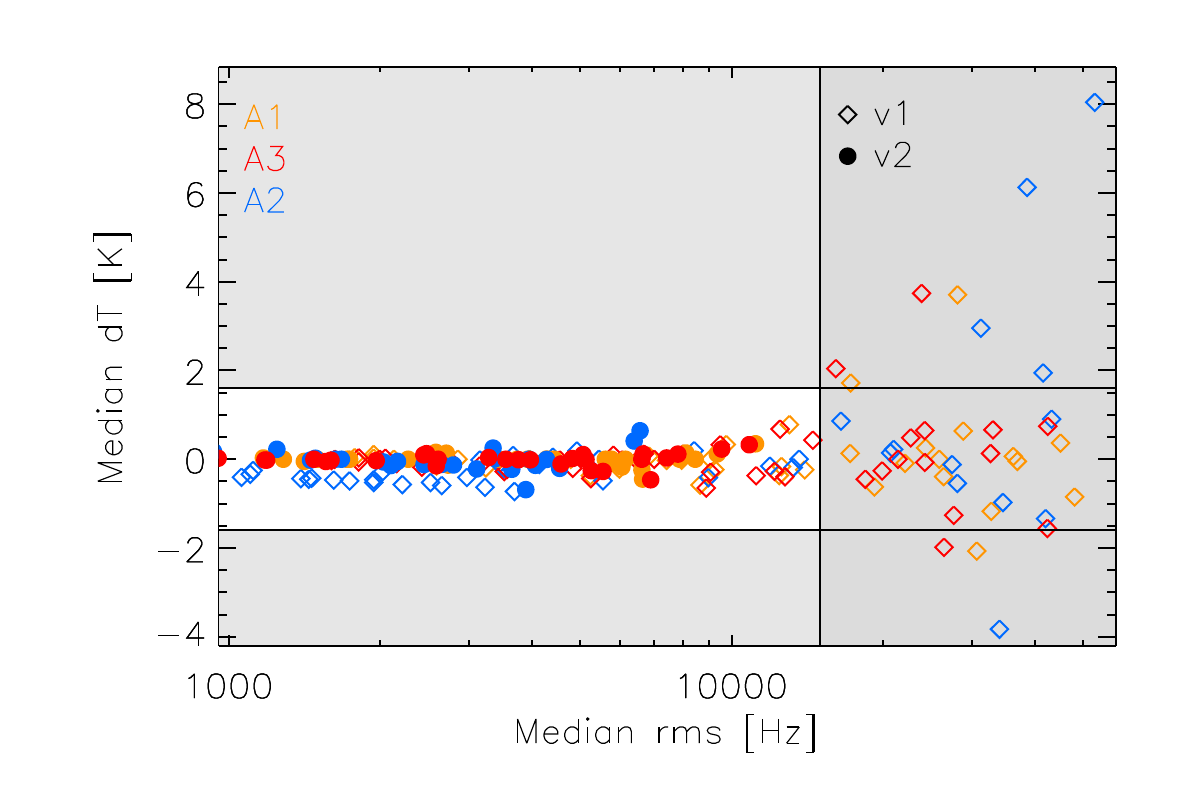}
\caption[N2R9 skydip scan selection.]{ Median dT quality-fit criterion
is plotted as a function of the median rms criterion for each skydip
scan of the N2R9 campaign and for the three arrays. {\lp The {\tt
skydips} that yield a poor fit of the KID resonance frequencies and hence
do not met Median dT criterion, are also discarded using
Median rms. The latter criterion further discards noisy {\tt skydips}.}
Empty diamonds show the results of the first
iteration of the skydip coefficient estimation, labelled 'v1', whereas
filled circles show the second iteration, labelled 'v2', for which only the skydips
that met both fit-quality criteria are included.
After the second iteration, all the remaining skydips met the criteria.}
\label{fig:skydipselection}
\end{center}
\end{figure}
This fit quality indicator is also sensitive to the noise
level during the {\tt skydip}. We therefore devise a second fit quality
indicator to further measure the bias between the data and the
best-fit model.
Namely, for each {\tt skydip}, we compute the average
$df_{\rm{reso}}^k$ of each KID $k$ and convert this quantity from Hertz to Kelvin
using the corresponding $c_1^k$ parameter. This cross-calibration
allows us to compare the $df_{\rm{reso}}^k$ estimates from different KIDs.
Median $dT$ is the median of the average $df_{\rm{reso}}^k$ in Kelvin over all the KIDs of an
array. With these two indicators in hands, we discard the {\tt skydip} scans
that are noisy or that yield a poor fit by applying the following selection
criteria:

\begin{itemize}
\item Median $\rm{rms} < 1.5 \times 10^{4}~\rm{Hz}$
\item Median $dT < 1.6~\rm{K}$
\end{itemize}

The threshold values have been determined using the set of 44 {\tt skydip}
scans of N2R9. The Median rms cut corresponds to twice the median of
this quantity per {\tt skydip} scan, whereas the Median $dT$ cut is twice
the standard deviation of Median $dT$ over the {\tt skydip} scans.
N2R9 {\tt skydip} scan selection is illustrated in
{\lp Fig.~\ref{fig:skydipselection},
which shows the complementarity between the two fit-quality
criteria. After selection, 15 skydips are kept for the final step of
the ($c_0^k$, $c_1^k$) fit in the case of the N2R9 campaign.}

The ($c_0^k$, $c_1^k$) estimation proceeds in two steps: first the
parameters are estimated using all the available {\tt skydip} scans for a
given campaign, then the estimation is repeated using only
{\tt skydip} scans that met the fit-quality criteria. After the second
iteration, we check that no extra {\tt skydip} scan outliers are left, as shown by
the 'v2' label data points in Fig.~\ref{fig:skydipselection}.
\begin{figure*}[!thbp]
  \begin{center}
    \begin{overpic}[clip=true, trim={0, -0.3cm, -0.3cm, 0}, width=0.3\textwidth]{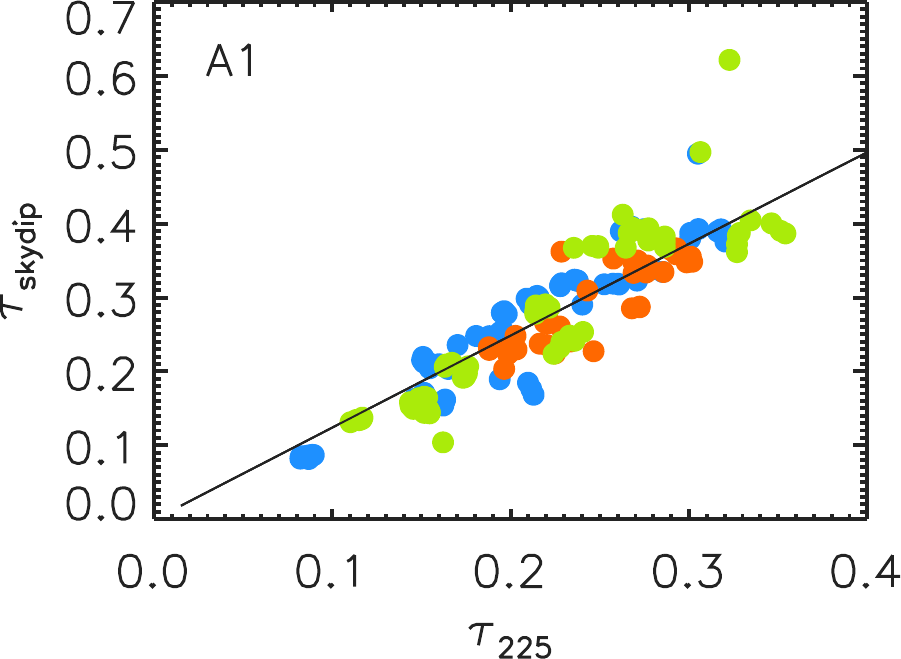}
      \put(0,70){\footnotesize a)}
    \end{overpic}
    \includegraphics[clip=true, trim={0, -0.3cm, -0.3cm, 0}, width=0.3\textwidth]{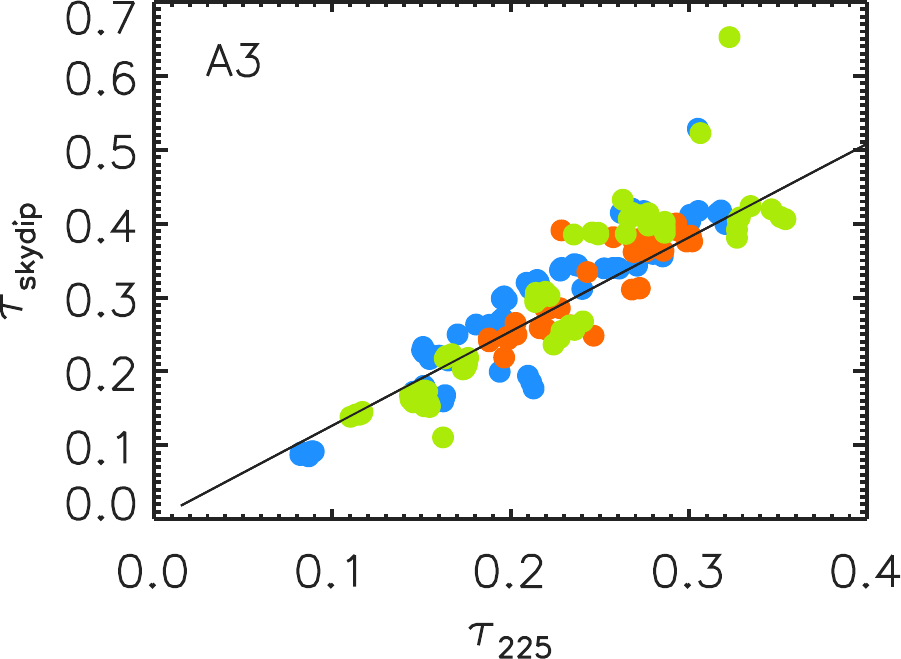}
    \begin{overpic}[clip=true, trim={-0.3cm, -0.3cm, 0, 0}, width=0.3\textwidth]{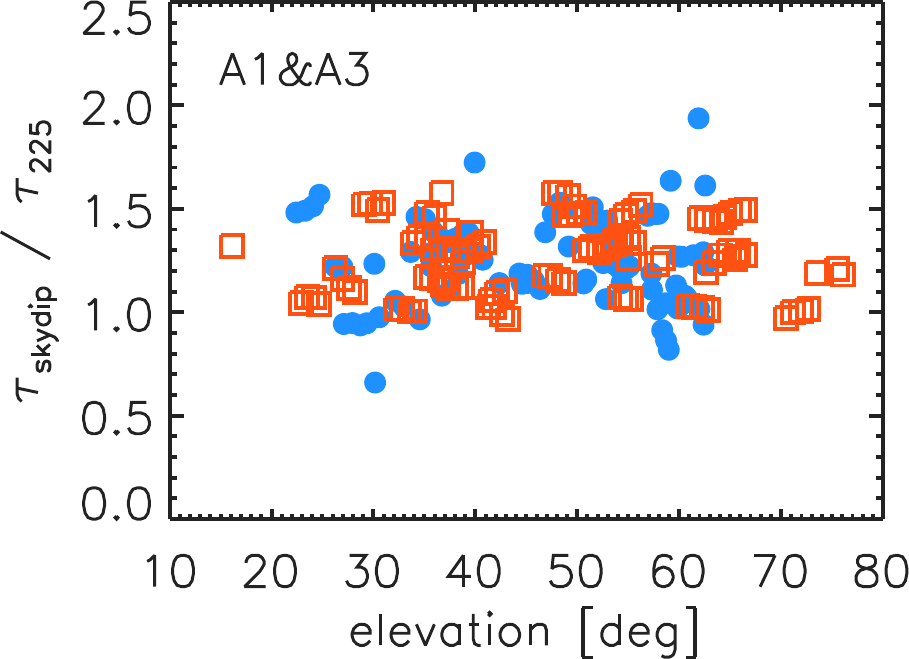}
      \put(0,70){\footnotesize b)}
    \end{overpic}
    \includegraphics[clip=true, trim={0, -0.3cm, -0.3cm, 0}, width=0.3\textwidth]{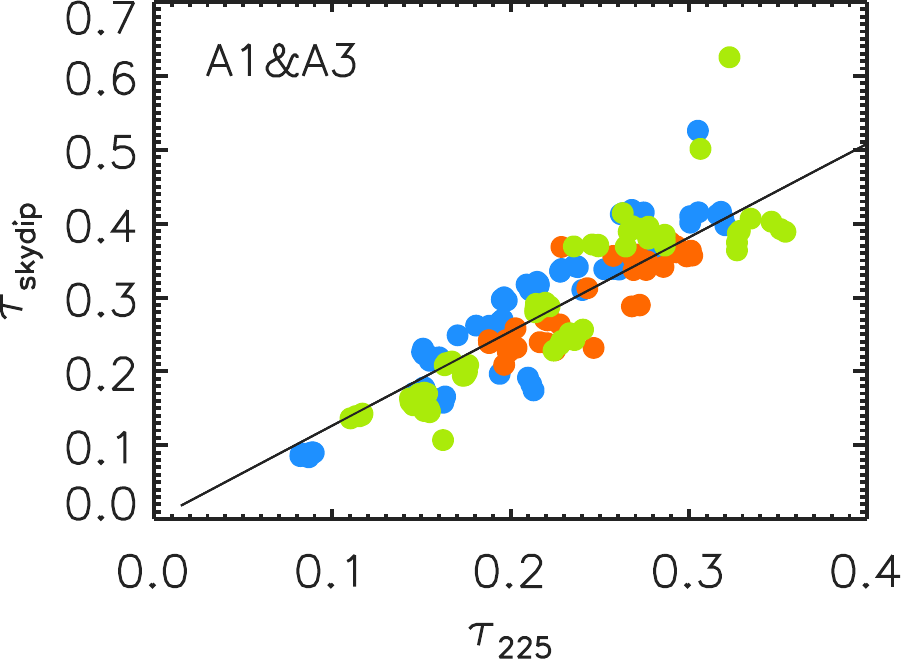}
    \includegraphics[clip=true, trim={0, -0.3cm, -0.3cm, 0}, width=0.3\textwidth]{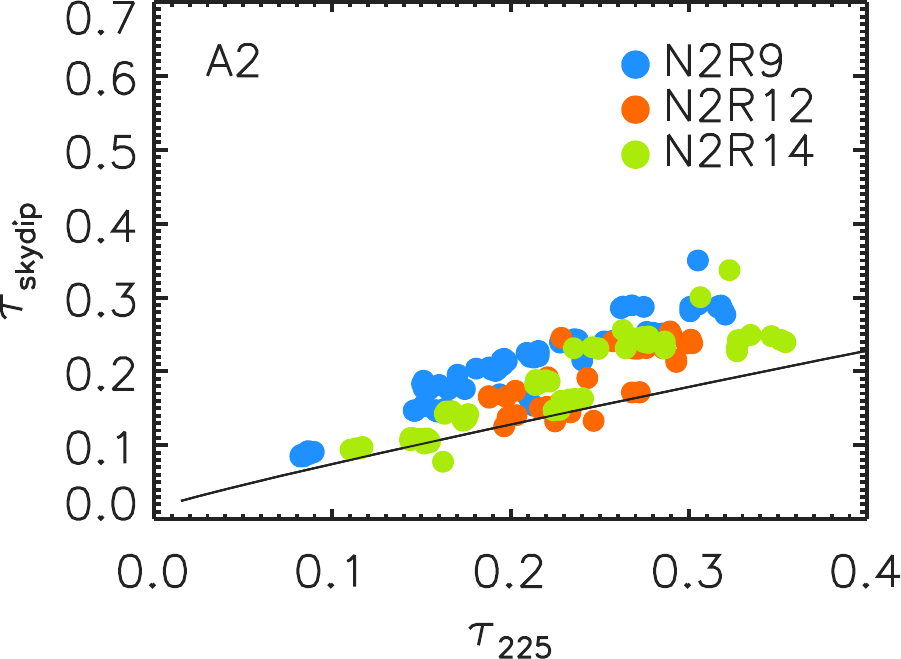}
    \includegraphics[clip=true, trim={-0.3cm, -0.3cm, 0, 0}, width=0.3\textwidth]{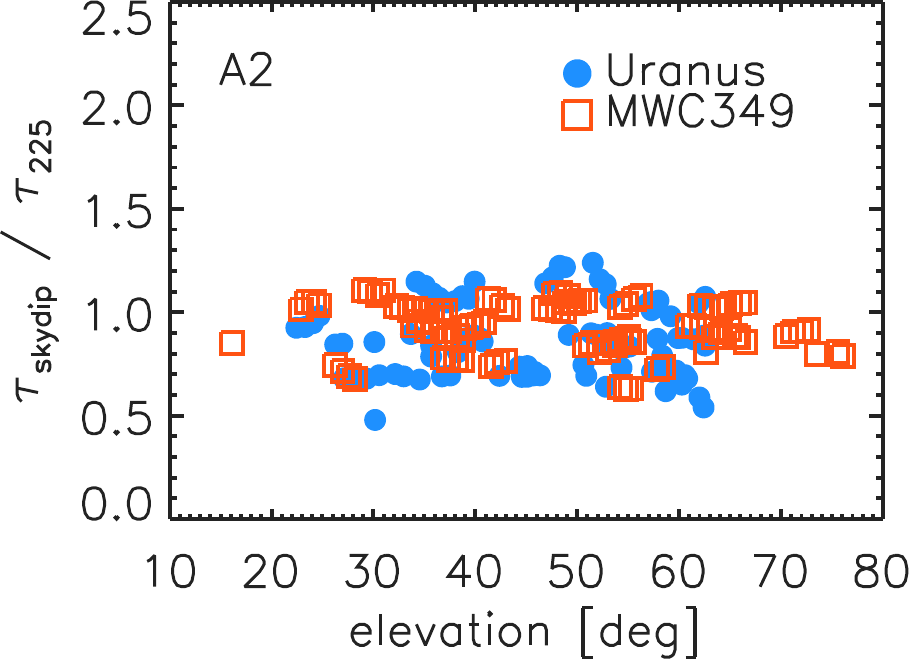}
   \caption[]{NIKA2
     skydip-based opacities $\taunu^{\rm{skydip}}$ consistency checks.
     a) $\taunu^{\rm{skydip}}$ vs median-filtered
    time-stamped IRAM 225\,GHz \taumeter\ opacities (see
    Sect.~\ref{se:taumeter-method}).
    For illustration purpose, the modelled correlations relying on an ATM model integrated in
    NIKA2 frequency bands are shown in black. b) $\taunu^{\rm{skydip}}$ stability against the observing
    elevation. The ratio between the skydip-based opacities and the
    \taumeter-derived opacities is shown as a function of the observing
    elevation as blue points for Uranus scans and empty red square for
    MWC349 scans. See discussion in Sect.~\ref{se:opacity_tests}. } 
\label{fig:skydip-to-taumeter-correl}
\end{center}
\end{figure*}
The stability of the ($c_0^k$, $c_1^k$) parameters and hence the
{\tt skydip} opacity estimates, have been tested against the
choice of the selection criteria. We found that the $\taunu$
estimates are robust against the {\tt skydip}-scan selection as long as the
selection includes good {\tt skydip} scans in high opacity condition
$(\tau_{1\rm{mm}} > 0.44)$ and as the poor
fitting {\tt skydip} scans, which mostly correspond to deviation from the
atmosphere plane-parallel model in high opacity {\tt skydip} scans (as seen in
Fig.~\ref{fig:skydipselection}), are excluded.

\subsubsection{{\tt Corrected skydip}}
\label{se:corrected-skydip}
The opacity estimates are ultimately tested by assessing the
stability of the top-of-the-atmosphere flux densities of bright sources for a large
range of atmospheric conditions, as will be addressed in
Sect.~\ref{se:photometry}. After opacity correction using the
{\tt skydip} $\taunu$ estimate, the calibration flux density
measurements show a residual dependency on the atmospheric
transmission, as discussed in Sect.~\ref{se:photometry}. This has
motivated the development of a corrected version of the {\tt skydip}
method that ensures the robustness of the flux densities against
atmospheric conditions.

{\lp As already noticed in Sect.~\ref{se:skydip-method}, for small values of the
line-of-sight opacity $\taunu x$, only the parameter combination
$c_1^k T_{\rm{atm}}\taunu$ is constrained in
Eq.~\ref{eq:skydip}. Degeneracies between $c_1^k$ parameters, the
atmospheric temperature and $\taunu$ can translate into a
scaling factor in the fitted skydip opacities
$\taunu^{\rm{skydip}}$. To take this effect into account, }  
we use the flux stability estimators described in
Sect.~\ref{se:taumeter-method} to fit a correction to $\taunu^{\rm{skydip}}$ as
\begin{equation}  
  \taunu =  a_\nu^{\rm{skydip}}\taunu^{\rm{skydip}}.
  \label{eq:corrected_skydip}
\end{equation}

We find $a_\nu^{\rm{skydip}}$ of
$1.36 \pm 0.04$,
$1.23 \pm 0.02$,
$1.27 \pm 0.03$ and
$1.03 \pm 0.03$ for A1, A3, A1$\&$A3 and A2 respectively.

Moreover, we test for an additional offset in the
correcting relation of the {\tt skydip} opacities given in
Eq.~\ref{eq:corrected_skydip}. We find best-fitting correcting factors
in agreement with the best-fit values estimated using the single-parameter
correcting relation, whereas the best-fit offsets are compatible with
zero at both wavelengths. We conclude that correcting the {\tt skydip}
opacity estimates for a normalisation as given in
Eq.~\ref{eq:corrected_skydip} suffices for ensuring flux density
robustness against atmospheric opacity conditions.

The exact physical origin for the discrepancy of the empirical factor
$a_\nu^{\rm{skydip}}$ from the expected unity value {\lp for the
$1\,\rm{mm}$ wavebands} is currently under investigation.
{\rev Beside the $c_1^k$-to-$\taunu$ degeneracy and the effect
of the variation of the atmospheric temperature, other explanations
include effects that are not directly related to the atmospheric
opacity derivation, as the empirical factor is measured on the flux
densities.} 
However, the stability of the flux densities corrected using the
{\tt corrected skydip} opacities {\lp and the results consistency using
several observation campaigns}, as discussed in
Sect.~\ref{se:photometry}, constitutes a validation of this approach.

\subsection{Opacity estimate consistency checks}
\label{se:opacity_tests}

First, we test the stability of the {\tt skydip} opacities from one
observation campaign to
another. Panels a) in Fig.~\ref{fig:skydip-to-taumeter-correl} show the
correlation between the {\tt skydip} $\taunu$ estimates $\taunu^{\rm{skydip}}$
and the median-filtered time-stamped IRAM $225\,\rm{GHz}$ \taumeter\ 
zenith opacities $\tau_{225}$, as described in
Sect.~\ref{se:taumeter-method}, for a series of scans
of Uranus and MWC349 acquired during the \emph{reference} observation
campaigns. As guidelines,
we also show the predicted correlations using an ATM model integrated
in the NIKA2 bandpasses and in the $225\,\rm{GHz}$ band. {\lp The
$\taunu^{\rm{skydip}}$ to $\tau_{225}$ correlation relations are
consistent within statistical errors for the three campaigns.}
At 1\,mm they are also in agreement with the ATM model expectations,
while at 2\,mm the ATM model underestimates the measured {\tt skydip}
$\taunu$. {\rev Possible explanations include a ground pick-up effect,
which would have comparatively more impact on the opacity derivation
at 2\,mm, where the atmosphere is more transparent, than at
1\,mm. Indeed a fraction of the beam detects radiation from the
ground, as evidenced by forward efficiency measurements using
heterodyne front-ends (see Sect.~\ref{se:beam_efficiency}).}
This mild discrepancy with the ATM model predictions is yet to be
understood, but has no impact on
our opacity measurements, which do not rely on this model nor on
the precision with which the observing bandpasses are known. {\lp
Further consistency test with ATM expectations will be performed in the
future using \emph{in-situ} bandpass measurements with a
dedicated Martin-Pupplett spectrometer.}

We further check the robustness of $\taunu^{\rm{skydip}}$ against the
observing elevation. 
Panel b) in Fig.~\ref{fig:skydip-to-taumeter-correl} shows the ratio of NIKA2
{\tt skydip} opacities to the $225\,\rm{GHz}$ \taumeter\ opacities as a function of
the average scan elevation. The {\tt skydip} opacity measurements have
no significant dependency on the elevation. Moreover, we observe that
the results are consistent for Uranus \bm\ scans and for
the shorter raster scans of the secondary calibrator MWC39, indicating
that the {\tt skydip} opacity estimates do not depend on the type of
observation scans.

As a summary, NIKA2 {\tt skydip} opacity estimates i) have reproducible
correlation coefficients with the $225\,\rm{GHz}$ \taumeter\ opacities from a
campaign to another, ii) are robust against the observing conditions,
and iii) are stable for various sources and scanning strategies. {\lp
In addition to these properties, the {\tt corrected skydip} opacities
further ensure flux density measurements that are immune to the
atmospheric and observing elevation conditions. We use them in
the \emph{baseline} calibration.}

\section{Calibration}
\label{se:calibration}

In this section, we present the absolute calibration of the flux densities. We
use Uranus as the main primary calibrator. Sect.~\ref{se:calibration_method}
describes the absolute calibration method, Sect.~\ref{se:flat_field} presents
the inter-calibration of all the KIDs and the flat fields. While
gathering several observations of calibrators, we have evidenced a
daily variation of the absolute calibration
coefficients related to daily variation of the beam
size induced by weather temperature. If left uncorrected, it leads to
a sizeable increase of the calibration uncertainties. To
overcome this issue, we primarily flag the most impacted observation
times of the day and exclude them from further analysis.
We discuss this effect in
Sect.~\ref{se:beam_variation}. In Sect.~\ref{se:baseline_calibration},
the \emph{baseline} calibration procedure is summarized and stability
tests are performed.


\subsection{Absolute calibration procedure and photometric system}
\label{se:calibration_method}

We detail here the procedure for calibrating the absolute scale of
the flux density and the chosen photometric system.

\subsubsection{Photometric system}
\label{se:photometric_system}

The main primary calibrators of NIKA2 are the giant planets Uranus and
Neptune. The latter is used when the former is not visible in the most
stable observing conditions. The flux density expectations of the
primary calibrators are derived in Appendix~\ref{ap:ref_flux_calibrator}. 
\begin{table}[!htbp]
\caption{NIKA2 reference frequencies and FWHM}
\label{tab:definitions}
\centering     
\begin{tabular}{lcc}
\hline\hline
      \noalign{\smallskip}
      & 1 mm & 2 mm \\
      \noalign{\smallskip}
      \hline
      \noalign{\smallskip}
      Reference frequency, $\nu_{0}$ & 260 GHz & 150 GHz \\
      Reference FWHM,  FWHM$_{0}$    & 12.5'' & 18.5'' \\
      \noalign{\smallskip}
      \hline
\end{tabular}
\end{table}

We parametrize the primary calibrator flux density as 
$S_{\rm{c}}(\nu) = S_{\rm{c}}(\nu_0)\, f(\nu/\nu_{0})$, where $f(\nu/\nu_{0})$
encloses the spectral dependence, 
as a function of a reference frequency $\nu_{0}$ that we choose
arbitrarily to be: $\nu_{0} = 150$~GHz for the $2\,\rm{mm}$ array and
$\nu_{0}= 260$~GHz for both $1\,\rm{mm}$ arrays. After projecting the raw
data (in units of the KID resonance frequency shift, $\rm Hz$) of a
calibrator {\sl c} on the sky, we model the calibrator raw map as a
fixed-width 2D Gaussian
\begin{equation}
  R_{\rm{c}}(\theta, \phi)  = A_{\rm{c}} \, e^{-\frac{\theta^{2}}{2\sigma_{0}^{2}}},
  \label{eq:gaussian_amplitude}
\end{equation}
{\lp where $A_{\rm{c}}$ is the amplitude of the 
Gaussian in Hz,} and $\sigma_{0}$ is derived from the
reference FWHM, labelled FWHM$_{0}$, which is $12.5''$ for the $1\, \rm{mm}$
arrays and $18.5''$ for the $2\,\rm{mm}$ array. These values have
been chosen larger than the main beam values, as reported in
Sect.~\ref{se:beam}, to account for a fraction of the signal stemming from
the first error beam and first side lobes.
Both the reference frequency, $\nu_0$, and the reference FWHM, FWHM$_{0}$, define
our reference photometric system, as summarized in Table~\ref{tab:definitions}.

The absolute calibration coefficients are estimated from observations
of primary calibrators $c$, as the ratio of
the calibrator flux density expectations at the reference frequency
$S_{\rm{c}}(\nu_0)$ and $A_{\rm{c}}$. Then, for any observed point-like source
{\sl s} of projected map $R_{\rm{s}}(\theta,\phi)$ in Hz, the map
\begin{equation}
  M_{\rm{s}}(\theta, \phi) = \frac{S_{\rm{c}} (\nu_{0})}{A_{\rm{c}}}
  R_{\rm{s}}(\theta,\phi),
  \label{eq:pointsourcephot}
\end{equation}
is calibrated in Jy/beam. {\lp The best-fit amplitude estimate
of the fixed-width FWHM$_{0}$ Gaussian on this map directly gives an
estimate of the flux density of the source at the reference
frequency $S(\nu_{0})$, excluding colour corrections.}

\begin{table*}[!thbp]
\caption{Colour correction factors for a target source  $S \propto \nu^{\alpha_{\rm{s}}}$, as defined using Eq.~\ref{eq:color_correction}.}
\label{tab:mod}
\centering 
\begin{tabular}{lrrrrrrrr}
\hline\hline
\noalign{\smallskip}
Array     & \multicolumn{8}{c}{$\alpha_{\rm{s}}$} \\
\noalign{\smallskip}
\hline
\noalign{\smallskip}
         &  -2 &  -1    &    0  & + 0.6 & +1  &  +2  & +3 & +4  \\       
\noalign{\smallskip}
\hline
\noalign{\smallskip}
          A1   & 0.876  &  0.916   &   0.951  & 0.969 &  0.981   &  1.005  &    1.024  &  1.037   \\
          A2   & 0.945  &  0.972   &   0.990  & 0.996 &  0.998   &  0.997  &    0.986  &  0.966      \\ 
          A3   & 0.907  &  0.940   &   0.967  & 0.980 &  0.987   &  1.001  &    1.009  &  1.011     \\
            \noalign{\smallskip}
            \hline
\end{tabular}
\end{table*}

\subsubsection{Colour correction}

The flux density estimate $S(\nu_{0})$ gives the
flux of the source at the reference frequency only if the source has
the same spectral behaviour as the calibrator. In general, to retrieve the
flux of the source at the reference frequency, a colour correction
$C_{\rm{s}}$ has to be applied
\begin{equation}
S_{\rm{s}}(\nu_{0}) = S(\nu_{0}) \,  C_{\rm{s}}(\nu_{0}, I_\nu^{\rm{s}}),
\end{equation}
which depends on the reference frequency $\nu_{0}$, the source
SED $I_\nu^{\rm{s}}$ and the NIKA2 bandpasses.
Neglecting the effect of the atmosphere on the NIKA2 transmission, we
compute the colour correction factor for target sources of SED that are
different from Uranus using
\begin{equation}
  C_{\rm{s}}(\nu_{0}, I_\nu^{\rm{s}}) = \frac{\int_{0}^{+\infty} I_\nu^{\rm{c}}~T({\nu})\, d\nu}{ \int_{0}^{+\infty} I_\nu^{\rm{s}}~ T({\nu})\, d\nu},
    \label{eq:color_correction}
\end{equation}
where $T({\nu})$ is the NIKA2 transmission
(Sect.~\ref{se:instru_bandpass}).
Assuming Rayleigh-Jeans SED for the calibrator
($I_\nu^{\rm{c}} \propto (\nu/\nu_0)^{\alpha_{\rm{c}}}$) and the source
($I_\nu^{\rm{s}} \propto (\nu/\nu_0)^{\alpha_{\rm{s}}}$), and a
spectral index $\alpha_{\rm{c}} = 1.6$ for Uranus, we provide colour
correction factors for eight values of the spectral index of the
source $\alpha_{\rm{s}}$ in Table~\ref{tab:mod}.

\subsubsection{Extended sources}
\label{se:extended_source_calib}

{\rev While study of point sources are usually performed with maps given in
Jy/beam unit, diffuse emission studies or aperture photometry ones
require map expressed in Jy/sr. To that aim, we need to take into
account the full beam pattern, as discussed in Sect.~\ref{se:beam_efficiency}.}
First we correct with the solid angle enclosed in the
reference fixed-width Gaussian beam $\Omega_{0} = 2\pi \sigma_0^2$ to
obtain a map homogeneous to Jy/sr. Then, to account for the signal in
the total beam pattern, we further correct with the reference beam
efficiency BE$_{0}$. As defined as the ratio of $\Omega_{0}$ and the
total beam solid angle $\Omega_{\rm{tot}}$, BE$_{0}$ represents the
fraction of the full beam solid angle that is enclosed in the solid
angle sustained by the reference beam. To summarize, the map in Jy/sr
$M(\theta, \phi)$ relates to the map in Jy/beam
$M_{\rm{s}}(\theta, \phi)$ using
\begin{equation}
M(\theta, \phi) = \frac{\rm{BE}_{0}}{\Omega_{0}} \, M_{\rm{s}}(\theta, \phi).
\label{eq:jybeam_to_jysr}
\end{equation}
\begin{table}[!htbp]
  \caption[]{Reference beam efficiencies for Array 1, Array 3, Array 1\&3 and Array 2}
  \label{tab:reference_beam_efficiency}
  \centering    
  \begin{tabular}{lrrrr}
    \hline\hline
    \noalign{\smallskip}
    & A1 & A3  & A1\&3 & A2 \\
    \noalign{\smallskip}
    \hline
    \noalign{\smallskip}
    FWHM$_{0}$ [arcsec]          & $12.5$     &  $12.5$     & $12.5$     & $18.5$     \\
    BE$_{0}$   [\% ]             & $60 \pm 3$ &  $61 \pm 3$ & $61 \pm 3$ & $72 \pm 2$ \\
        \noalign{\smallskip}
    \hline
  \end{tabular}
\end{table}
{\rev In Table~\ref{tab:reference_beam_efficiency}, we provide the
estimates of the reference beam efficiency, which are derived from the
measured $\Omega_{\rm{tot}}$ as given in
Sect.~\ref{se:beam_efficiency}.}

\subsubsection{Calibration procedure}
\label{se:practical_calib}

{\lp In practice, the calibration procedure is performed in two steps. First
$S_{\rm{c}}(\nu_{0})$-to-$A_{\rm{c}}$ ratios per detector $G_k$
are estimated for each KID $k$ using the map per KID projected from
a \bm\ scan of a calibrator. The calibration coefficient $G_k$
for the KID $k$ is computed at zero atmospheric opacity as:
\begin{equation}
  G_k = \frac{S_{\rm{c}}(\nu_0)\, e^{-\taunu\,x}}{A_k},
  \label{eq:kid_gain}
\end{equation}
where $S_{\rm{c}}(\nu_0)$ is the expected flux density of the source at
the reference frequency $\nu_0$ (see
Appendix~\ref{se:ref_flux_uranus_neptune}), $\taunu\, x$ is the
line-of-sight opacity measured using the {\tt corrected skydip} method
(see Sect.~\ref{se:opacity}) and $A_k$ is the best-fit
amplitude of the reference FWHM$_0$ Gaussian, which is fitted in the
KID map. 
This first step accounts for both the relative calibration between KIDs and the absolute
calibration using a single calibrator scan.

Secondly, the absolute calibration is further refined by evaluating a
flux density rescaling factor using a series of observations of
Uranus or Neptune. After the first step of the calibration is
performed, the KID TOI are projected into a calibrated map $M_\nu$, as
described in Sect.~\ref{se:dataproc}, where $\nu$ stands for the three
arrays and the $1\,\rm{mm}$-array combination, and the atmospheric attenuation
is corrected. For each of the calibrator observation scans, we
compute the ratio between the expected calibrator flux density
$S_{\rm{c}}(\nu_{0})$ and the measured calibrator flux density in $M_\nu$, which
is estimated as described in Sect.~\ref{se:photometric_system}. The
flux density rescaling factor per array is the average
expected-to-measured flux density ratio over all the selected calibrator scans.} 

The primary calibrator scans are first selected as discussed in
Sect.~\ref{se:data_selection}. Then, in addition to
the \emph{baseline} scan selection cuts, we use a Gaussian beam size
criterion. The FWHM estimated from the planet
observation map is required to be lower than $12.5''$ at $1\,\rm{mm}$ and lower
than $18''$ at $2\,\rm{mm}$. In further mitigating the flux scatter
due to beam broadening, we ensure better accuracy of the absolute
calibration.

\subsection{Relative calibration \& flat fields}
\label{se:flat_field}
While absolute calibration of each KID also \emph{de facto} provides
relative calibration, the latter is interesting in itself to
characterize the instrument. We focus on this aspect in this
section.

\begin{figure*}[!thbp] 
\begin{center}
  \includegraphics[width=0.95\textwidth]{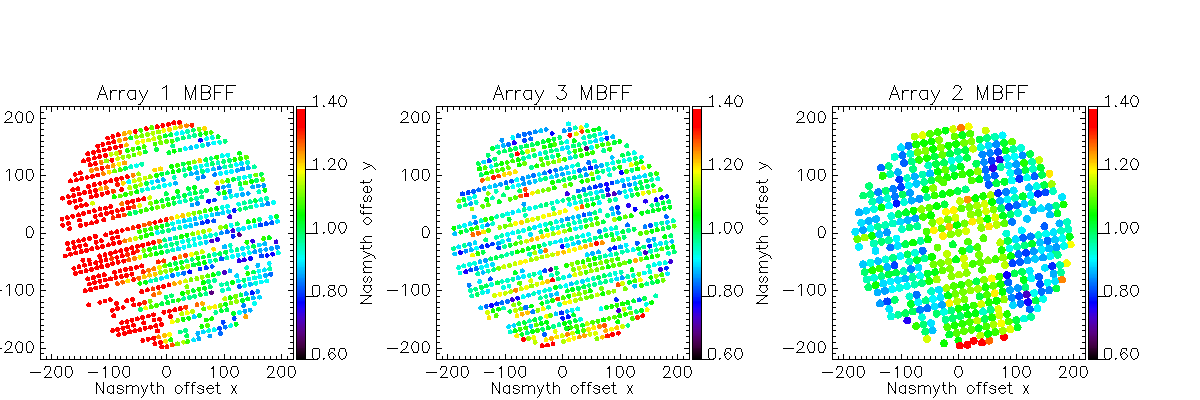}
  \includegraphics[width=0.8\textwidth]{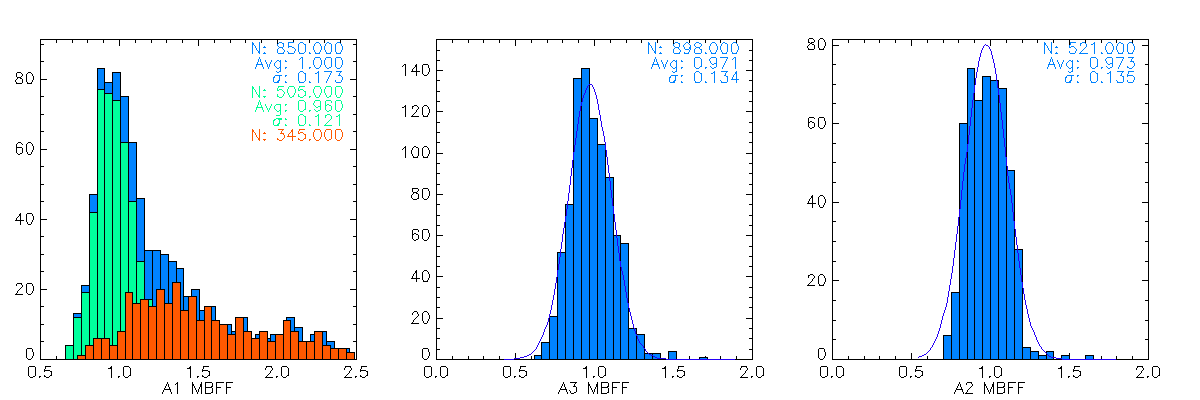}
\caption[Average main beam flat fields]{Average main beam flat fields
  obtained by combining the flat fields of five
  \bm\ scans. The top row plots show the normalized average flat fields of Array
  1, 3 and 2, respectively. {\lp The offset positions with respect to the center of
  the array are given in arcsecond in the Nasmyth coordinate
  system. The colour code gives the value of the KID calibration
  coefficients, as defined in Eq.~\ref{eq:kid_gain}, normalized by the
  average calibration coefficient over all the
  KIDs of the array.} The bottom plots
  show the average flat field distributions using all KIDs (blue),
  using Array 1 KIDs that are positioned out of the shadow zone
  (green) and using Array 1 KIDs inside the shadow zone, which is
  defined in the text.}
 \label{fig:avg_mbff}
\end{center}
\end{figure*}

The dispersion of the detector responsivity across the field of view (\aka\ flat
fields) has been characterized in the following ways.

\noindent \emph{Main beam flat fields.} These are the focal plane
distribution of the calibration coefficients per KID. {\lp They describe the
focal plane distribution of the point spread function (PSF) in
the far field of the telescope.} The calibration coefficients $G_k$ are
estimated using Eq.~\ref{eq:kid_gain}, as discussed in
Sect.~\ref{se:practical_calib}. 

\noindent \emph{Forward beam flat fields.} These are the focal plane
distributions of the relative response of
each KID to the near field atmospheric background. They are estimated
using the correlation factor of each detector TOI 
to a median common mode estimated off-source (see Sect.~\ref{se:toi_proc} for
more details on common modes).

Figure~\ref{fig:avg_mbff} 
shows the average main beam 
flat field for the three arrays. These have been constructed by
combining the normalized flat fields of five \bms\ acquired during two
technical observation campaigns. These data were
selected by thresholding the line-of-sight opacity measured in the
1\,mm band to $\taunu\,x \leq 0.85$. The distributions for the average flat
fields are shown in the bottom panel of Fig.~\ref{fig:avg_mbff}.

We observe a significant variation of the flat fields for A1 from the left-most side
to the right-most side of the FOV. This reveals a significant change of A1
detector responsivities depending on their position in the focal plane. Namely, this
effect mainly impacts the left-most third of the array, which is
referred to as the "shadow-zone''. This variation of the
flat field translates into a broadening of the distribution shown in
the lower panel of Fig.~\ref{fig:avg_mbff}.  However,
we verified that A1's flat field dispersions are in line with the ones of A3 after the
detectors within the shadow-zone were flagged out using a
crescent-shaped mask. The masked flat field distributions are shown in
green in Fig.~\ref{fig:avg_mbff}, 
whereas shadow-zone distributions are in red. The same FOV patterning
is also observed in the forward beam flat fields, which excludes a
main beam related issue. 

The shadow zone effect is caused by {\lp a misbehaving of the dichroic in
the polarized transmission which is out of specifications. As a
result, the $1\,\rm{mm}$ polarisation that illuminates A1 is
attenuated.}
This effect,
which implies {\lp a dependence of the frequency cut-off on the radiation
incidence angle and linear polarisation,}
was reproduced using optical simulations. Furthermore, this hypothesis was
verified using observations at the technical campaign of September
2018. During this test campaign, a new \emph{hot-pressed} dichroic had
been installed in place of the current \emph{air-gap} dichroic.
The shadow zone variations of the flat field for A1 were
not observed during the September 2018 campaign, while huge distortions
across the field of view of A2 were reported. These distortions are
due to the bending of the hot-pressed dichroic at low temperature.
This test has confirmed that the shadow zone effect was due
to incoming radiation absorption by the current air-gap dichroic.
Further efforts are currently conducted to the design of a dichroic that
combines robustness against bending induced by low temperature and
optimal transmission.
{\lp The air-gap dichroic, which is immune to low temperature-induced
deformation, has been re-installed at the end of the September 2018
run coming back to the instrumental set up discussed in this paper. We
have checked that, as expected, the performance of the instrument
after this intervention was consistent with the one reported in this
paper.}

\subsection{The temperature-induced variation effect}
\label{se:beam_variation}

We evidenced a daily variation of the flux density estimates that correlates to the
measured beam size. This beam broadening happens mostly in afternoons and around
sunrise and sunset. It is also reproducible from one campaign to another. It most certainly
comes from the combination of two different effects.

First inhomogeneous solar illumination leads to large scale deformations of the
\trentemetre\ primary mirror, which in turn lead to variable de-focussing of the
telescope. {\lp To mitigate this effect, the telescope is equipped with an
  active thermal ventilation system of the primary mirror and an active
  temperature control of the secondary support legs. This system is, however,
  challenged when the Sun partially illuminates the telescope}
  {\lp This is a known effect that also impacts
  observations with the heterodyne instruments operated at the IRAM
  \trentemetre\ telescope. It had also been already observed with the previous
  generation total-power instruments MAMBO-2 \citep{Kreysa1999}.} However, the
magnitude of this effect is likely to have increased with the slow disappearance
of the surface painting of the primary mirror in the past few years.

Second, on short time scales, atmospheric anomalous refraction {\lp also often plays a
  role.}. As far as the \trentemetre\ telescope is concerned, it has first
been described in~\citet{Altenhoff1987}. Based on experience with
  the heterodyne receivers, afternoon hours are
  often affected with an unstable atmosphere when rising moist air moves
  through the beam of the telescope and causes random refraction. The pointing
  is then observed to change within few seconds by few arcseconds {\lp resulting in an average
    enlargement of the beam pattern.} This effect has been confirmed by
  measuring several arcsec displacements of the source position when projected using
  different subsets of subscans of a single observation, as described
  in Appendix~\ref{ap:beam_monitoring}. We find that the apparent beam broadening during
  the afternoon is due to anomalous refraction for between one third
  and one half of the scans over the period studied here.

{\lp As both effects (the primary mirror deformations and the anomalous
  refraction) are due to ambient temperature variations, we refer to them as
  \emph{temperature-induced variation effects} in the following.}\\

\begin{figure}[ht!]
  \begin{center}
    \includegraphics[clip=true, trim={0.9cm, 0.5cm, 0.5cm, 0.5cm}, width=0.4725\textwidth]{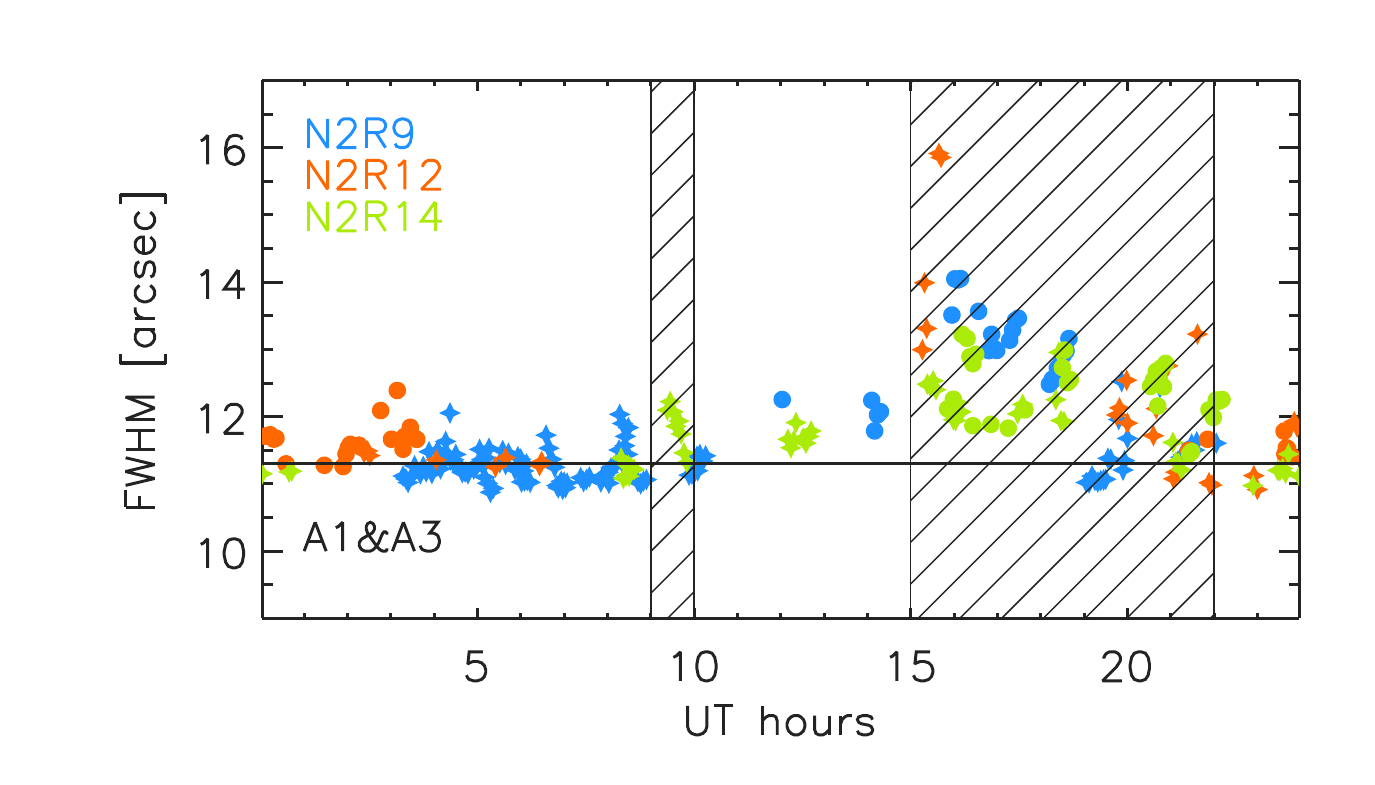}
    \includegraphics[clip=true, trim={0.5cm, 0.5cm, 0.5cm, 0.5cm}, width=0.4875\textwidth]{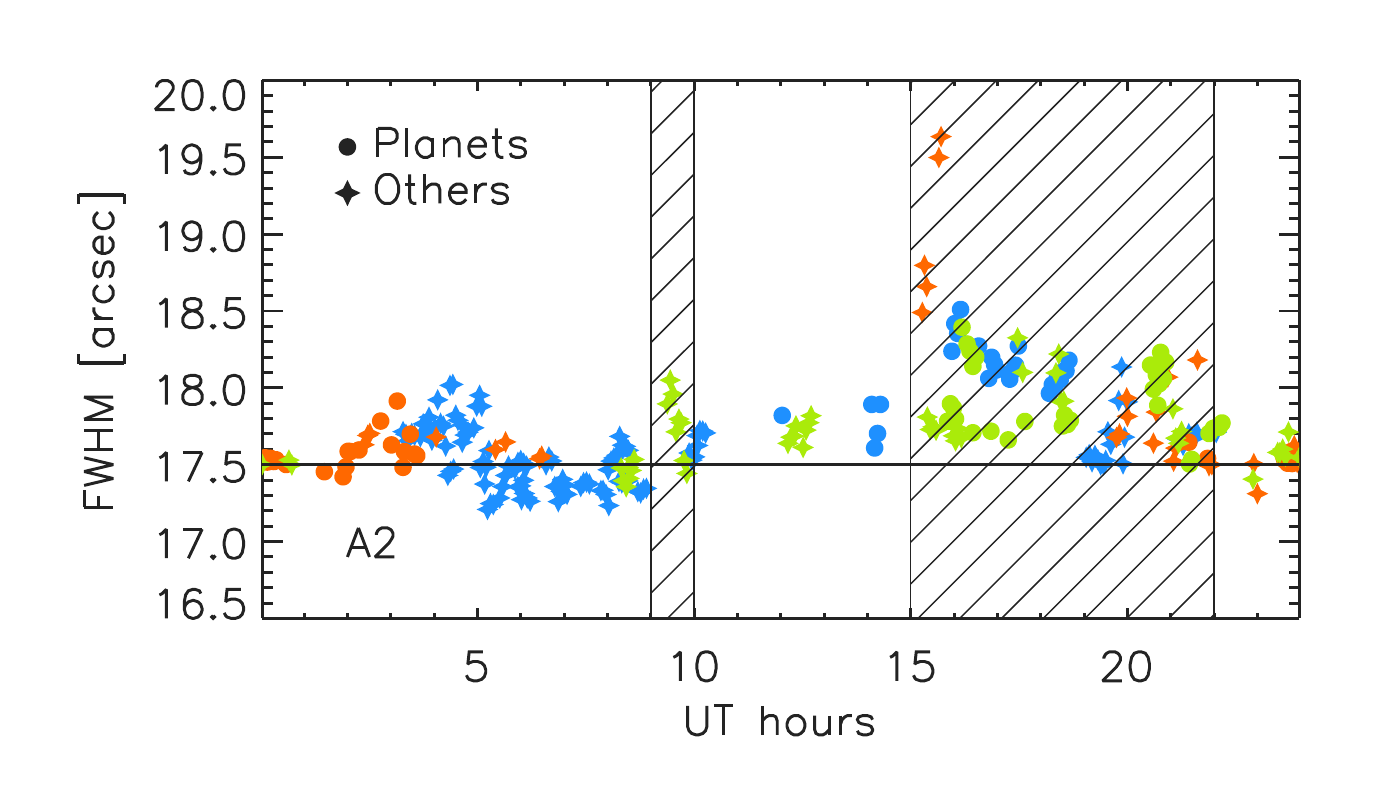}
    \caption[Beam size monitoring using OTF scans]{Beam size
      monitoring using OTF raster scans. Geometrical FWHM at $1\,\rm{mm}$ (top panel)
      and $2\,\rm{mm}$ (bottom panel), as a function of the
      observation time in UT hours, are shown using scans of giant
      planets (filled circles) and bright point-like sources with a
      flux density higher than $1\,\rm{Jy}$ (filled stars) for the three \emph{reference}
      observation campaigns (N2R9,N2R12 and N2R14). The cross-hatched areas
      correspond to the observing time periods that are discarded using
      the \emph{baseline} scan selection, as described in Sect.~\ref{se:data_selection}.} 
\label{fig:beam_monitoring_otf}
  \end{center}
\end{figure}

The temperature-induced variation of the beam size as a function of
the UT hours is shown in Fig.~\ref{fig:beam_monitoring_otf} for the
three reference campaigns using bright sources. These are
selected by thresholding the flux density estimates above $1\,\rm{Jy}$
at both wavelengths. The beam size is estimated by fitting a 2D
elliptical Gaussian to the map {\lp and computing the geometrical FWHM
using Eq.~\ref{eq:fwhm_geom}.} For the resolved planet as Uranus, the
FWHM estimates are corrected for the beam broadening due to the finite
extension of the apparent disc of the planet, as in
Sect.~\ref{se:mainbeam}.
We observe the same evolution of
the FWHM for all campaigns. 
This goes from a plateau at a median value of $11.3''$ at $1\,\rm{mm}$
and $17.5''$ at $2\,\rm{mm}$ during the night, to a smooth rise that
reaches a maximum of about $14''$ at $1\,\rm{mm}$ and $18.5''$ at
$2\,\rm{mm}$ around 16:00 UT hours. The beam broadening becomes large
around 15:00 UT and returns to the plateau only around 22:00 UT.
To mitigate the impact of the temperature-induced variation,
observation scans acquired during this time interval must be
discarded. The UT ranges that are discarded
using the \emph{baseline} scan selection (see
Sect.~\ref{se:data_selection}) are shown as cross-hatched areas in
Fig.~\ref{fig:beam_monitoring_otf}.
They consist of the afternoon
period between 15:00 and 22:00 UT as well as the period from 9:00 UT to 10:00 UT while the Sun
rises. 
{\lp Whereas the global trend of the beam variations is the
same for all campaigns, we observe some variability in the amplitude
of the effect over the campaigns. This supports the assumption of the
important role of the partial illumination of the primary mirror by
direct sun light in the temperature-induced effect. This in turn
induces a variability of the amplitude of the effect depending on the
angle between the telescope bore-sight 
and the Sun all along the
observations.}

The same beam size variations in time are observed using scans of giant planets
(Uranus and Neptune) or other bright
sources (mainly quasars). However, planets lead slightly larger FWHM
estimates than quasars, because of
the larger contribution of the error beams to the fitted 2D Gaussian,
as their flux are measured with a higher signal-to-noise.


\subsection{Baseline calibration}
\label{se:baseline_calibration}

To assess NIKA2 performance, we rely on a baseline calibration that
resorts to the following steps: i) the calibration in FWHM$_0$ Gaussian
as detailed in Sect.~\ref{se:calibration_method} is implemented, ii)
the effect of the temperature-induced variation of the beam size is
mitigated using the \emph{baseline} scan selection described in
Sect.~\ref{se:data_selection} and iii) the
atmospheric attenuation is corrected using the {\tt corrected skydip}
opacity estimation described in Sect.~\ref{se:corrected-skydip}.

The \emph{baseline} calibration is validated below by checking the
stability of Uranus flux density estimates against the beam size
(Sect.~\ref{se:baseline_calibration_scans}) and against the
atmospheric transmission
(Sect.~\ref{se:baseline_calibration_atm}). The \emph{baseline}
calibration results are compared on alternative calibration methods
using other opacity correction (Sect.~\ref{se:baseline_calibration_opacity}).

\subsubsection{Flux stability against the beam size}
\label{se:baseline_calibration_scans}

\begin{figure}[!htbp]
\begin{center}
\includegraphics[clip=true, trim={0, -0.3cm, -0.3cm, 0}, width=0.72\linewidth]{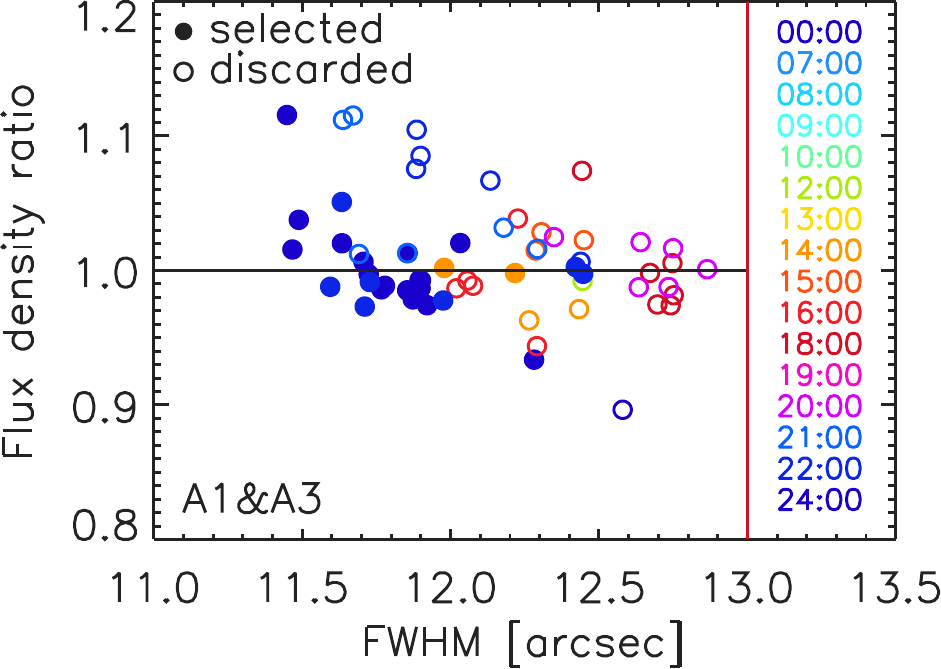}
\includegraphics[clip=true, trim={0cm, -0.3cm, -0.6cm, 0}, width=0.707\linewidth]{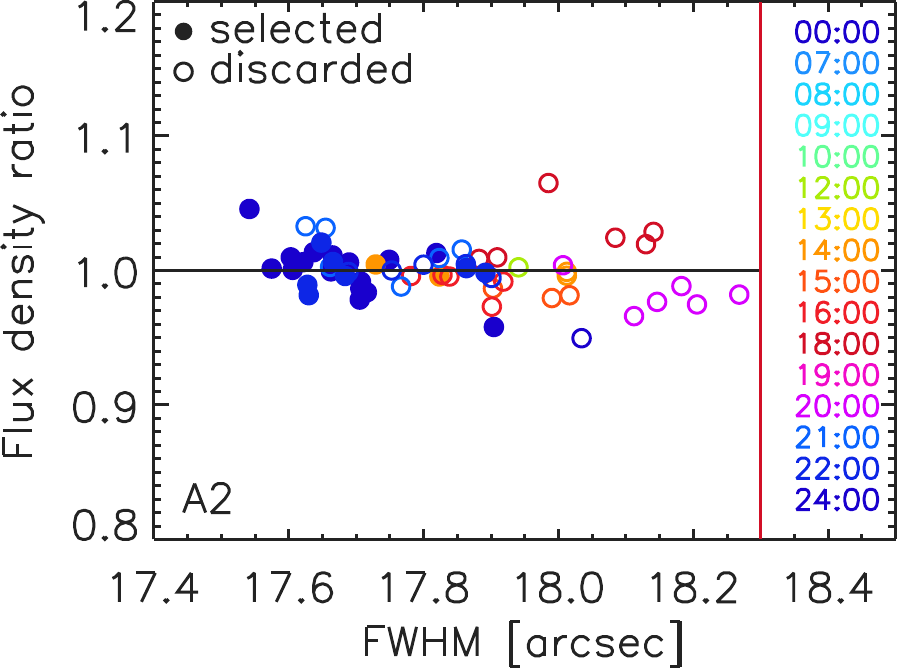}
\caption[Uranus flux density stability against FWHM]{ Uranus flux
density ratio vs beam size after baseline calibration. The ratio
of Uranus measured flux densities to expectations as a function of the
measured 2D Gaussian beam FWHM is shown for the $1$-mm array
combination (top panel) and for array 2 (bottom panel) after absolute calibration using the
\emph{baseline} method. These plots include all Uranus scans acquired during the 
N2R9, N2R12 and N2R14 campaigns and whose beam FWHMs are below the threshold indicated
by the vertical red lines (open circles), as well as the scans that
met the \emph{baseline} selection criteria (filled circles).}
\label{fig:calib_uranus_vs_fwhm_all}
\end{center}
\end{figure}

We present the Uranus measured-to-predicted flux density ratio as a
function of the 2D Gaussian FWHM estimates and colour-coded from the
observation times given in UT hours in
Fig.~\ref{fig:calib_uranus_vs_fwhm_all}. {\lp As expected from the
discussion in Sect.~\ref{se:beam_variation}, the largest FWHMs are
measured on scans acquired in late afternoon, especially from 16:00 UT
to 21:00 UT.}

The flux density estimates have been calibrated beforehand, so that
the flux density ratios are equal to unity in average by construction.
We observe no significant dependence of the selected scan flux ratios
(shown as filled circles in Fig.~\ref{fig:calib_uranus_vs_fwhm_all})
on the beam FWHM. {\lp This gives a first indication of the efficiency
  of the \emph{baseline} scan selection to mitigate the
temperature-induced beam variation effect. The flux stability against
the beam FWHM is further assessed in Sect.~\ref{se:photometry}.}

\subsubsection{Flux stability against the atmospheric transmission}
\label{se:baseline_calibration_atm}

\begin{figure}[!htbp]
\begin{center}
  \begin{overpic}[clip=true, trim={0, -0.3cm, -0.3cm, 0}, width=0.49\linewidth]{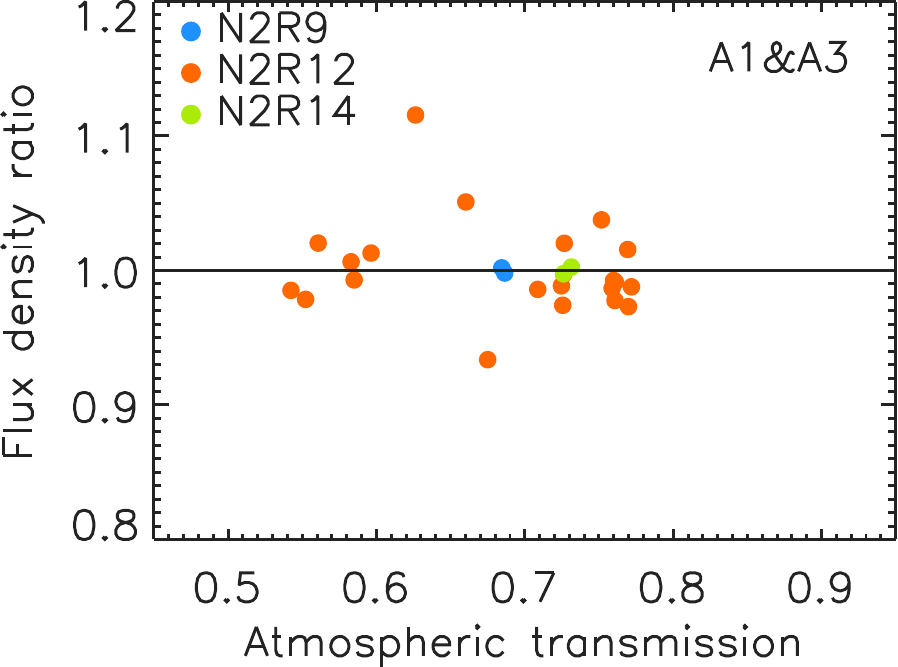}
    \put(20,23){\footnotesize Baseline}
  \end{overpic}
  \includegraphics[clip=true, trim={0, -0.3cm, -0.3cm, 0}, width=0.49\linewidth]{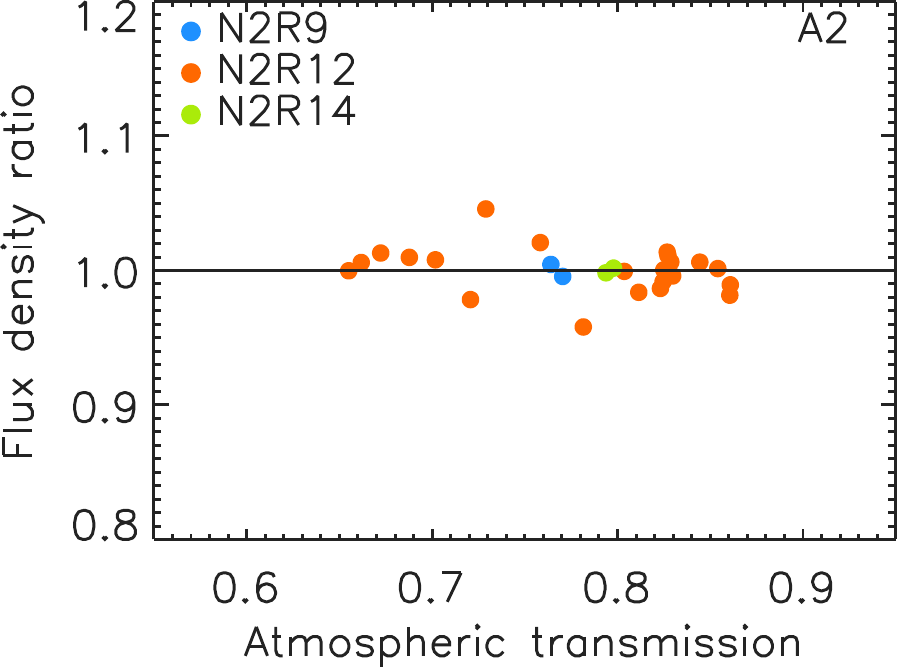}
  \begin{overpic}[clip=true, trim={0, -0.3cm, -0.3cm, 0}, width=0.49\linewidth]{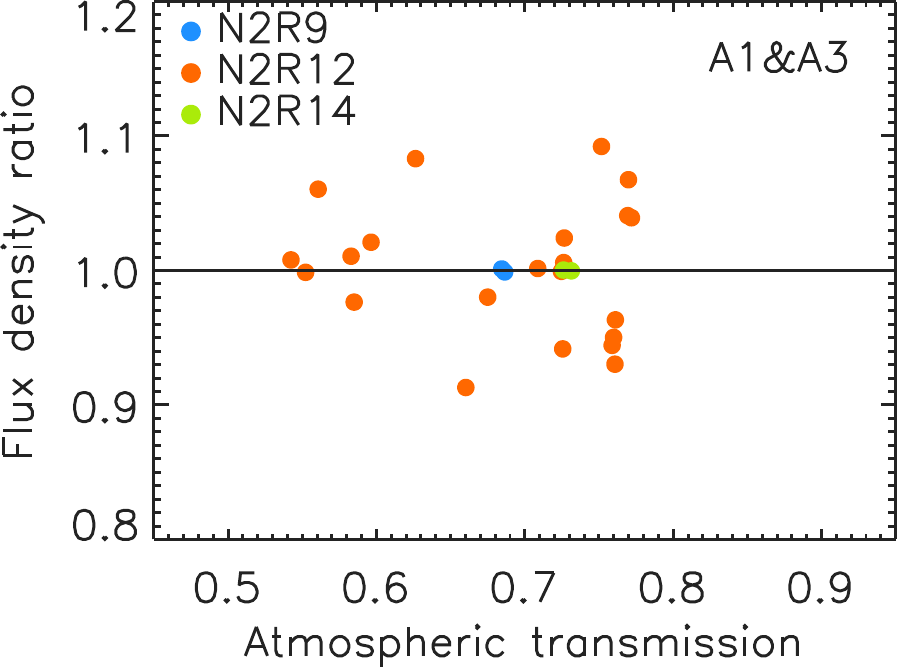}
    \put(20,23){\footnotesize Taumeter}
  \end{overpic}
  \includegraphics[clip=true, trim={0, -0.3cm, -0.3cm, 0}, width=0.49\linewidth]{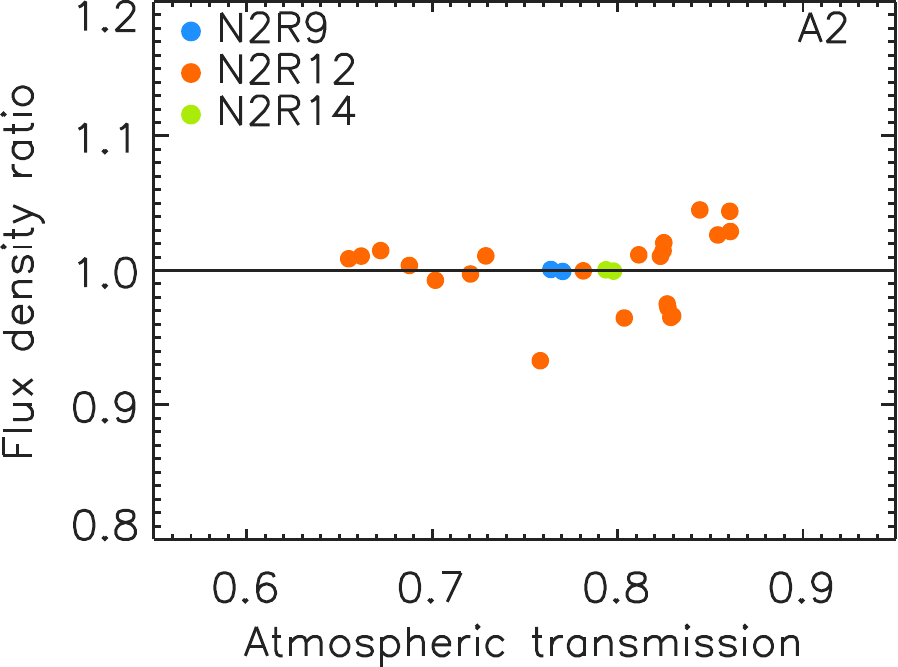}
  \begin{overpic}[clip=true, trim={0, -0.3cm, -0.3cm, 0}, width=0.49\linewidth]{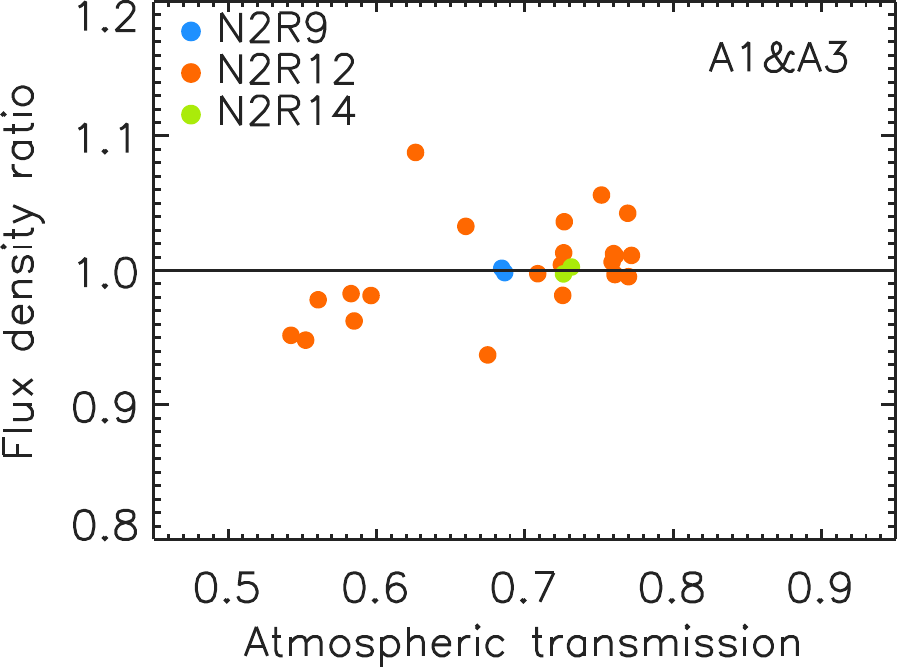}
    \put(20,23){\footnotesize Skydip}
  \end{overpic}
  \includegraphics[clip=true, trim={0, -0.3cm, -0.3cm, 0}, width=0.49\linewidth]{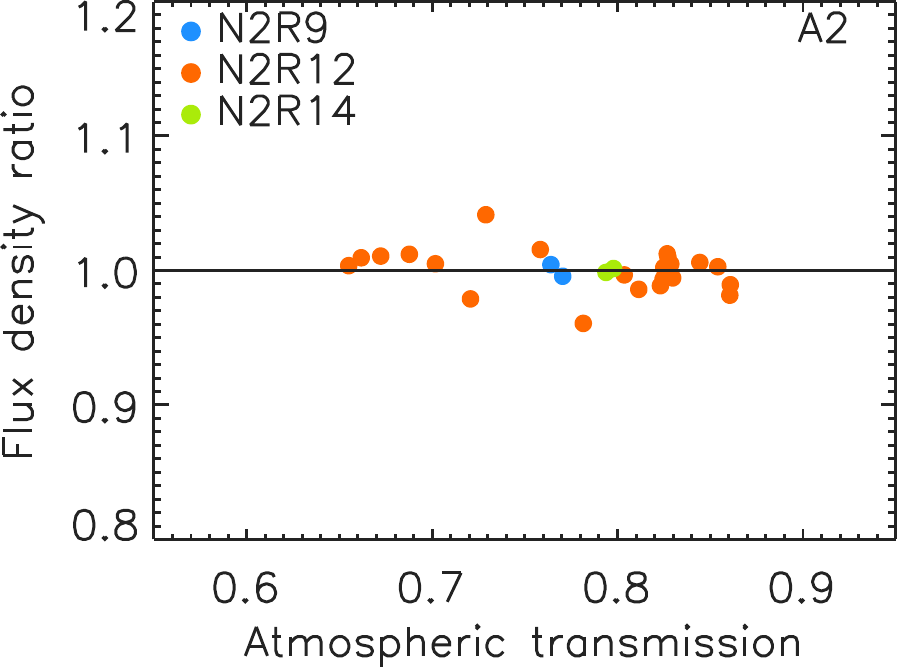}
  \caption[Uranus flux density stability against atmospheric
    transmission]{Uranus flux density ratio vs atmospheric transmission
    shown for the $1$-mm array
    combination (left column) and for array 2 (right column) after absolute
    calibration using (\emph{first row}) the baseline method, (\emph{second row}) the 'taumeter'-based and
    (\emph{third row}) the 'skydip'-based methods. These plots
    include all Uranus scans acquired during N2R9, N2R12 and N2R14
    campaigns. }
  \label{fig:calib_uranus_vs_atmtrans}
\end{center}
\end{figure}

We test the stability of Uranus flux densities calibrated using the
\emph{baseline} method against the atmospheric transmission. The latter
depends on the measured zenith opacity $\taunu$ and the scan
average \airmass\ x as $\exp{(-\taunu \, x)}$. In
the first row of Fig.~\ref{fig:calib_uranus_vs_atmtrans}, Uranus flux ratio
is shown as a function of the atmospheric
transmission for the $1$-mm array combination and Array 2 and for the
three \emph{reference} observation campaigns (N2R9, N2R12 $\&$ N2R14). We
observe no significant correlation of the flux ratio with the atmospheric
transmission, which gives a first
indication of the robustness of the flux density estimates against the
atmospheric conditions using the baseline calibration. This will be
further tested using a larger number of scans towards other sources in
Sect.~\ref{se:photometry}.

\subsubsection{Comparison with other opacity correction methods}
\label{se:baseline_calibration_opacity}

As a cross-check we have derived the absolute
calibration factors using the {\tt taumeter}
(Sect.~\ref{se:taumeter-method}) and the {\tt skydip}
(Sect.~\ref{se:skydip-method}) atmospheric opacity
correction methods. We then compare Uranus
flux density estimates after absolute calibration using the baseline
calibration and these two alternative corrections. Figure~\ref{fig:calib_uranus_vs_atmtrans}
shows the Uranus measured-to-modelled
flux ratio as a function of the atmospheric transmission for A1$\&$A3
and for A2 after the {\tt taumeter} correction (second row) and
after the {\tt skydip} correction (third row). We observe more
dispersion for the {\tt taumeter}-based flux ratio, whereas the {\tt
skydip}-based ratio is very similar as the baseline ratio except
for a slight decrease of the flux at low atmospheric
transmission. Thus the {\tt taumeter} and {\tt skydip} atmospheric
opacity correction methods can be used for
the absolute calibration in complement to {\tt corrected skydip}, e.g. to
perform robustness tests as in Sect.~\ref{se:photometry}.

%
%
\section{Photometry \& Stability Assessment}
\label{se:photometry}

NIKA2 photometric capabilities after the calibration presented in
Sect.~\ref{se:calibration}, are assessed in this section. Firstly,
we use observation of secondary calibrators (planetary nebulae NGC7027, CRL2688, and
MWC349A) to test the consistency of the flux density estimates with
expectations. The flux density expectations
in NIKA2 bands for these calibrators are given in
Appendix~\ref{se:ref_flux_secondaries}. Then,
we verify the stability of the photometry with
respect to the atmospheric conditions using a large amount of
observations towards a large variety of {\rev point-like}
sources. 

The quality criteria used to assess the photometric capabilities and
calibration results are defined in Sect.~\ref{se:photometry_criteria}.
In Sect.~\ref{se:photometry_baseline}, these criteria are evaluated
for the baseline calibration, and in Sect.~\ref{se:photometry_others},
we compare these results with other calibration method results before
summarizing the main results in Sect.~\ref{se:photometry_summary}.

\subsection{Calibration accuracy and uncertainty assessment}
\label{se:photometry_criteria}

We assess the photometric performance by evaluating two
quality criteria: first, the calibration bias checks the accuracy of
the absolute calibration and then the {\rev point-source} rms
calibration uncertainties test the stability of the flux
densities. {\rev The stability of the full beam pattern at large
angular scales, which impacts the stability of the diffuse emission
flux density, is not tested here}. {\lp Finally, we review here the
systematic uncertainties on the flux density.}

\subsubsection{Calibration bias}
\label{se:def_calibration_bias}

We define the calibration bias
$b_{\nu}$, where $\nu$ stands for Array 1, 2, 3 and the
$1\,\rm{mm}$ array combination, as
the ratio between the measured flux density $\hat{S}_{\nu}$ using the
reference photometry
(Sect.~\ref{se:photometric_system}) and the flux density
expectations $S_{\rm{s}}(\nu_0)$ as given in
Appendix~\ref{se:ref_flux_secondaries}. From a series of
secondary calibrator scans, we evaluate the average calibration bias
$b_{\nu}$, which by construction, should be equal to
unity within uncertainties.
Moreover, we check the stability of the calibration bias against
the observed opacities as a robustness test of the
opacity derivation method. Likewise, we verify that the photometry is
insensitive to optical variations by checking the stability of the
calibration bias against the measured beam size.

\subsubsection{{\rev Point-source} rms calibration uncertainties}
\label{se:def_calibration_rms_error}
We evaluate the standard deviation of bright {\rev point-like} source
measured-to-median flux density ratio $\sigma_{\nu}$ per array or array combination $\nu$.
As the flux density of most of the considered sources is unknown a priori, we
compare the flux density estimate in a single observation scan to the
average flux density throughout an observation campaign. This method
requires the selection of sources that are bright enough to be
detected with a high signal-to-noise ratio with a single repetition of an usual
$8'\times 5'$ OTF raster scan. Namely, we perform a source
selection by setting a threshold on the flux estimate to $800\,\rm{mJy}$ at
$1\,\rm{mm}$ and $400\,\rm{mJy}$ at $2\,\rm{mm}$. Moreover, we consider
only the sources for which a minimum of $10$ scans are available after
selection to ensure a precise average flux density
estimation. Finally, the selected source scans must meet the \emph{baseline}
scan selection criteria given in Sect.~\ref{se:data_selection}.

{\lp Since the flux density ratios are not Gaussian distributed, we
evaluate the 68 and 95\% confidence level (C.L.) contours using the
measured distributions in order to further characterise the
uncertainties. We check that the rms
errors are larger than the 68\% C.L. contours and thus provide
conservative $1\sigma$-like errors.}    

{\lp As the rms of the flux density ratio is estimated on a scan set
that is representative of the observing conditions encountered at
the \trentemetre\ telescope, this is an estimate of the
calibration uncertainties that encloses errors of
optical, atmospheric, instrumental noise and data processing
origins. This includes the errors sourced by the \afternoon\ beam 
variations, the effect of the elevation, the uncertainties of the
atmospheric opacity correction using either the {\tt skydip} or
the {\tt taumeter} method and the atmospheric and instrumental noise
residuals after the data reduction (Sect.~\ref{se:dataproc}). {\rev
However, as the data set comprises point-like sources and because flux
density measurements in the reference photometric system are immune to
beam variations at large angular scales, we refer to the rms of the
flux density ratios as point-source rms calibration uncertainties.}

{\rev For extended sources and diffuse emission studies, the maps in
Jy/beam units, where the beam refers to the reference beam, are
converted into Jy/sr units using
Eq.~\ref{eq:jybeam_to_jysr}. Propagating the uncertainty on the
total beam solid angles, the reference beam efficiencies are estimated
with a precision of 5$\%$ and 3$\%$ at 1 and 2\,mm,
respectively, as reported in Table~\ref{tab:reference_beam_efficiency}
in Sect.~\ref{se:extended_source_calib}. These uncertainties must be
further accounted for in the error budget of diffuse emission studies.}

\begin{figure}[!thbp]
  \begin{center}
    \begin{overpic}[clip=true, trim={0.9cm, 0.2cm, 0, 0.6cm},width=0.532\linewidth]{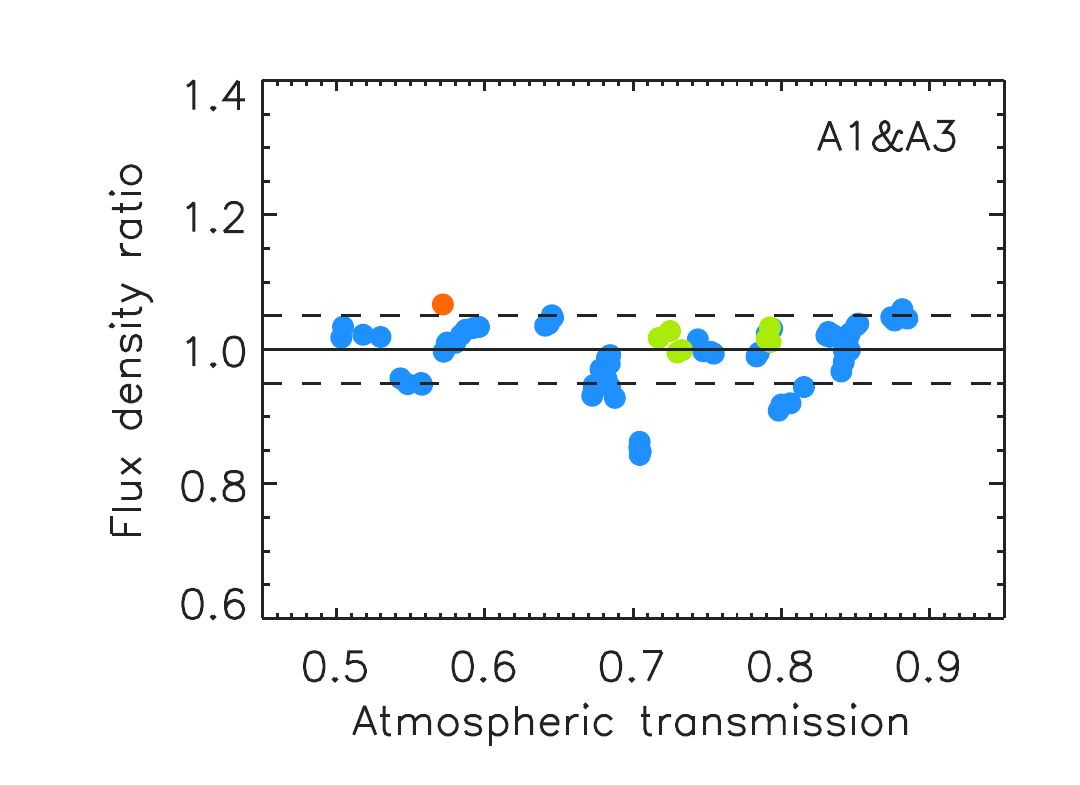}
      \put(20,60){\footnotesize {\tt Baseline}}
    \end{overpic}
    \includegraphics[clip=true, trim={1.8cm, 0.2cm, 0.5cm, 0.7cm},width=0.457\linewidth]{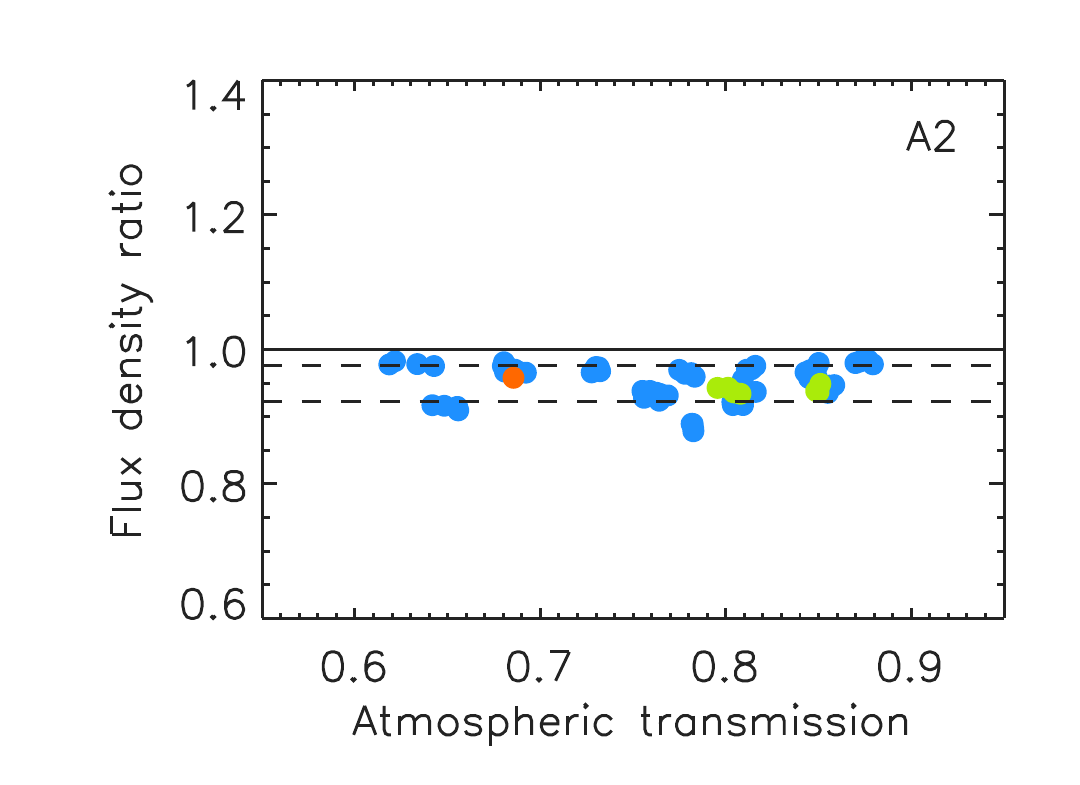}
    \begin{overpic}[clip=true, trim={0.9cm, 0.2cm, 0, 0.6cm},width=0.532\linewidth]{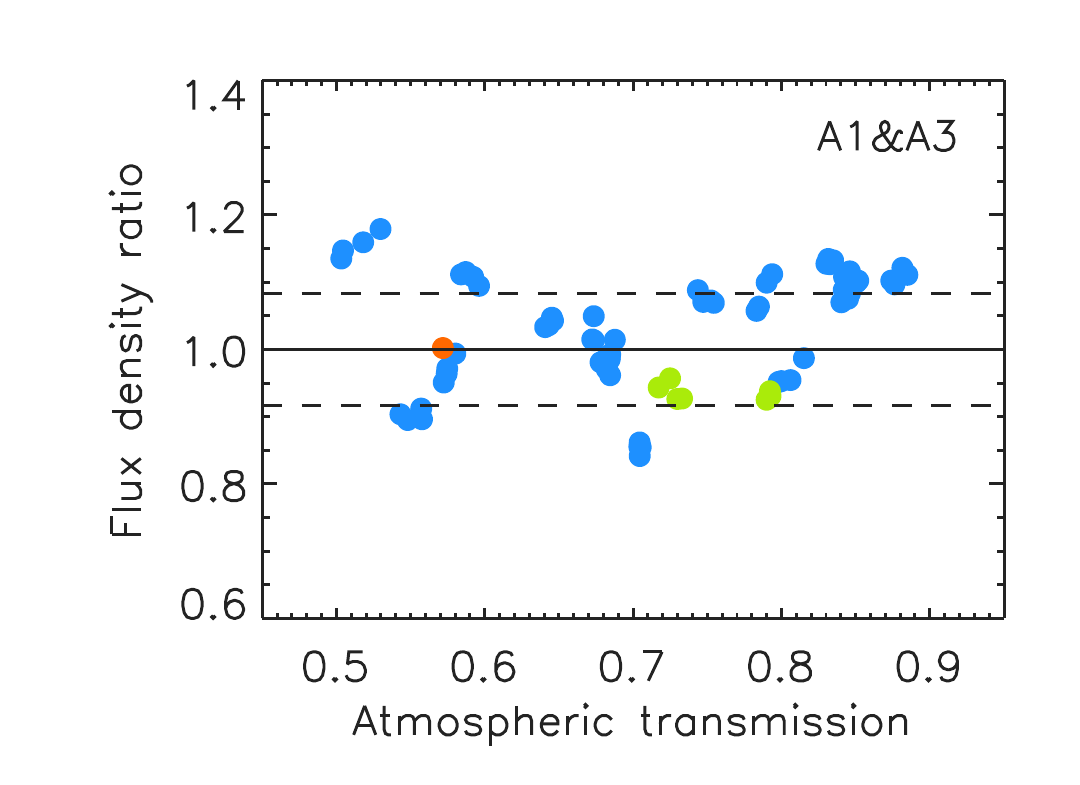}
      \put(20,60){\footnotesize {\tt Taumeter}}
    \end{overpic}
    \includegraphics[clip=true, trim={1.8cm, 0.2cm, 0.5cm, 0.7cm},width=0.457\linewidth]{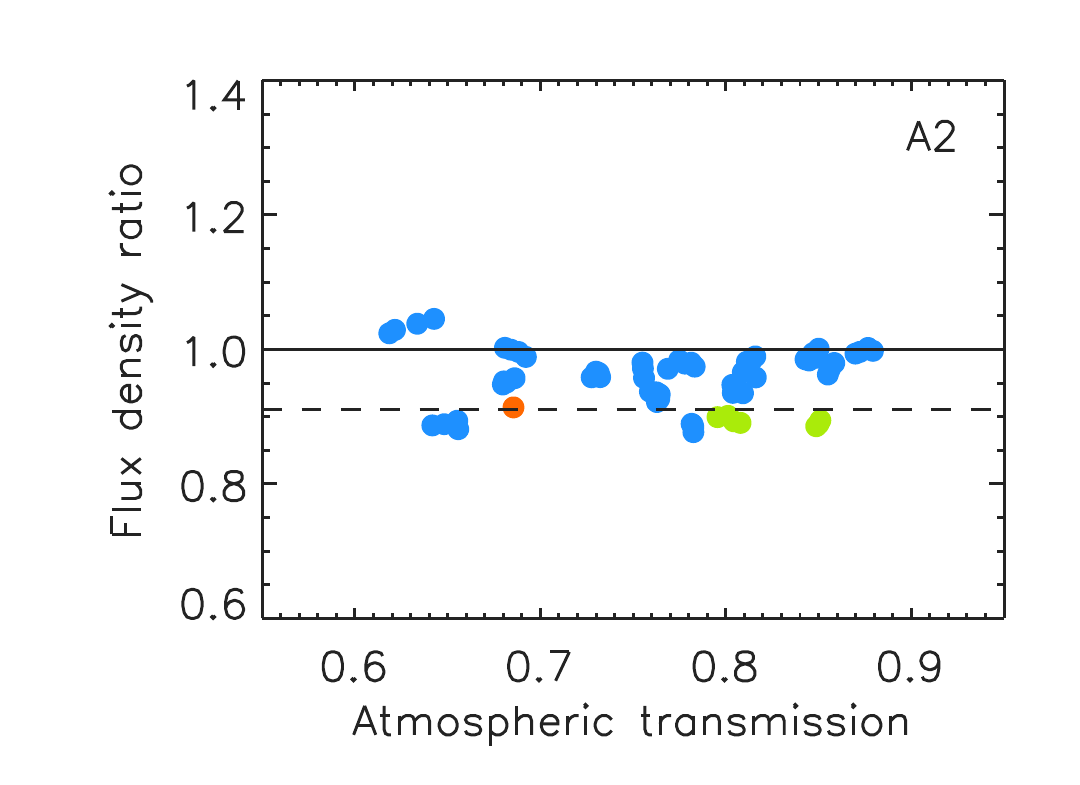}
    \begin{overpic}[clip=true, trim={0.9cm, 0.2cm, 0, 0.6cm},width=0.532\linewidth]{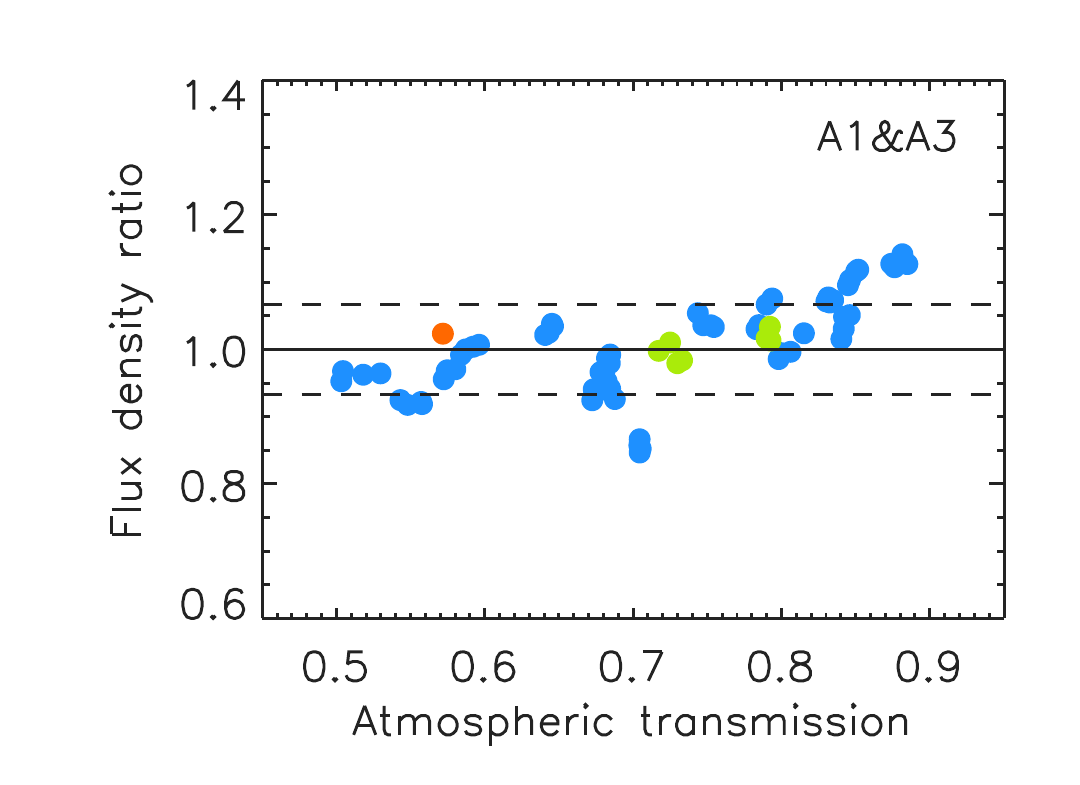}
      \put(20,60){\footnotesize {\tt Skydip}}
    \end{overpic}
    \includegraphics[clip=true, trim={1.8cm, 0.2cm, 0.5cm, 0.7cm},width=0.457\linewidth]{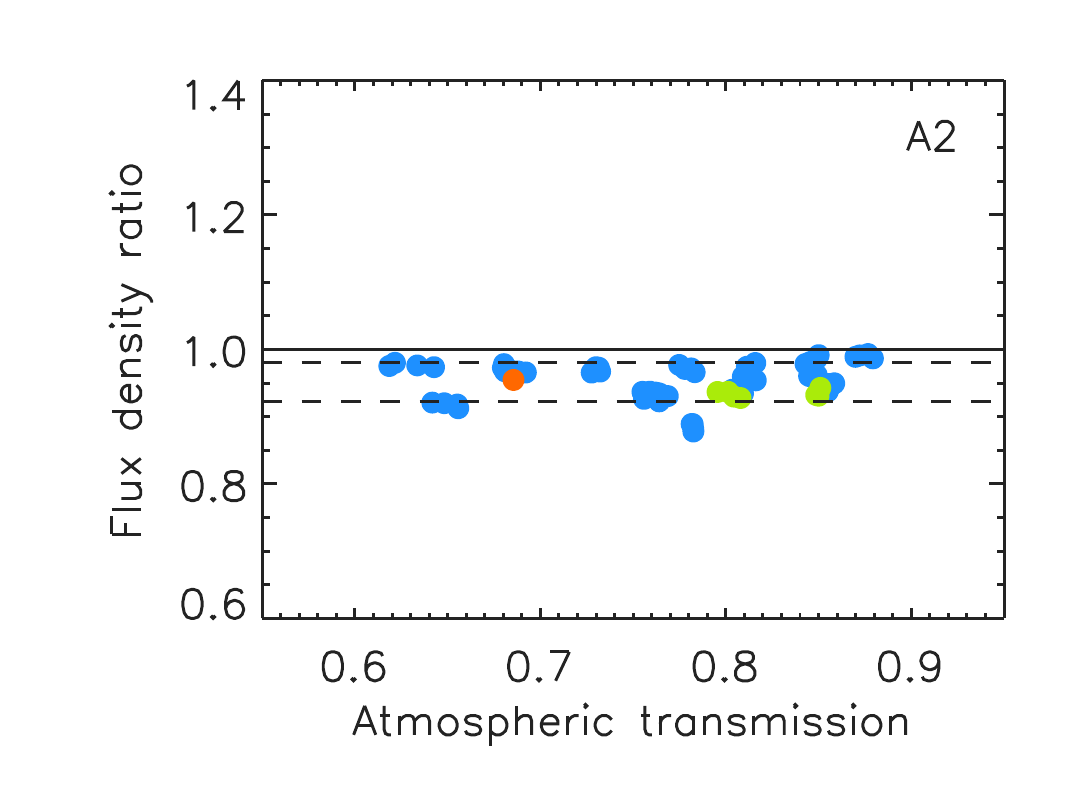}
    \begin{overpic}[clip=true, trim={0.9cm, 0.2cm, 0, 0.6cm},width=0.532\linewidth]{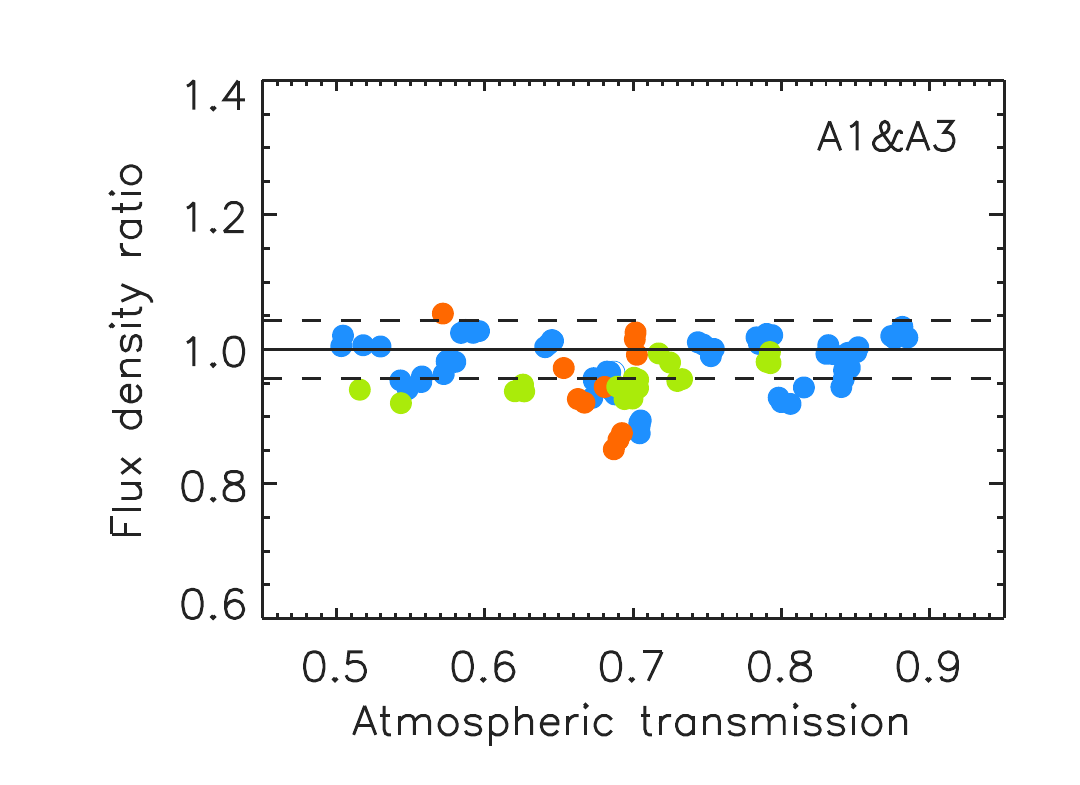}
      \put(20,60){\footnotesize {\tt PC-demo}}
    \end{overpic}
    \includegraphics[clip=true, trim={1.8cm, 0.2cm, 0.5cm, 0.7cm},width=0.457\linewidth]{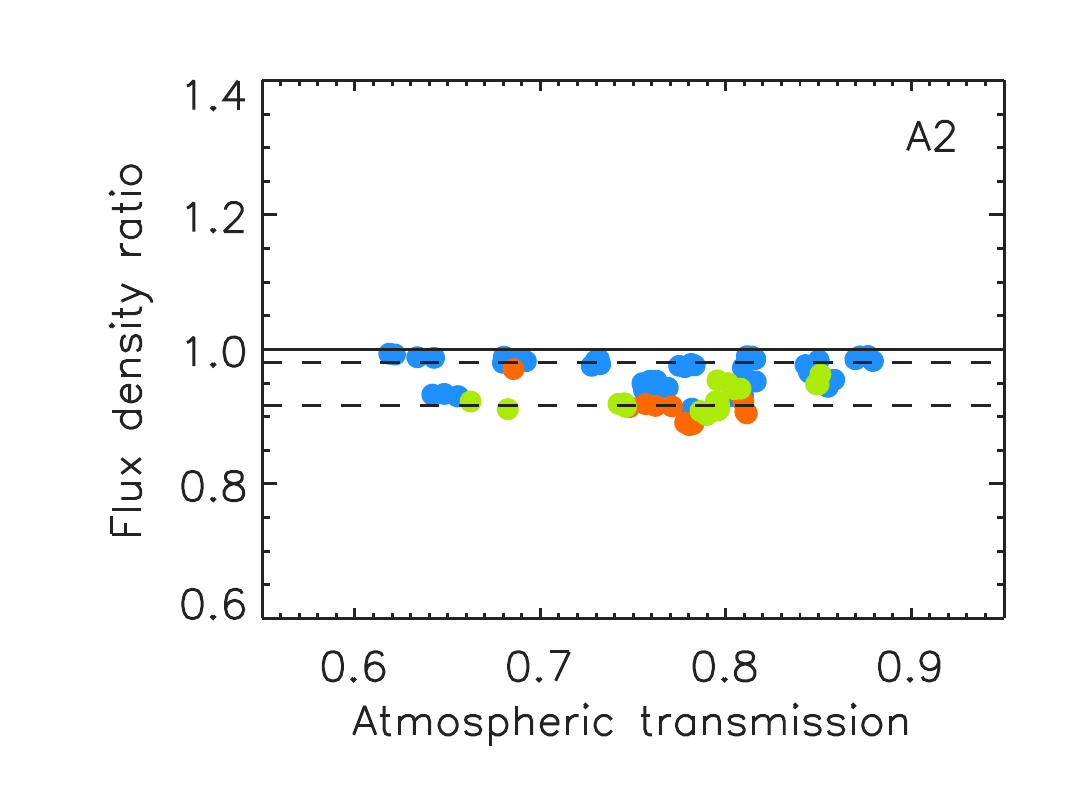}
    \begin{overpic}[clip=true, trim={0.9cm, 0.4cm, 0, 0.6cm},width=0.532\linewidth]{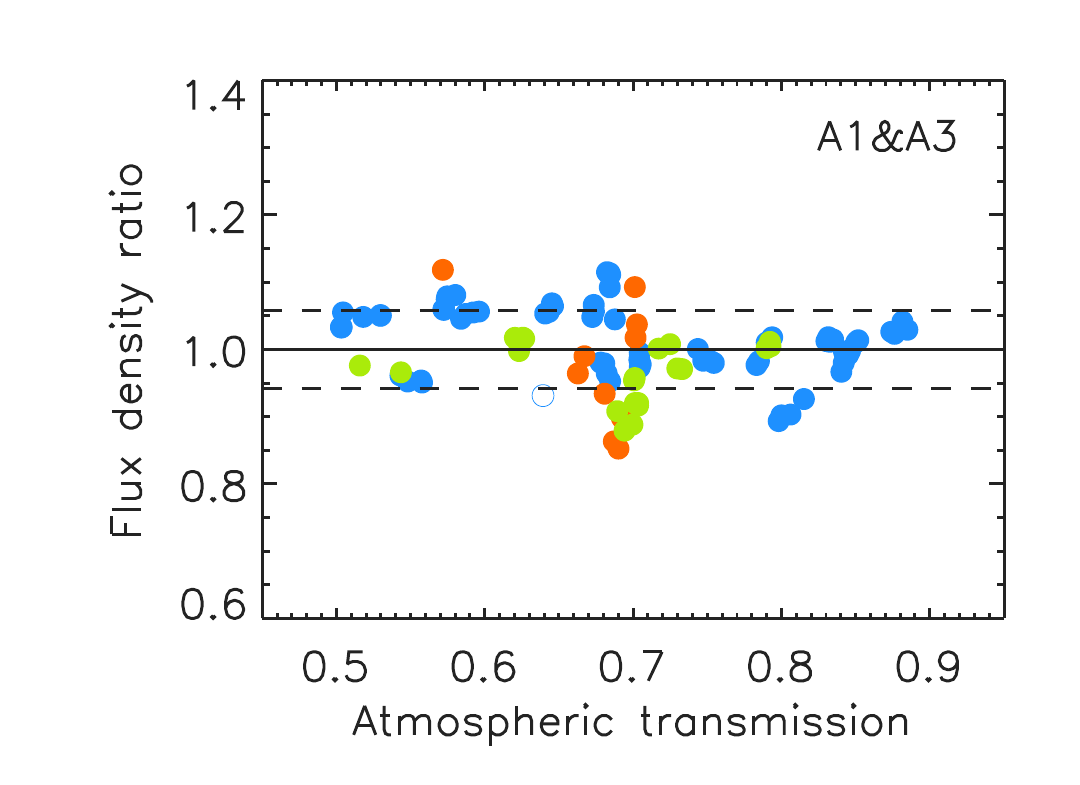}
      \put(20,60){\footnotesize {\tt PC-point}}
    \end{overpic}
    \includegraphics[clip=true, trim={1.8cm, 0.4cm, 0.5cm, 0.7cm},width=0.457\linewidth]{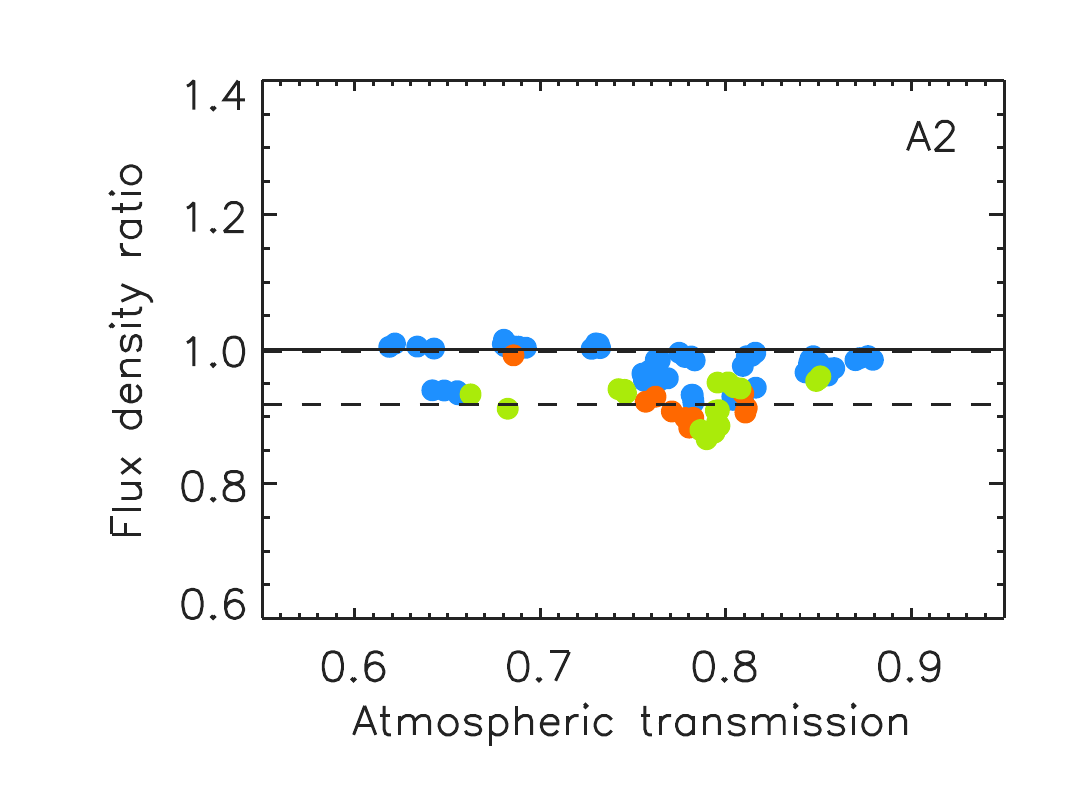}
    \vspace{-0.3cm}
    \caption[Calibration bias comparison]{Comparison of the
        calibration bias for five calibration methods using
          observations of MWC349.
       The measured-to-expected flux density ratio is shown as a
      function of the atmospheric transmission for the baseline method
      (first row) as well as for methods using the {\tt taumeter} (second
      row) and {\tt skydip} (third) opacity correction, and for methods
      resorting to the {\tt PC-demo} (fourth) and {\tt PC-point} (fifth)
      photometric corrections. Dashed lines
      show the flux density ratio rms errors.}
    \label{fig:mwc349_obstau_others}
  \end{center}
\end{figure}

\subsubsection{Absolute and systematic uncertainties}
\label{se:def_systematic_errors}

To account for all uncertainties, we must add to the {\rev
point-source} rms calibration uncertainties the absolute calibration
uncertainties and the systematic errors. The absolute uncertainty is
the uncertainty on the primary calibrator flux density expectations. 
In the case of Uranus, \citet{Morenothesis} and \citet{Bendo2013} report 
uncertainties of about $5\%$ at both wavelengths.

For the \emph{baseline} calibration method, which resorts to the {\tt
corrected skydip} method for the atmospheric opacity correction (see
Sect.~\ref{se:corrected-skydip}), the uncertainties on the
correcting factor $a_\nu^{\rm{skydip}}$, as defined in
Eq.~\ref{eq:corrected_skydip}, must be propagated to the flux
uncertainties. These uncertainties, which are referred to as the
{\tt corrected skydip} uncertainties, depend on the line-of-sight atmospheric
opacity $\taunu x$. Precisely, because the {\tt corrected skydip}
opacity correction is consistently used for both the primary
calibrator and the target source flux measurement, the {\tt corrected skydip}
uncertainties depends on the difference between the average
line-of-sight opacity of the primary calibrator scans and the
line-of-sight opacity of the source scan.   
We evaluate the {\tt corrected skydip} uncertainties for two
different $\taunu x$ values. 1) For the reference IRAM \trentemetre\
winter observing conditions, defined as $2\,\rm{mm}$ of pwv and an
elevation of $60^{\rm{o}}$, the {\tt corrected skydip} uncertainties are of
0.6\% at $1\,\rm{mm}$ and 0.3\% at $2\,\rm{mm}$. 2) In the worst
observing conditions allowed by the \emph{baseline} scan selection
(Sect.~\ref{se:data_selection}),
which are $\taunu x$ of 0.7 at $1\,\rm{mm}$ and of 0.5 at
$2\,\rm{mm}$, we find {\tt corrected skydip} uncertainties of 2\% and
$1.5\%$ at 1 and $2\,\rm{mm}$, respectively. These constitute
conservative upper limits on the {\tt corrected skydip}
uncertainties.

The uncertainties on NIKA2 bandpass measurements (see
Sect.~\ref{se:instru_bandpass}) propagate into
uncertainties on the flux densities after the colour correction using
Eq.~\ref{eq:color_correction}. These
uncertainties depend on the source SED but are negligible in most of
the cases. In particular, for MWC349, we find uncertainties below 0.1\% at
both wavelengths.}

\subsection{Baseline calibration photometry results}
\label{se:photometry_baseline}

We measure the calibration bias and rms uncertainties, as defined
in the previous section (Sect.~\ref{se:photometry_criteria}) using the
\emph{baseline} calibration method (Sect.~\ref{se:baseline_calibration}).

The calibration bias is evaluated using a
series of scans of MWC349 acquired during the
 reference observation campaigns. Namely, we use the 72 scans that met
 the \emph{baseline} selection
 criteria (see Sect.~\ref{se:data_selection}) over the 109 available
scans for MWC349. The first row of
Fig.~\ref{fig:mwc349_obstau_others}, labelled 'baseline', shows the
calibration bias $b_{\nu}$ for the combination of the $1\,\rm{mm}$ arrays and
Array 2 as a function of the atmospheric transmission 
$\exp \left( - \taunu \, x \right)$. No significant dependency of the
calibration bias on the atmospheric transmission is observed.

Table~\ref{tab:baseline-photometry} gathers the calibration bias
estimates for the three observation campaigns and for all the scans.
In the $1\,\rm{mm}$ band, we find
$b_\nu$ in agreement with unity within the statistical dispersion for
the three campaigns,
whereas a $5\%$ lack of flux with respect to expectations is observed
at $2\,\rm{mm}$, consistently for the three campaigns. This bias has a
low significance with respect to the absolute calibration precision of
NIKA2 (see Sect.~\ref{se:photometry_criteria}).
This will be further investigated by using other calibration methods
in Sect.~\ref{se:photometry_others}.

\begin{table}[!thbp]
  \begin{center}
    \caption[Baseline calibration results]{Baseline calibration results:
  photometry accuracy and uncertainties. The first sub-panel labelled 'Bias' gives the
  calibration bias $b_{\nu}$ and the second sub-panel labelled 'Rms' the calibration
  rms error $\sigma_{\nu}$, as defined in
  Sect.~\ref{se:photometry_criteria},
  using observations during N2R9, N2R12, N2R14 and the combination of
  the three campaigns. {\lp In each sub-panel, the first row indicates the
    number of acquired scans, while the second row gives the
    number of selected scans using the \emph{baseline} scan selection.}}
\label{tab:baseline-photometry}
\begin{tabular}{clrrrr}
  \hline\hline
  \noalign{\smallskip}
   \multicolumn{2}{c}{Characteristics} &  N2R9  & N2R12   &  N2R14 & Combined \\
  \noalign{\smallskip}
  \hline
  \noalign{\smallskip}
  Bias &  $\#$ total    &  68    &  14     &   27     &    109    \\
       &  $\#$ selected &  64    &   1     &   7      &     72    \\
       &  A1            &  0.95  &  1.03   &   0.94   &   0.95    \\
       &  A3            &  0.99  &  1.07   &   1.00   &   1.00    \\
       &  1mm           &  0.97  &  1.05   &   0.97   &   0.98    \\
       &  2mm           &  0.95  &  0.95   &   0.93   &   0.95    \\
  \hline
  \noalign{\smallskip}
  Rms  &  $\#$ total    &  303   &  72     &   112    &    487   \\
  $[\%]$ &  $\#$ selected &  219   &  33     &    12    &    264   \\
       &  A1            &  5.7   &  4.6    &   2.9    &    5.5   \\
       &  A3            &  6.2   &  5.7    &   2.4    &    6.0   \\
       &  1mm           &  5.9   &  5.0    &   2.5    &    5.7   \\
       &  2mm           &  3.2   &  2.1    &   1.1    &    3.0   \\  
\hline
\end{tabular}
\end{center}
\end{table}

\begin{figure}[!thbp]
  \begin{center}
      \includegraphics[clip=true, trim={0.9cm, 0, 0.5cm, 0.6cm},width=0.75\linewidth]{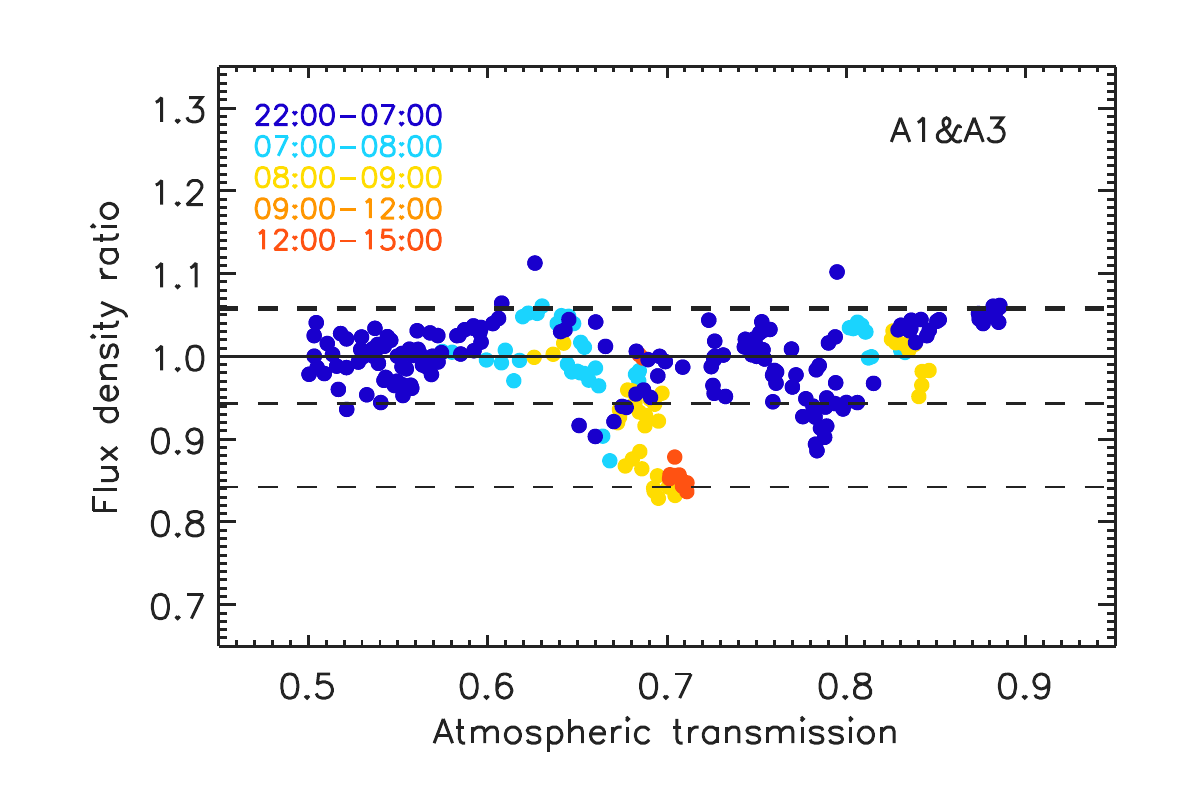}
     \includegraphics[clip=true, trim={0.9cm, 0, 0.5cm, 0.6cm},width=0.75\linewidth]{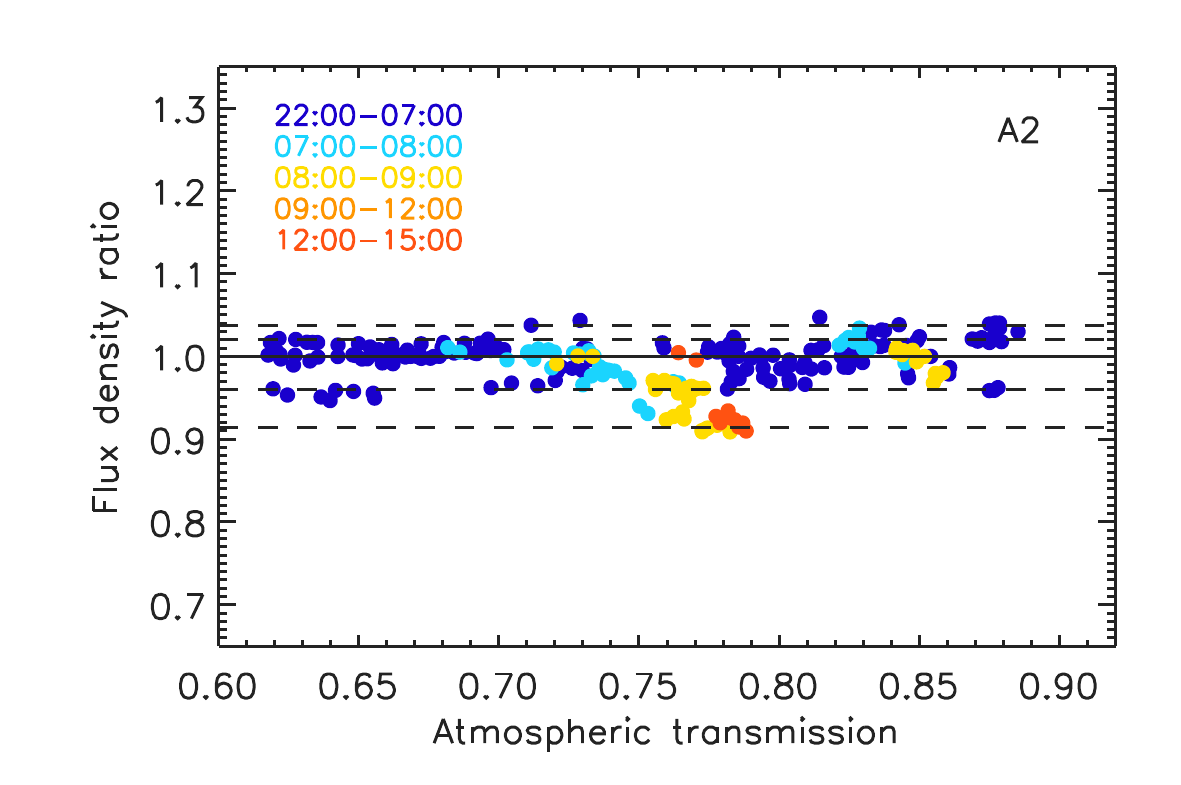} 
    \caption[Baseline calibration rms error estimate]{Baseline
      rms calibration uncertainties. The
      measured-to-median flux density ratio of bright sources is
      plotted as a function of the atmospheric transmission
      colour-coded according to the UT
      observation time of the scans for the combination of A1$\&$A3
      (top panel)
      and for A2 (bottom panel).
      The inner dashed lines from either sides of the
      unity-ratio line show the rms errors, {\lp which
      are less than 6\% at $1\,\rm{mm}$ and 3\% at $2\,\rm{mm}$, while
      the outer dashed lines show the $95\%$ confidence level contours.}
      The lowest flux ratio data points correspond to some of the
      scans acquired during daytime between 8:00 UT and 15:00 UT
      hours (yellow and red), while the scans acquired during night-time
      between 22:00 UT and 7:00 UT yield data points (dark blue)
      well distributed within the rms error with a few outliers.}
    \label{fig:allbright_rms_corrected_skydip}
  \end{center}
\end{figure}
Figure~\ref{fig:allbright_rms_corrected_skydip} shows the
measured-to-median flux densities evaluated from bright source scans
for the combination of Array $1\&3$ and Array 2 as a function of the
atmospheric transmission and colour-coded as a function of the
observation time. From a total of 487 scans towards
flux-selected sources, acquired during N2R9, N2R12 and N2R14, 264 met
the baseline selection criteria and are included in
Fig.~\ref{fig:allbright_rms_corrected_skydip} for testing the
calibration stability. The calibration uncertainties are
estimated using the standard deviation of the flux density ratios for
the three campaigns. Results are gathered in
Table~\ref{tab:baseline-photometry}.
Combining all the scans, we find rms uncertainties of $5.5\%$ for A1,
$6.0\%$ for A3, $5.7\%$ for the $1\,\rm{mm}$ band and $3.0\%$ for A2.
{\lp Using the flux ratio distributions, we construct the 68 and 95\%
C.L. intervals. The 68\% C.L. intervals are $-6.4\%$ and $+3.4\%$ at
$1\,\rm{mm}$ and $-3.8\%$ and $+1.5\%$ at $2\, \rm{mm}$. Hence in average
the 68\% C. L. errors are of $4.9\%$ and $2.7\%$ at 1 and $2\, \rm{mm}$,
respectively. We conclude that the rms errors are conservative
estimates of the 68\% C. L. errors at both wavelengths. The 95\%
C.L. contours are $-15.8\%$ and $+5.9\%$ at $1\,\rm{mm}$ and $-8.6\%$ and
$+3.8\%$ at $2\, \rm{mm}$.
The rms errors and the 95\%
C.L. interval are shown in
Fig.~\ref{fig:allbright_rms_corrected_skydip} with the inner and
outer dashed lines, respectively.} 

The flux density ratio is constant within the rms errors along the
wide range of tested atmospheric transmission, ranging from 0.5 to 0.9
at $1\,\rm{mm}$.
However, some scans at atmospheric transmissions of about 0.7 at
$1\,\rm{mm}$ show a mild lack of flux density with respect to the
median within the 95\% C. L. contours. The scans
affected by the lack of flux have all been observed
either between 12:00 and 14:00 UT or between 8:00 and 9:00 UT,
that are close to the threshold of the observation time cuts of
the \emph{baseline} scan selection (see
Sect.~\ref{se:data_selection}). These scans are likely
to be affected by the \afternoon\ beam broadening or by the
sunrise focus drift, respectively. Furthermore, we find that restricting the
used observation time to the 10 more stable hours (from 22:00 to
08:00 UT) would result in rms calibration uncertainties of
$3.6\%$ at $1\,\rm{mm}$ and $1.2\%$ at $2\,\rm{mm}$, which constitute an
improvement of about $60\%$ at $1\,\rm{mm}$ and $40\%$ at $2\,\rm{mm}$
of the rms errors.  
The \emph{baseline} scan selection, which i)
retains 16 hours of observation time
a day and ii) 
{\lp results in state-of-the-art rms calibration
uncertainties,} constitutes {\lp an useful trade-off representative of most
of the observations with NIKA2.} 

%
%
\subsection{Comparison with other calibration methods}
\label{se:photometry_others}
\begin{table*}[!htbp]
\begin{center}
\caption[Comparison of calibration results using five
  methods]{Comparison of results using five calibration
  methods: (first column) the \emph{baseline} calibration, (second and
  third) the calibration methods using other opacity correction, {\tt
  taumeter} and {\tt skydip}, respectively, and
  (fourth and fifth) the calibration methods using a photometric correction, {\tt
  PC-demo} and {PC-point}, respectively (see Appendix~\ref{se:photometric_correction}).  
  The calibration biases, as defined in
  Sect.~\ref{se:def_calibration_bias}, are reported in the sub-panel
  labelled 'Bias'. The sub-panel labelled 'Rms' gathers (first row) the total number of
  observation scans of bright sources (see
  Sect.~\ref{se:def_calibration_rms_error}) acquired during
  the \emph{reference} campaigns, (second row) the number of selected
  scans, and (third to sixth rows) the rms calibration uncertainties
  (Sect.~\ref{se:def_calibration_rms_error}) for Array 1, Array 3, the
  combination of A1 and A3, and Array 2, respectively. }
\label{tab:Calibration_results_all}
\begin{tabular}{clrrrrr}
  \hline\hline
  \noalign{\smallskip}
  \multicolumn{2}{c}{}  &  \multicolumn{5}{c}{Methods} \\\cline{3-7}
  \noalign{\smallskip}
  \multicolumn{2}{c}{Characteristics} &  baseline  & {\small {\tt taumeter}}  & {\small {\tt skydip}}  &  {\small {\tt PC-demo}} & {\small {\tt PC-point}} \\
  \hline
  \noalign{\smallskip}
  Bias &  A1            &   0.95   &  0.98    &  0.97    &   0.95    &  0.97  \\
       &  A3            &   1.00   &  1.02    &  1.02    &   0.99    &  1.00  \\
       &  1mm           &   0.98   &  1.01    &  1.00    &   0.97    &  0.99  \\
       &  2mm           &   0.95   &  0.95    &  0.95    &   0.95    &  0.95  \\
  \hline
  \noalign{\smallskip}
  Rms  &  $\#$ total      &   487    &    487   &    487    &    396    &  396 \\
  $[\%]$ &  $\#$ selected &   264    &    264   &    264    &    291    &  283 \\
       &  A1            &   5.5    &    7.5   &    7.3    &    4.0    &  4.9 \\
       &  A3            &   6.0    &    8.1   &    7.1    &    4.1    &  5.2 \\
       &  1mm           &   5.7    &    7.9   &    7.1    &    3.8    &  4.9 \\
       &  2mm           &   3.0    &    3.8   &    3.0    &    2.2    &  2.4 \\
\hline
\end{tabular}
\end{center}
\end{table*}

In this section, the \emph{baseline} calibration results are compared to
results drawn either using other calibration methods obtained from different
opacity corrections ({\tt taumeter} and {\tt skydip} as
discussed in Sect.~\ref{se:baseline_calibration_opacity}), or
including a photometric correction ({\tt PC-demo} and {\tt PC-point},
as described in Appendix~\ref{se:photometric_correction}) to mitigate
the \afternoon\ variation effect (Sect.~\ref{se:beam_variation}).
For robustness test, we evaluate and compare the photometry quality criteria of
Sect.~\ref{se:photometry_criteria} for the five calibration methods.

\subsubsection{Calibration bias}
\label{se:calibration_bias_all}

We present the calibration bias as a function of the atmospheric
transmission for the five calibration methods in
Fig.~\ref{fig:mwc349_obstau_others} and report the results in the row
labelled 'Bias' of Table~\ref{tab:Calibration_results_all}.

At $1\,\rm{mm}$, all methods lead to flux density estimates in
agreement with expectations within the rms dispersion. However,
{\tt taumeter} flux ratios have more dispersion than
the \emph{baseline} flux ratios whereas {\tt skydip} shows some
dependency on the atmospheric
transmission, with a 10 to $15\%$ excess of the flux density with
respect to expectations at high transmission. This residual
systematic effect has motivated the development of the {\tt corrected
  skydip} method, as discussed in
Sect.~\ref{se:corrected-skydip}. These features, which are
already noticeable from Fig.~\ref{fig:mwc349_obstau_others}, will be
confirmed and further discussed later using more observation scans. On
the other hand, the calibration methods based on photometric
correction (Appendix~\ref{se:photometric_correction}) yield an
unbiased photometry (calibration bias in agreement with unity
within the rms error) while allowing the use of $30\%$ more
scans. These results are encouraging for the exploitation of scans
acquired during the observing periods
impacted by the \afternoon\ beam variation effect
(Sect.~\ref{se:beam_variation}).

At $2\,\rm{mm}$, all methods result in a similar calibration bias
of $0.95$ with a rms error of $0.05$ estimated on the MWC349 scans. 
This corresponds to a low-significance $5\%$ lack of flux density
towards MWC349.
To summarize, the calibration bias at $2\, \rm{mm}$ is stable against
i) a large range of atmospheric conditions, ii) the observation campaign, iii) the
opacity correction method, iv) the method to treat the
temperature-induced beam variation effect.
This $5\%$ lack of flux density is thus probably due to
uncertainties on the flux density expectations for this source.
They come in two flavours.
{\lp Firstly the uncertainties on the flux expectation, as reported in
Appendix~\ref{se:ref_flux_secondaries}, consist in the propagation of
the errors on the fitted SED from the \emph{Plateau de Bure Interferometre}
(PdBI) and the \emph{Very Large Array} (VLA) observations. Systematic
uncertainties that may also impact the SED are not included.}  
Secondly, the NIKA2 flux density extrapolation from
interferometer data may be not straightforward for MWC349. {\lp In
particular, NIKA2 flux extrapolation ignores the contamination by
strong masers in the radio recombination lines~\citep{masingRRL},
while strong maser emission lines are masked in PdBI observations to
measure the continuum. In addition, the resulting continuum shows
indications of variability.}

\subsubsection{Calibration rms uncertainties}

The calibration rms uncertainties for the five methods evaluated using
flux-selected source scans are gathered in the sub-panel labelled 'Rms'
of Table~\ref{tab:Calibration_results_all}.

Compared to the \emph{baseline} method, the {\tt taumeter} method leads to 
rms errors increased of about $40$ and $30\%$ at 1 and
$2\,\rm{mm}$, respectively. The {\tt skydip} method shows lower
dispersion but a mild correlation with the atmospheric transmission, as
discussed in Sect.~\ref{se:calibration_bias_all}.

In addition, we have checked the flux density ratios for the bunch of
scans with an atmospheric transmission of about 0.7, which were
discussed in Sect.~\ref{se:photometry_baseline}, by comparing
calibration methods with or without photometric correction. The
flux density ratios are low (within the 95\% C.L. interval) for the scans
observed within 12:00 and 15:00 UT in the three first methods, as shown in
Fig.~\ref{fig:allbright_rms_corrected_skydip} for
the \emph{baseline} method. By contrast, they are within the 68\% C.L. interval
when using a photometric correction. This further validates the
hypothesis that the low flux
density of these scans is due to \afternoon\ beam effect, as assumed in
Sect.~\ref{se:photometry_baseline}. This also constitutes an example
of the calibration improvement obtained by using a
photometric correction.

Moreover, results based on the {\tt PC-demo} method show that rms
calibration uncertainties as low as $3.8$ and $2.2\%$
at 1 and $2\,\rm{mm}$ are within the reach of NIKA2 {\lp without any
selection based on the observation time.}
However, we recall this method relies on 
accurate beam estimates. Using {\tt PC-point}, which is the
practical case, still improves the calibration uncertainties
w.r.t. the \emph{baseline} results but by a factor of about $20\%$ in both
bands. Furthermore, the differences between
the flux density ratios from {\tt PC-demo} and {\tt PC-point}, which are
seen e.g. from the corresponding panels of
Fig.~\ref{fig:mwc349_obstau_others}, are likely
to be due to the photometric correction noise
when monitoring the beam from pointing scans (see
Appendix~\ref{ap:beam_monitoring}). We conclude that more
control on the beam monitoring is needed before routinely using a calibration
based on photometry correction. By contrast, the baseline method
combines good performance with robustness.

\subsection{Summary}
\label{se:photometry_summary}
Among the methods that rely on the UT hour-based
scan selection to mitigate the effect of beam size variations, the
\emph{baseline} method shows the best performance in terms of calibration
bias and uncertainties. The methods that rely on a photometric correction
show good calibration results, and thus represent a promising lead
to further improve the calibration uncertainties. However, their
robustness depends on the accuracy of the beam monitoring. {\lp The
proposed beam monitoring based on {\tt pointing} scans induces some
extra dispersion of the flux densities. A more accurate beam monitoring is
feasible but requires using dedicated observation scans.} 
From the \emph{baseline} method results discussed in
Sect.~\ref{se:photometry_baseline}, we have found that the measured
flux density of MWC349 is in agreement with expectations within $5\%$
for both wavelengths. Moreover, the {\rev point-source} rms calibration
uncertainties are of $5.7\%$ at $1\,\rm{mm}$ and of $3\%$ at $2\,\rm{mm}$
using a series of 264 scans of sources of flux density above
$1\, \rm{Jy}$.
These results demonstrate the excellent accuracy and stability of the
NIKA2 {\rev point-source} photometric capabilities.

\section{Sensitivity}
\label{se:sensitivity}

In this section, we derive the on-sky sensitivity of the instrument using a
large amount of observation scans, including deep integration on faint sources,
and assess the stability of our results against the observing
conditions.
We evaluate the noise equivalent flux density, which is referred to as
NEFD hereafter. To further represent the mapping capabilities, we also
derive the mapping speed.
We first discuss the definitions and the technical derivation of these quantities in
Sect.\ref{se:integration_time} from measurements. Then, we briefly present
several estimation methods that have been considered and the data sets that have
been selected in Sect.~\ref{se:nefd_method}. Results, together with robustness
tests, are reported in Sect.~\ref{se:nefd_results}.

\subsection{NEFD and mapping speed definitions}
\label{se:integration_time}

The NEFD is the $1\,\sigma$ error on the flux
density of a point source in one second of integration time,
considered at zero atmospheric opacity.
At this stage, we consider a map of a point-like source located at its
center and observed with zero atmospheric opacity. The estimation of
the flux density uncertainty $\sigma$ is described in Sect.~\ref{se:dataproc}.
We derive here the integration time from actual observations. Indeed,
it is not simply the duration of a scan, it depends on the KID
distribution in the FOV and the scanning strategy.
The flux density map, which has a resolution of $\Delta r$, also comes
along with a hit map $H_{\rm{p}}$, which counts the number of
data samples per pixel (see Sect.~\ref{se:dataproc}).
From this, we derive the integration time at the center
of the map as
\begin{equation}
t_{\rm{center}} = \frac{\langle H_{\rm{p}}\rangle_{\rm{center}}}{f_{\rm{sam}}}\,,
\end{equation}

where $f_{\rm{sam}}$ is the sampling frequency and $\langle
H_{\rm{p}}\rangle_{\rm{center}}$ is the average of the hit map taken in a disk
of radius of one FWHM to be immune to shot noise statistics in individual map
pixels. Correcting this quantity from the density of detectors per map pixel at
the same time, gives the on-source integration time per beam, in other
words, the time when the source is actually being observed, by at least one
detector:

\begin{equation}
  t_{\rm{beam}} = t_{\rm{center}}\, \frac{g^2}{\Delta r^2}\,,
\label{eq:t_det}
\end{equation}

where $g$ is the center-to-center distance between adjacent KIDs in the FOV
(Sect.~\ref{se:grid_distortion}). $t_{\rm{beam}}$ matches the total duration of
the scan if the scanning strategy is designed so that the source is always in
the FOV and if the FOV is full of valid KIDs. The NEFD is defined
using the previous quantities as 

\begin{equation}
  \rm{NEFD} = \sigma \sqrt{t_{\rm{beam}}}\, 
\label{eq:nefd_def}
\end{equation}

and is given in $\rm{mJy}\cdot s^{1/2}$.

The mapping speed, $M_{\rm{s}}$, is the sky area $\mathcal{A}_{\rm{scan}}$ that
can be mapped at a noise level $\Delta_\sigma$ of $1\,\rm{mJy}$ in an
integration time $\Delta_t$ of one hour. Noting $d_{\rm{FOV}} =
6.5\,\rm{arcmin}$ the FOV diameter and $\eta$ the fraction of valid KIDs for an
observation (Sect.~\ref{se:fov_geometry}), the mapping speed is

\begin{equation}
M_{\rm{s}} = \frac{\mathcal{A}_{\rm{scan}}}{\Delta_\sigma\Delta_t} = 
\eta \, \frac{\pi}{4} d_{\rm{FOV}}^2 \, \frac{1}{\rm{NEFD}^2}\, ,
\label{eq:mapping_speed}
\end{equation}

and has units of arcmin$^2 \cdot \rm{mJy}^{-2} \cdot \rm{h}^{-1}$. 
In real observation conditions, characterized with a given atmospheric
opacity $\taunu$ and a given \airmass\ $x$, correcting the flux
density for atmospheric attenuation using Eq.~\ref{eq:uncorr_flux}   
increases the NEFD as 
\begin{equation}
\rm{NEFD}_{\taunu,\, x} = \rm{NEFD}\, e^{\taunu x}
\label{eq:nefd_tau_el}
\end{equation}

Using these definitions, the integration time required to reach a target flux
density uncertainty $\sigma_{obs}$ over an observing area
$\mathcal{A}_{\rm{scan}}$ for an airmass $x$ and for an atmospheric opacity $\taunu$ is
\begin{equation}
  t_{\rm{obs}} = \frac{\mathcal{A}_{\rm{scan}}}{ M_{\rm{s}}} \, \left(\frac{e^{\taunu\, x}}{\sigma_{obs}}\right)^2.
\end{equation}

\begin{figure*}[!thbp]
  \begin{center}
    \includegraphics[trim={0.5cm, 0, 0, 0.5cm}, clip, angle=0, width=0.495\textwidth]{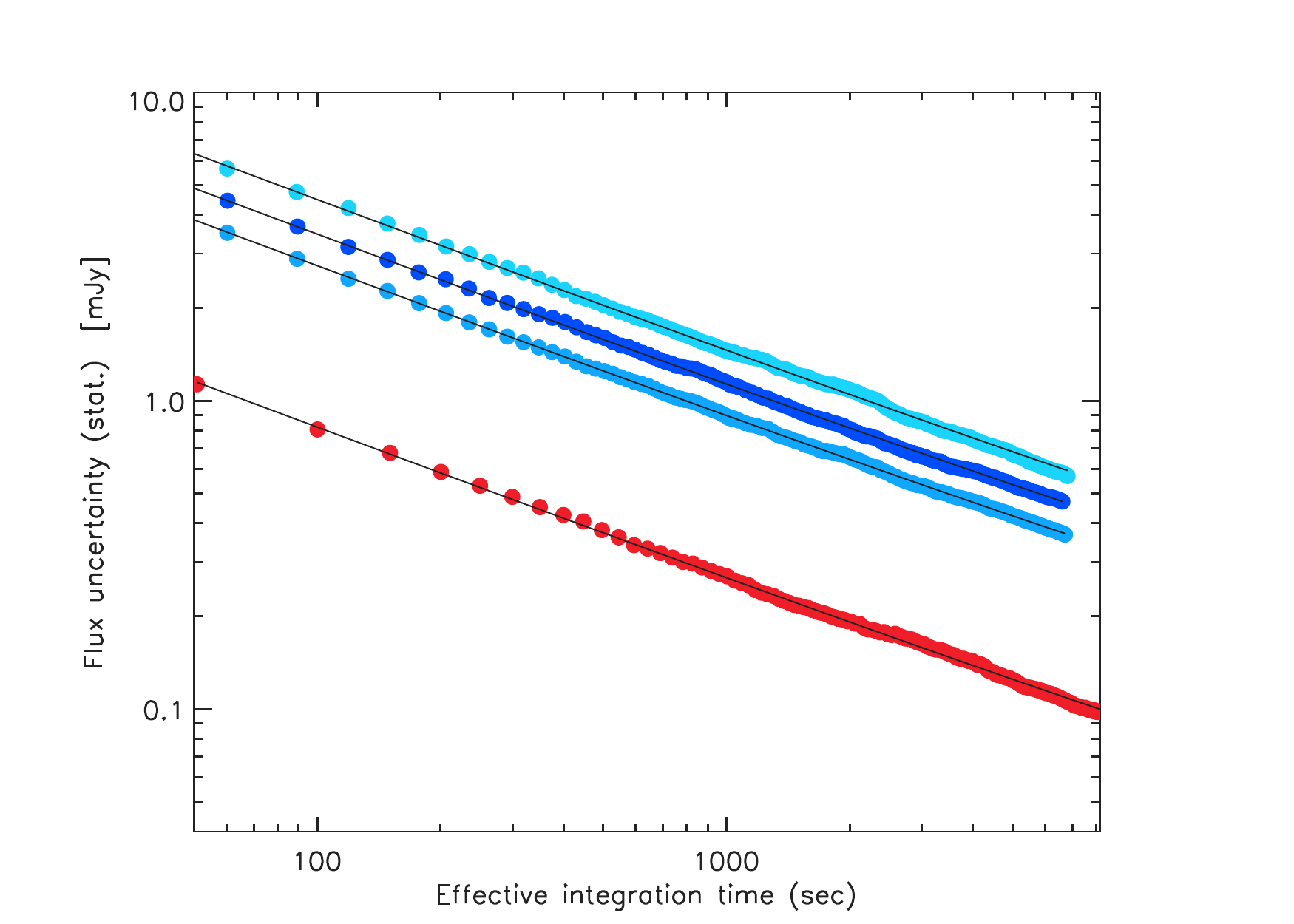}
    \includegraphics[trim={0.5cm, 0, 0.2cm, 0.5cm}, clip, angle=0, width=0.485\textwidth]{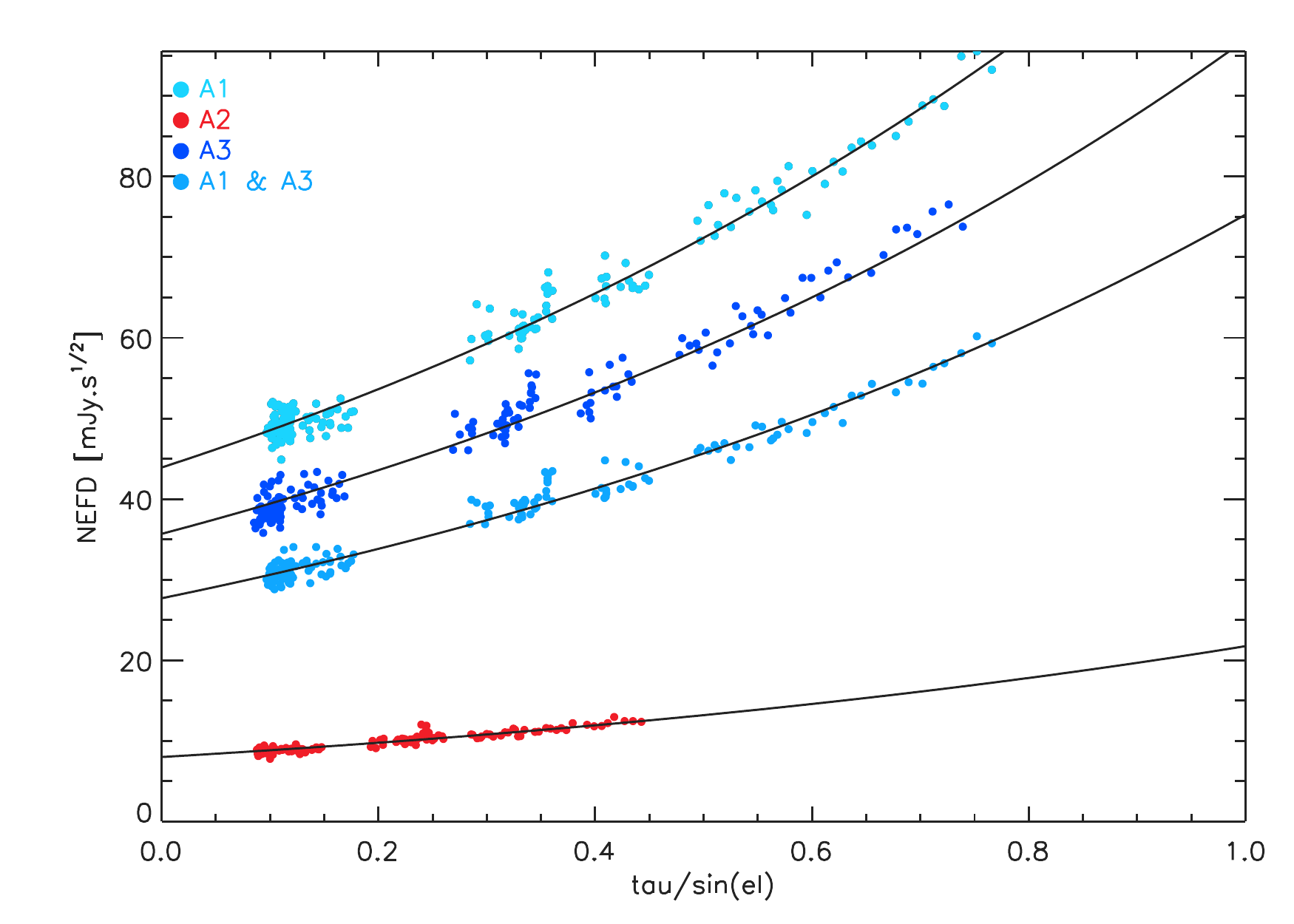}
    \caption{Comparison of NEFD estimates using two methods on observations of
      \hls. \emph{Left panel:} Evolution of the 1\,$\sigma$ flux density
      uncertainties as a function of the effective integration time
      $t_{\rm{eff}}$, as defined in Eq.~\ref{eq:sigma_tau_w8}, for A1
    (cyan), A3 (dark blue), the combination of A1\&A3 (medium blue),
    and A2 (red). The solid black lines are the best-fit models using
    $\sigma(t_{\rm{eff}}) =  \rm{NEFD}/sqrt(t_{\rm{eff}})$. \emph{Right panel:} NEFD as a function of the
    measured line-of-sight opacity using the same colour code as in the
    left panel. The solid black lines are the theoretical
    fits of $\rm{NEFD}_{\taunu,\, x } = \rm{NEFD}\, e^{\taunu\,x}$ and give the
    NEFD when extrapolated to $\taunu/\sin(\elev)= 0$. }
    \label{fig:nefd_twomethods}
  \end{center}
\end{figure*}

\subsection{NEFD estimation methods and scan selection}
\label{se:nefd_method}

We have developed several methods for the NEFD estimation. First,
we use deep integrations on faint sources. Second, we resort to joint
analysis of multiple scans without combining them.

\subsubsection{Deep integration method}
According to Eq.~\ref{eq:nefd_def} and Eq.~\ref{eq:nefd_tau_el}, if a
source was observed under stable atmospheric conditions, the flux
uncertainty would scale directly like
$t_{\rm{beam}}^{-1/2}$. Using long-time integration observation on a
source, this relation provides both a way to estimate the NEFD and to
check that the noise does integrate down as expected with the integration
time. To that aim, we produce a series of maps, as described in Sect.~\ref{se:map_projection},
using an inverse-variance co-addition of an increasing number of observation
scans, and perform a photometric analysis on each map according to
Sect.~\ref{se:photometry}. However, in practice, in particular for
integrations of several hours, observing conditions do change.
Since all the scans are not acquired in the same conditions of
atmospheric opacity and observing elevation, {\rev they will not need
the same atmospheric opacity correction, nor have
the same level of atmospheric noise residuals after the noise
decorrelation, as described in Sect.~\ref{se:toi_proc}}, and hence,
they do not contribute with the same weight to
the co-addition. In such case, an effective integration time for the
co-addition of $n$ scans is defined as
\begin{equation}
t_{\rm{eff}}(n) = \sum_{i=1}^{n}\, t_i\,  e^{-2\taunu^i\, x_i},
\label{eq:sigma_tau_w8}
\end{equation}
where $t_i$, $\taunu^i$ and $x_i$ are the integration time, the zenith
opacity and the \airmass\ of the $i$-th scan of the $n$-scans
co-addition.
Generalizing Eq.~\ref{eq:nefd_def}, the flux density uncertainties on
the co-addition of $n$ scans is $\sigma(n) = \rm{NEFD}/\sqrt{t_{\rm{eff}}(n)}$.
An estimate of the
NEFD can therefore be obtained in fitting
$\sigma(n)$ as a function of the corresponding $t_{\rm{eff}}(n)$.

\subsubsection{Scatter method}
For any scan, we derive $\rm{NEFD}_{\taunu,\, x}$.
Using Eq.~\ref{eq:nefd_tau_el}, the joint analysis of a series of
scans acquired with various observing conditions provides an estimate
of the NEFD. The scan sample can gather different sources. 
The selection of the source target for the NEFD derivation is
primarily based on the flux density. Indeed, noise
characterization may be biased {\lp by the signal of a bright source
stemming from the most extended error beams and far side lobes.}
We therefore restrict the analysis to sources with estimated flux below 1\,Jy.

\begin{table}[!htbp]
  \centering
  \caption[]{Stability of the NEFD estimates. Top-of-atmosphere NEFD
    in $\rm{mJy}.s^{1/2}$ for the two methods described in the text, which
  are the deep integration (labelled Deep int) and the scatter method
  (Scatter), and using two different data sets, \hls\
  and all sub-Jy sources acquired during the reference observation
  campaigns. The results given in the last row are based on more than a thousand
  scans (202, 481 and 430 scans during N2R9, N2R12 and N2R14, respectively).}
  \label{tab:nefd_summary}
  \begin{tabular}{llrrrr}
    \hline\hline
    \noalign{\smallskip}
    Data set   & Method   & A1      &   A3    &   A1\&A3 &    A2 \\
    \noalign{\smallskip}
    \hline
    \noalign{\smallskip}
    \hls &     Deep int.  &  46.6  &    38.4  &    30.4  &   8.5  \\
         &     Scatter    &  45.7  &    36.3  &    28.5  &   8.2  \\
    \hline
    \noalign{\smallskip}
    N2R9     & Scatter    & 47.0 &  36.9  & 28.8  & 8.4 \\
    N2R12    &            & 47.3 &  36.4  & 30.2  & 8.5 \\
    N2R14    &            & 47.3 &  39.8  & 30.9  & 9.3 \\
    Combined &            & 47.2 &  37.9  & 30.1  & 8.8 \\
    \hline
  \end{tabular}
\end{table}

\subsection{Results and robustness tests}
\label{se:nefd_results}

\begin{figure*}[!thbp]
\begin{center}
\includegraphics[clip=true,width=0.47\textwidth]{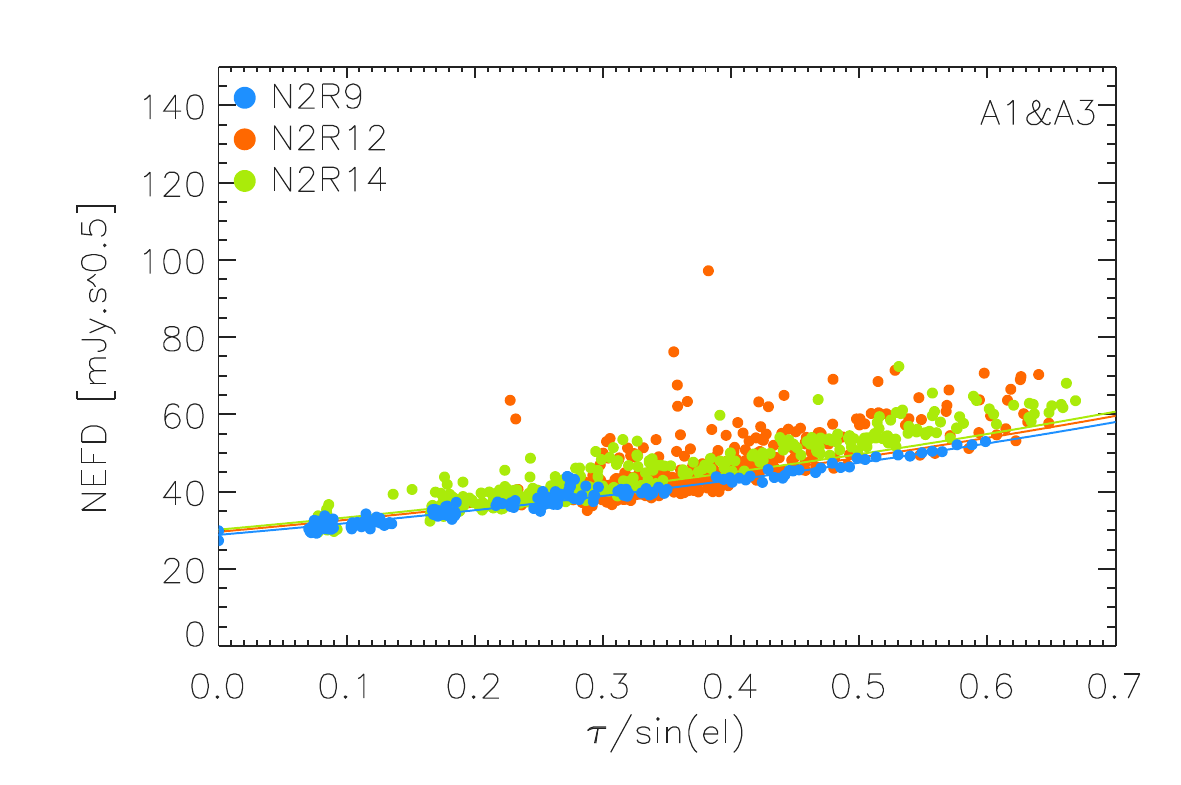}
\includegraphics[clip=true,width=0.47\textwidth]{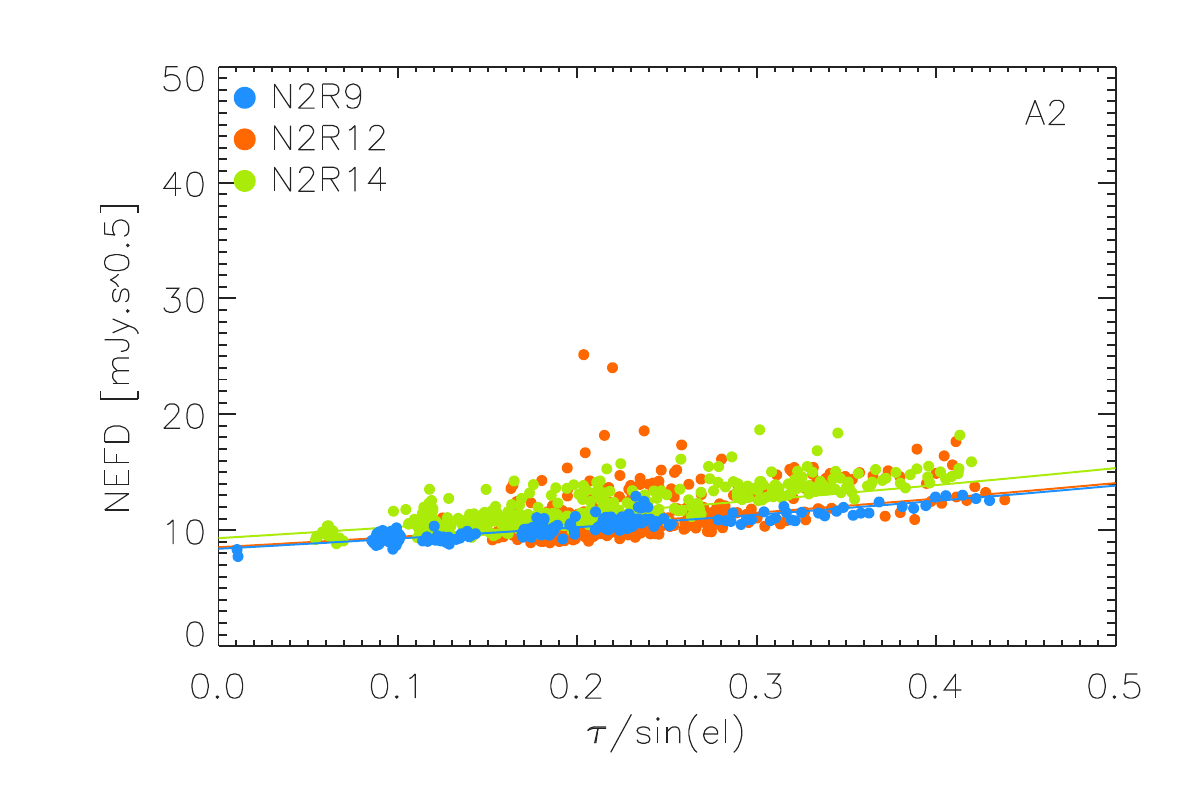}
\caption{Comparison of the NEFD estimates for three observation
  campaigns. The measured NEFD using the scatter method is plotted as a function of
  line-of-sight opacity ($\taunu\,x$) for the $1\,\rm{mm}$ (left) and $2\,\rm{mm}$ (right)
  channels. Data points are NEFD estimates in $\rm{mJy}\cdot s^{1/2}$ for N2R9 (blue), N2R12 (orange)
  and N2R14 (chartreuse). We also show in the plots the expected NEFD evolution
  with the line-of-sight opacity as solid curves using the median
  zenith opacity derived from all the scans acquired during a campaign.}
\label{fig:nefdvsbackground_below_1Jy}
\end{center}
\end{figure*}

\begin{table}[!thbp]
  \begin{center}
    \caption[NEFD estimates on all sub-Jy sources]{Median NEFD and rms
      uncertainties in $\rm{mJy}\cdot s^{1/2}$, as well as the derived mapping
      speed and mapping speed rms uncertainties in
    $\rm{arcmin}^2\cdot \rm{mJy}^{-2}\cdot\rm{h}^{-1}$, evaluated
      in using the scatter method on all sub-Jy sources of runs N2R9, N2R12
      and N2R14, given at pwv=0 and 90 degrees elevation (first three rows) and extrapolated at the
      reference Winter observing conditions at the IRAM
      \trentemetre\ telescope site (last three rows), which are defined
      as $2\,\rm{mm}$ pwv and 60 degrees elevation.}
    \label{tab:nefd_astro}
    \begin{tabular}{lrrrr}
      \hline\hline
      \noalign{\smallskip}
                    & A1      &   A3    &   A1\&A3 &    A2 \\
      \noalign{\smallskip}
      \hline
      \noalign{\smallskip}
      NEFD\, \small{($0\,\rm{mm}$ pwv, $90^{\rm{o}}$)}             & 47.2    & 37.9    &    30.1  &    8.8   \\
      Rms NEFD\, \small{($0\,\rm{mm}$ pwv, $90^{\rm{o}}$)}         &  3.9    &  3.5    &     2.9  &    1.1   \\
      M$_{\rm{s}}$\, \small{($0\,\rm{mm}$ pwv, $90^{\rm{o}}$)}      & 45      &  70     &    111   &   1388   \\
      Rms M$_{\rm{s}}$\, \small{($0\,\rm{mm}$ pwv, $90^{\rm{o}}$)}  &  4      &   6     &     11   &    174   \\
      \hline
      \noalign{\smallskip}
      NEFD\, \small{($2\,\rm{mm}$ pwv, $60^{\rm{o}}$)}             & 56.6    & 45.6    &    36.1  &    9.8   \\
      Rms NEFD\, \small{($2\,\rm{mm}$ pwv, $60^{\rm{o}}$)}         &  4.7    & 4.2     &     3.5  &    1.2   \\
      M$_{\rm{s}}$\, \small{($2\,\rm{mm}$ pwv, $60^{\rm{o}}$)}      &  31    & 48       &    77   &   1119   \\
      Rms M$_{\rm{s}}$\, \small{($2\,\rm{mm}$ pwv, $60^{\rm{o}}$)}  &   3    &  4       &     7     &  137   \\
      \hline
    \end{tabular}
\end{center}
\end{table}

First we test the stability of the NEFD estimates using the two methods on the
same data set. With this goal, during the N2R9 run, we selected \hls, a
moderately faint source \citep{2012A&A...538L...4C}, expected to have
flux densities of 74.5\,mJy at $1\,\rm{mm}$
and 15.7\,mJy at $2\, \rm{mm}$. 
It was observed for about 9\,h in total over three nights using
$8 \times 5$~arcmin$^2$ OTF raster scans of various orientations.\\

The left panel of Fig.~\ref{fig:nefd_twomethods} shows the flux density
uncertainties as a function of the integration time. The effective
integration time is different between the 1\,mm and 2\,mm arrays
because they have different KID spacings $g$ in the FOV and different
atmospheric opacities (see Eq.~\ref{eq:sigma_tau_w8}). Black lines show
the fit with the inverse of the square root of the integration time,
confirming that the noise integration is well consistent with the expected
scaling law. The observed small variations around the theoretical fit
correspond to variations of line-of-sight opacity during the
integration. These variations are taken into account for evaluating the NEFD as
discussed in the previous section. The right panel of
Fig.~\ref{fig:nefd_twomethods} shows the
$\rm{NEFD}_{\taunu,\, x}$ per scan, along with best-fit models using
Eq.~\ref{eq:nefd_tau_el}, from which are derived the \rm{NEFD} estimates. 
Results from these analyses, per array and for the
combined A1 and A3, are presented in Table~\ref{tab:nefd_summary}.

We observe systematically higher NEFD for A1 compared to A3, which is a mainly
due to the dichroic-induced 'shadow effect' that also impacts the flat fields,
as discussed in Sect.~\ref{se:flat_field}.  The dichroic-induced effect, which
also impacts A3, explains part of the NEFD difference of performance between the $1\, \rm{mm}$
and $2\, \rm{mm}$ channels, together with the higher atmospheric noise and lower
telescope beam efficiency at $1\, \rm{mm}$ with respect to $2\,\rm{mm}$
wavelength. Moreover, the NEFD evolution with sky noise is well
consistent with expectations for each array and each observing wavelength.

As a second robustness test, we check the stability of the NEFD for
three observation campaigns. Figure~\ref{fig:nefdvsbackground_below_1Jy} shows the
measured NEFD using the scatter method (see Sect.~\ref{se:nefd_method}) for the
sub-Jansky sources acquired at the N2R9, N2R12 and N2R14
campaigns. The NEFD estimates for the
three campaigns are in agreement within uncertainties for the whole
range of line-of-sight opacities that have been tested.
The solid lines show the expected dependence with
exp${[\taunu\,x]}$ as given in Eq.~\ref{eq:nefd_tau_el}. 
Since the measured NEFD$_{\tau,\, x}$ are not Gaussian distributed, we
derive the NEFD as the median of the NEFD$_{\tau,\, x}$
per scans after correction of the atmospheric attenuation, which provides us
with a more robust estimate compared to a fit. The NEFD estimates
are given in Table~\ref{tab:nefd_summary}.

Combining the data set of N2R9, N2R12 and N2R14 campaigns, more than one
thousand observations scans of sub-Jy sources meet the baseline selection
criteria (see Sect.~\ref{se:data_selection}), providing robust NEFD estimates
that are representative of the average NIKA2 performance. The rms
uncertainties are evaluated as the rms scatter of the individual
NEFD$_{\taunu,\, x}$ estimates after correction with
$e^{\taunu\,x}$. 
These values are then extrapolated using the IRAM
\trentemetre\ telescope reference Winter observing conditions: $2\,\rm{mm}$ of
precipitable water vapour (pwv) and $60$ degrees elevation. The NEFD
estimates, as well as the rms uncertainties, are gathered in
Table~\ref{tab:nefd_astro}.
From these estimates, we also derive the corresponding mapping speeds,
which are also given in Table~\ref{tab:nefd_astro}.
We report mapping speeds of $1388 \pm 174$ and
$111 \pm 11 \, \rm{arcmin}^2\cdot \rm{mJy}^{-2}\cdot \rm{h}^{-1}$ at 1
and 2\,mm, respectively.

\section{Summary and conclusions}
\label{se:summary}

\begin{table*}[!thbp]
\caption{Summary of the main characteristics describing NIKA2 measured performance}
\label{tab:nika2summary}
\centering    
\begin{tabular}{rrrcl}
  \hline\hline
  \noalign{\smallskip}
    & Array 1\&3 & Array 2 & & Reference \\
    \noalign{\smallskip}
    \hline
    \noalign{\smallskip}
    Reference Wavelength  [mm]  & 1.15  & 2.00   &  & \\
    Reference Frequency  [GHz]  & 260  & 150   &  & Sect.~\ref{se:photometric_system}  \\
    Frequency [GHz]             &  254.7\&257.4  & 150.9 &  & Sect.~\ref{se:instru_bandpass}  \\
    Bandwidth         [GHz]     &   49.2\&48.0   & 40.7  &  & \\
    \hline
    \noalign{\smallskip}
    Number of designed detectors                   &  1140\&1140 &    616  & & Sect.~\ref{se:array}\\
    Number of valid detectors\tablefootmark{a}     &  952\&961   &    553  & & Sect.~\ref{se:avg_kidpar}\\
    Fraction of valid detectors [$\%$]             &  84        &     90  & & \\
    Pixel size in beam sampling unit\tablefootmark{b}\hspace{3mm} [$\lambda/D$] & 1.1 &  0.87 & & Sect.~\ref{se:grid_distortion} \\
    \hline
    \noalign{\smallskip}
    FWHM\tablefootmark{c}\hspace{3mm} [arcsec]    &  $11.1 \pm 0.2$  &  $17.6 \pm 0.1$  & & Sect.~\ref{se:mainbeam}\\
    Main beam efficiency\tablefootmark{d}\hspace{3mm} [$\%$] & {\rev  $47 \pm 3$}   &  {\rev $64 \pm 3$}  &  & Sect.~\ref{se:beam_efficiency}\\
    Rms FWHM across the FOV [arcsec]              &    0.6        &      0.6        & & \citet{Adam2018} \\
    \hline
    \noalign{\smallskip}
    Reference FWHM\tablefootmark{e}\hspace{3mm} [arcsec]          & $12.5$     &   $18.5$  &  & Sect.~\ref{se:photometric_system}\\
    Reference beam efficiency\tablefootmark{f}\hspace{3mm}  [\% ] & {\rev $61 \pm 3$} & {\rev $72 \pm 2$} &  & Sect.~\ref{se:extended_source_calib}\\
    \hline
    \noalign{\smallskip}
    Rms pointing error    [arcsec]               & $<3$ &  $<3$  & & Sect.~\ref{se:pointing} \\
    \hline
    \noalign{\smallskip}
    Absolute calibration uncertainty [\%]      &   5         & 5 & & Sect.~\ref{se:photometry_criteria}, App.~\ref{se:ref_flux_uranus_neptune} \\
    Systematic calibration uncertainty\tablefootmark{g}\hspace{3mm}  [\%]      &    0.6        & 0.3 & & Sect.~\ref{se:def_systematic_errors} \\
    {\rev Point-source} rms calibration uncertainty [\%]                   &   5.7       &     3.0       & & Sect.~\ref{se:photometry_baseline} \\
    \hline
    \noalign{\smallskip}
    $\alpha$ noise integration in time\tablefootmark{h}\hspace{3mm}  & 0.5 & 0.5 & & Sect.~\ref{se:nefd_results} \\
    \hline
    \noalign{\smallskip}
    NEFD\tablefootmark{i}\hspace{3mm} [$\rm{mJy} \cdot \rm{s}^{1/2}$]  & $30 \pm 3$  & $9 \pm 1$ &  & Sect.~\ref{se:nefd_results}\\
    M$_{\rm{s}}$\tablefootmark{j}\hspace{3mm} [arcmin$^2 \cdot
      \rm{mJy}^{-2} \cdot \rm{h}^{-1}$] & $111 \pm 11$  &  $1388 \pm 174$ &  & \\
    \hline
  \end{tabular}
  \tablefoot{ \\
    \tablefoottext{a}{Number of usable detectors, which have been selected in at least two FOV reconstructions}
    \tablefoottext{b}{Calculated from real array pixel size [2.75\,mm / 2.0\,mm] and unvignetted entrance pupil diameter [27\,m]}
    \tablefoottext{c}{Full-width at half-maximum of the main beam using the combined results of three methods}
    \tablefoottext{d}{Ratio between the main beam and the total
      beam solid angles including large angular scale error beams and far
      side lobes}
    \tablefoottext{e}{Full-width at half-maximum of the beam used in our reference photometric system}
    \tablefoottext{f}{Ratio between the reference FWHM beam and the
      total beam solid angles including large angular scale error beams and far
      side lobes}
    \tablefoottext{f}{Systematic calibration uncertainties due to the opacity correction using the {\tt corrected skydip} method estimated at the reference IRAM \trentemetre\ winter observing conditions: 2\,mm pwv, $60^o$ elevation} 
    \tablefoottext{h}{Effective power law of noise reduction with integration time}
    \tablefoottext{i}{NEFD at zero opacity}
    \tablefoottext{j}{Mapping speed at zero opacity}
    
  }
\end{table*}

{\lp We have presented NIKA2 performance at the IRAM
  \trentemetre\ telescope. It has been evaluated with a
  \baseline\ calibration method that goes from
  observations and raw data to measured flux densities.
  In the \baseline\ calibration photometric system, the flux
  density of point-like sources are measured as the amplitude estimates
  of a fixed-width Gaussian of FWHM of $12.5''$ and $18.5''$ at the
$150$ and $260\,\rm{GHz}$ reference frequency, respectively.

  This method relies on three main analysis choices. First the
  atmospheric and instrumental correlated noises are corrected using a
  simple and robust method based on the subtraction of a series of
  the average temporal signals of the most correlated detectors. The
  atmospheric opacity is estimated using an improved version of 
  the method proposed in \citet{Catalano2014} and used in
  \citet{Adam2018}, which is based on the use of
  NIKA2 as an in-band total-power taumeter. Finally a scan selection
  based on the observation time is performed and retains 16 hours
  of observation a day in order to mitigate the effect beam size
  variations due to anomalous refraction of the atmosphere and
  partial illumination of the \trentemetre\ telescope that mainly
  impact the afternoon observations. We have also considered the pros
  and cons of alternative methods, which may in the future lead to even
  better calibration accuracy and stability.}

The performance of the NIKA2 camera has been assessed using a large
number of observations of primary and secondary calibrators and faint
 sources that have been acquired during three
observational campaigns over one year. The data set spans the whole
range of observing
elevation and atmospheric conditions encountered on-site.
The main characteristics that define the NIKA2 performance are
summarized in Table~\ref{tab:nika2summary}. We highlight the main
points in the following.

\begin{enumerate}
\item 
  All designed KIDs detect the
  signal at least in some observation
  scans. We conservatively retain only the most stable KIDs, which are
  immune to the cross-talking effect and yield good signal-to-noise
  measurement. 
  We report valid KID fractions of $84\%$
  for the $1\,\rm{mm}$ channel arrays and $90\%$ for Array 2. The other
  KIDs, which do not meet the validity criterion, are randomly
  distributed across the FOV, so that the whole $6.5\,\rm{arcmin}$ FOV is
  covered.
  \vspace{1mm}
\item 
  The main beam is well described with a 2D
  Gaussian of FWHM of $11.1''$ for the $1\,\rm{mm}$ channel arrays
  and $17.6''$ for Array 2, with uncertainties of $0.2''$ for the
  combination of Array 1$\&$3 and of $0.1''$ for Array 2.
  {\rev Comparing the main beam fit to the measured full beam, and 
    including large angular scale contributions from IRAM
    \trentemetre\ telescope heterodyne measurements, we have derived the
  main beam efficiency.}
  We found main beam
  efficiencies of {\rev $48 \pm 4 \%$} at $1\,\rm{mm}$ and {\rev $63 \pm 3 \%$} at
  $2\,\rm{mm}$.
  {\lp These results show that a significant fraction of the power is received
    outside the main beam. This underlying,
    extended, low-level beam pattern shows a complex structure of error 
    beams, rings, spokes, {\rev and far side lobes}.}
  Using individual maps per KID, \citet{Adam2018} reported an rms
  dispersion of the main beam FWHM across the FOV of about $0.6''$ at
  both wavelengths. This is consistent with the measured curvature of
  the best focus surface across the FOV. 
  {\lp We also provide the reference beam efficiencies, which are the
    fixed-width Gaussian beam efficiencies, that allow taking into
    account the power stemming from outside the reference beam, {\rev
      as needed for studying the diffuse emission.}} 
  \vspace{1mm}
\item 
  We have evaluated the rms calibration uncertainties using 264 
  scans of {\rev point-like} sources whose flux density is above about
  one Jy. We find {\rev point-source} rms calibration
  uncertainties of about $6\%$ at $1\,\rm{mm}$ and about $3\%$ at
  $2\,\rm{mm}$, which are state-of-the-art performance for a
  ground-based millimetre-wave instrument. {\rev The rms calibration
    uncertainties for the diffuse emission must further include the
    uncertainties on the reference beam efficiencies, as given in
    Table~\ref{tab:nika2summary}}. The absolute
  calibration uncertainties are of $5\%$ and the systematic
  calibration uncertainties evaluated at the IRAM
  \trentemetre\ reference Winter observing conditions are
  below $1\%$ in both channels.
   \vspace{1mm}
\item 
  The noise does well integrate as the square root of the integration time. We
  have derived robust estimate of the NEFD using more than a thousand scans
  encompassing a large range of observing conditions. We found NEFD at zero
  atmospheric opacity of $30 \pm 3\,\rm{mJy}\cdot s^{1/2}$ at $1\, \rm{mm}$ and
  $9 \pm 1\,\rm{mJy}\cdot s^{1/2}$ at $2\, \rm{mm}$.
  The NEFD estimates demonstrates the high-sensitivity of
  the KID arrays of NIKA2. 
  The instrumental sensitivity at $1\,\rm{mm}$
  is however currently mainly limited by the non-optimal transmission of
  the air-gap dichroic plate
  , mostly prominent in one polarisation component
  (A1) but affecting the other (A3) as well.
  In addition to the dichroic upgrade, further possible areas of
  improvements for the $1\,\rm{mm}$ observation channel are: 1)
  improve the data processing and in particular the noise
  decorrelation methods, 2) increase the bandwidth of the $1\,\rm{mm}$
  arrays (subjected to the improvement of the dichroic) and 3)
  upgrade the surface of the telescope.
   \vspace{1mm}
\item  NIKA2 mapping capabilities are better represented by evaluating the
  mapping speed, which is defined as the sky area that is covered in one
  hour of observation with a noise level of $1\,\rm{mJy}$. We found
  mapping speeds at zero atmospheric opacity of $111$ and
  $1388\, \rm{arcmin}^2 \cdot \rm{mJy}^{-2} \cdot \rm{h}^{-1}$ at
  $1\, \rm{mm}$ and  $2\, \rm{mm}$, respectively. NIKA2 mapping speed is thus at
  least an order of magnitude better than the previous generation of IRAM
 \trentemetre\ telescope resident continuum cameras~\citep{Catalano2014, Staguhn2011_GISMO, Kreysa1999}.
  
\end{enumerate}

We conclude that NIKA2 has unique capabilities in fast dual-band
mapping at tens arcsecond resolution. {\lp It is currently available to the
whole community and will operate at the IRAM \trentemetre\ telescope
at least for the next decade.}
NIKA2 performance meet the requirements
to address a large range of science topics in astrophysics and
cosmology.

\begin{acknowledgements}
  {\rev We warmly thank Attila Kov\'acs for his in-depth reading,
    constructive inputs and interesting discussions, which have
    contributed to improve the quality of the paper.}
  We would like to thank the IRAM staff for their support during the
  numerous campaigns. 
  The NIKA2 dilution cryostat has been designed and built at the Institut N\'eel. 
  In particular, we acknowledge the crucial contribution of the Cryogenics Group, and 
  in particular Gregory Garde, Henri Rodenas, Jean Paul Leggeri, Philippe Camus. 
  This work has been partially funded by the Foundation Nanoscience
  Grenoble, the LabEx FOCUS ANR-11-LABX-0013 and the ANR under the
  contracts "MKIDS", "NIKA" and ANR-15-CE31-0017. This work is
  supported by the French National Research Agency in the framework of
  the "Investissements d’avenir” program (ANR-15-IDEX-02).
  We have benefited from the support of the European Research Council Advanced
  Grant ORISTARS under the European Union's Seventh Framework
  Programme (Grant Agreement no. 291294). We acknowledge financial
  support from the “Programme National de Cosmologie and Galaxies”
  (PNCG) funded by CNRS/INSU-IN2P3-INP, CEA and CNES, France.
  We acknowledge funding from the ENIGMASS French LabEx (F. R.), the FOCUS French
  LabEx doctoral fellowship programme (A.R.) and the CNES doctoral
  fellowship program (A.R.). R. A. acknowledges support from
  Spanish Ministerio de Econom\'ia and Competitividad (MINECO) through
  grant number AYA2015-66211-C2-2. M.D.P. acknowledges support from
  Sapienza Universita' di Roma thanks to Progetti di Ricerca Medi 2017,
  prot. RM11715C81C4AD67. 
\end{acknowledgements}

%
\bibliographystyle{bibtex/aa} 
\bibliography{bibtex/NIKA2_biblio} 
%

\begin{appendix}
  \section{Reference flux density of the calibrators}
  \label{ap:ref_flux_calibrator}

\subsection{Uranus and Neptune}
\label{se:ref_flux_uranus_neptune}

For the flux densities of the giant planets, we use the ESA calibrator models\footnote{The models used for the calibration of the \emph{Herschel} satellite are documented at the URL: \url{https://www.cosmos.esa.int/web/herschel/calibrator-models}}.
These models provide the planet brightness temperature in the
Rayleigh-Jeans approximation as a function of the frequency. The
resulting flux density of the primary calibrator is therefore: 
\begin{equation}
S_c(\nu_0) = \Omega \, \frac{2 \nu_0^{2} k T_{RJ}}{c^2}
\end{equation}
where $\Omega$ is the solid angle of the planet on the sky, 
which can be calculated using:
\begin{equation}
\Omega = \pi \frac{r_{e} r_{p-a}}{D^{2}} 
\label{eq:omega}
\end{equation}
where $r_{e}$ is the equatorial radius of the planet and $r_{p-a}$ is
its apparent polar radius, and $D$ the distance to the
planet. $r_{p-a}$ can be computed from the sub-observer latitude $\phi$
({\it e.g.} the latitude of the 30-m telescope as seen from the planet in the
planet equatorial reference frame) and the polar radius of the
planet $r_{p}$ as:
\begin{equation}
r_{p-a} = \sqrt{r_{p}^2 \cos^{2}\phi + r_{e}^2 \sin^{2} \phi}
\end{equation}

The planets ephemerides are obtained using the
Horizon system\footnote{An interface to the Horizon system is given at \url{https://ssd.jpl.nasa.gov/horizons.cgi}} with the quantities
listed in Table~\ref{tab:planetphysparam}. The planet flux densities for a given date are computed using a dedicated
photometry tool\footnote{This software is written in Python and available at \url{https://github.com/haussel/photometry/blob/master/notebooks/planet_fluxes.ipynb}}. The model spectra are linearly interpolated in log space at the
reference frequencies of the NIKA2 bandpasses. 

\begin{table}[!ht]
\begin{center}
\begin{tabular}{|c|c|c|}
\hline
     & Uranus & Neptune \\
\hline
$r_{e}$ [km]  & 25559 & 24764 \\ 
\hline
$r_{p}$ [km]  & 24973 & 24341  \\
\hline
$\phi$         & Ob-lat & Ob-lat \\
\hline
$D$   [AU]    & delta   & delta \\
\hline
\end{tabular}
\end{center}
\caption[Primary calibrator flux models]{Physical quantities used for the Uranus and Neptune fluxes
  computation (equation~\ref{eq:omega}. Ob-lat and delta are quantities 
  tabulated by NASA Horizons system as a function of the date}
\label{tab:planetphysparam}
\end{table}

The Uranus and Neptune models have been compared to 
observations of these planets with the \emph{Planck}
satellite~\citep{PLCK-LII}.

For Uranus, the model used in the comparison
is the ESA V2, and it is found to over-predict by $4\,\rm{K}$ (about 4\%) the
observed RJ temperature at 143\,GHz, to agree at 217\,GHz, and
to under-predict at 353\,GHz. We use for the NIKA2 calibration the ESA
model V4, that predict a flux respectively -3.3\%, 0.3\% and 4.7\% higher in the
143, 217 and 353\,GHz bands, that correspond to a few percent accuracy
with respect to \emph{Planck} observations.

For Neptune, the same study compared Planck observation with the ESA V5
model, {\it i. e.} the same one used for NIKA2 calibration. For this
planet, temperatures are found to disagree at most by $5\, \rm{K}$, that is 4.1\%,
with the same trend with frequency as observed for Uranus.

In summary, this study confirms that Uranus ESA V4 and Neptune ESA V5
models are accurate to 5\% for predicting planet flux densities. This
result agrees also with the accuracy estimated from Herschel SPIRE
and PACS observations~\citep{Mueller2016, Swinyard2014}. 
Furthermore, the variations of Uranus and Neptune fluxes over the duration of a typical
NIKA2 run are taken into account, although they are negligible
compared to the model accuracy. On the other hand, we
found\footnote{This effect is illustrated in the Python notebook
  distributed with the software at \url{https://github.com/haussel/photometry/blob/master/notebooks/planet_fluxes.ipynb}} that not
taking into account the planet shape and orientation with respect to
the observer in the computations of its solid angle can lead to errors
between 1 and 2\%.

\subsection{Secondary calibrators}
\label{se:ref_flux_secondaries}

The secondary calibrator MWC349 is a stellar
binary system, including the young Be star MWC349A, surrounded by a
disk. Its radio continuum emission
originates in an ionized bipolar outflow~\citep{Tafoya2004}. MWC349A has
been monitored with the PdBI and VLA, and
shown to be 
only slightly angularly resolved,
making it a point source for the \trentemetre\ telescope.
We have computed the flux densities at the NIKA2 reference frequencies 150 and
260 GHz with $S_\nu = 1.16\pm0.01 \times
(\nu/100 \rm{GHz})^{0.60\pm0.01}$ provided by this
monitoring\footnote{See \emph{e.g.} the talk given by M.~Krips at the 10th 
IRAM Millimetre Interferometry School in Grenoble (France) in October
2018 at: 
\url{https://www.iram-institute.org/medias/uploads/file/PDFs/IS-2018/krips_flux.pdf}}.


The secondary calibrator CRL2688 is an Asymptotic Giant Branch
star. Its radio continuum emission is mostly from circumstellar dust
and is somewhat extended~\citep{Knapp1994}.  Its flux densities at
$850\ \mu$m and $450 \ \mu$m have been stable at the 5\% level as
monitored by SCUBA2 in 2011-2012~\citep{Dempsey2013_SCUBA2}.
We have extrapolated these flux densities to 150 and 260 GHz
using the power law $S_s(\nu_0) \propto \nu_0^{\alpha}$ with an index
$\alpha=2.44\pm0.18$ derived from the SCUBA2 measurements.


The secondary calibrator NGC7027 is a young, dusty, carbon rich
Planetary Nebula with an ionized core.  It is extended in the
continuum and molecular lines~\citep{Bieging1991}, and is not a point
source for the \trentemetre\ telescope.  Its most recent flux densities are
reported at $1100\, \mu$m and $2000\, \mu$m in~\citet{Hoare1992}. It has
been reported to decrease by $\sim$ 0.145 percent/yr in the optically
thin part of its spectrum above $6\,\rm{GHz}$ from VLA
observations~\citep{Zijlstra2008, Hafez2008}. This makes
these flux densities uncertain by 3.6\% currently. Its SED from cm
wavelengths to optical is also presented in~\citet{Hafez2008}. {\lp
\citet{Perley2013} measured its flux density between 1 and $50\,
\rm{GHz}$. Moreover, flux density measurement at 90 GHz over 20 years
with the \trentemetre\ telescope are presented in \citet{Kramer2008}.}
The flux densities have been extrapolated to 150 and 260 GHz and the
modelled decrease since 1992 has been included.

All these expected flux densities extrapolated from the literature are
summarized in Table~\ref{tab:flux_ref_sec}.

\begin{table*}[!thbp]
  \caption[Reference flux densities of secondary calibrators]{Reference flux densities of secondary calibrators at the NIKA2 reference frequencies 150 and 260 GHz. Uncertainties of flux densities extrapolated
    at 150 and 260 GHz include contribution of the uncertainty on the spectral index $\alpha$, which is defined as $S_{\nu} \propto \nu^{\alpha}$.}
  \label{tab:flux_ref_sec}
  \centering    
  \begin{tabular}{|l|c|c|c|l|}
    \hline\hline
    \multicolumn{1}{|c}{}  & \multicolumn{3}{|c}{flux  densities (Jy)} & \multicolumn{1}{|c|}{}  \\
    \hline
    &    A1 \& A3       &  A2             &            &   Ref. \\
    &  260 GHz          &  150 GHz        & $\alpha$ &      \\
    \hline
    MWC349A   &   $2.06\pm0.04$  &  $1.48\pm0.02$ &  $+0.60\pm0.01$      &  PdBI \& NOEMA monitoring  \\
    NGC7027  &   $3.46\pm0.11$   &  $4.26\pm0.24$  &  $-0.34\pm0.10$     &  \citet{Hoare1992}      \\
    CRL2688  &   $2.91\pm0.23$   &  $0.76\pm0.14$  &  $+2.44\pm0.18$     &  \citet{Dempsey2013_SCUBA2} \\
    \hline
  \end{tabular}
\end{table*}

Measured flux densities however are determined over the broad
bandwidth of each array and so must be colour-corrected to be compared
to the expected flux densities of Table~\ref{tab:flux_ref_sec}.  For
this purpose, we have derived colour corrections for sources with spectral
indices $\alpha$ comprised between -2 and +4 in Table~\ref{tab:mod}. 
As it can be seen, this effect can be a few percent for MWC349, NGC7027, and CRL2688.

  \section{Reconstruction of the focal surfaces}
  \label{ap:focus_surfaces}

Owing to the NIKA2 $6.5~\rm{arcmin}$ FOV, the focus is expected to
slightly change across the FOV, defining curved focal surfaces at the
location of the three arrays. Therefore, beam patterns are expected to
show some scatter across the FOV accordingly to the focal
surfaces. Although all the detectors on a flat array cannot be
individually focalized, an optimal axial focus of the telescope can be
found to maximize the number of detectors at the best focus and hence,
maximize the resolution of the NIKA2 maps.
This optimal focus setting is obtained by measuring the focus at the
center of the arrays as described Sect.~\ref{se:axial_focus} and apply
a focus shift of $-0.2\,\rm{mm}$, which is 
predicted using ZEMAX simulations, and verified by measuring
the focus surfaces as described in the following.

\begin{figure*}[!thbp]
\begin{center}
  \includegraphics[trim={0, 9.5cm, 0, 9.5cm}, clip=true, width=\linewidth]{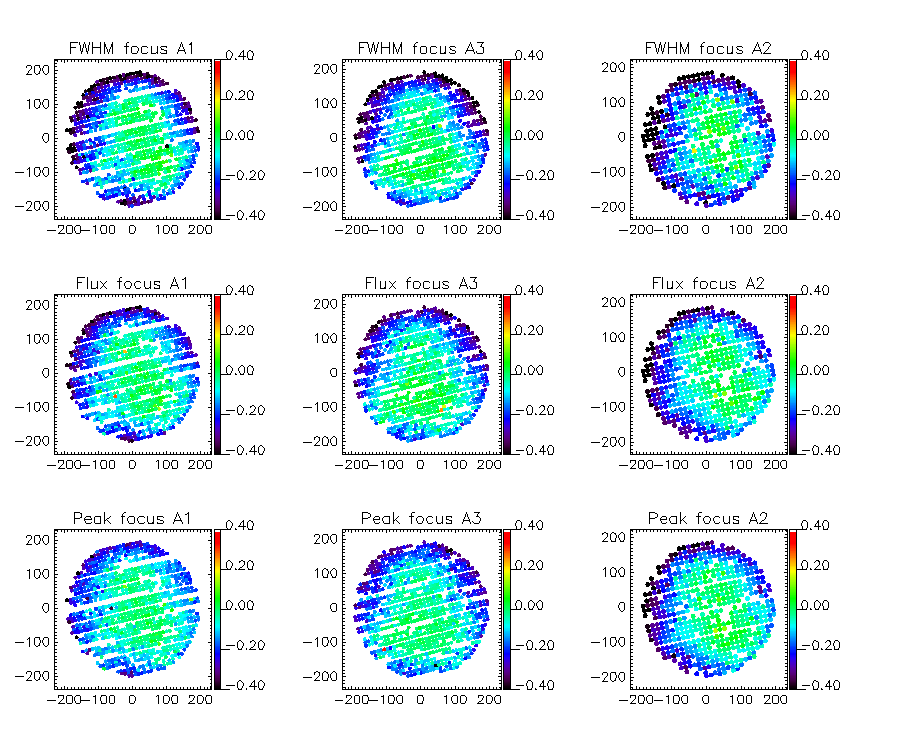}
\caption[Focus surfaces]{Focus surface of A1, A3 and A2 arrays from left to
  right. In this example, the focus estimates rely on the maximisation of the flux
  density. On each plot, the x and y axis are the Nasmyth offsets
  w.r.t. the center of the array in arcsec, while the color-code represents
  the relative focus estimate w.r.t. the central focus, given in mm.}
\label{fig:focus-surfaces}
\end{center}
\end{figure*}

\subsection{Method}

We estimate NIKA2 focal surfaces by means of a sequence of five \bms\ of bright
point-like sources, typically planets or bright quasars, for various
axial positions $z$ of the telescope sub-reflector around its nominal
position, which is the optimal axial focus $z_{\rm{opt}}$. 
The axial position is changed in step of $0.6~\rm{mm}$ to probe a large
focus range for measuring even the extreme variation of the focus surfaces,
namely $z \in \{-1.2, -0.6, 0, 0.6, 1.2 \} + z_{\rm{opt}}$.  Each
\bm\ is analysed using the data reduction pipeline, as described in
Sect.~\ref{se:dataproc}, and $4''$-resolution individual maps per KID
are produced. 
Therefore, a series of
five maps at various focus positions is available for each detector, from which
the best focus is estimated as described in Sect.~\ref{se:axial_focus}. The
ensemble of the relative focus estimate per KIDs with respect to the best focus
at the center of the array constitutes the focus surface. An accurate estimate
of the center focus is obtained as the weighted average focus estimate of the
KIDs lying in a $30''$ radius around the geometrical center of the array. This
average does not induce any sizeable bias thanks to the flatness of the focus
surface in the innermost regions. For robustness test, we consider three focus
estimates: the first two are the same as discussed in
Sect.~\ref{se:axial_focus} -- namely i) $\hat z_{\rm{fwhm}}$ the focus that
minimizes the geometrical FWHM and ii) $\hat z_{\rm{peak}}$ the focus that
maximizes the amplitude of the best-fitting elliptical Gaussian -- whereas the
third one is $\hat z_{\rm{flux}}$ the focus that maximizes the flux
density in the reference photometric system
(Sect.~\ref{se:photometric_system}). The comparison between the two
estimators based on Gaussian-amplitude fitting ($\hat z_{\rm{peak}}$
and $\hat z_{\rm{flux}}$), will test the stability of the focus
results against the exact choice of the beam fitting function.

\subsection{Data set}

Nine defocused \bm\ sequences have been acquired, including incomplete
sequences and sequences hindered by poor atmospheric conditions. To
check for systematic effect, a focus measurement is performed
immediately before and after the \bm\ sequence. Using these
measurements, the central focus drift between the starting time and the
end of the sequence is estimated. 
We select sequences that i) comprises at least four scans (four
z-focus steps), ii) have been observed with a zenith opacity at $225\,\rm{GHz}$ (as indicated by
the IRAM \taumeter) below 0.5 and iii) have a maximum central focus
drift of $0.5~\rm{mm}$. These criteria preserve five sequences from which focus
surfaces can be reconstructed, 
corresponding to observations of the bright quasar 3C84 and Neptune.

\subsection{Results}
For each detector $k$ and for each \bm\ sequence $s$, we obtain for
the array $\nu$, a focus measurement $z_\nu^{k, s} \pm \sigma_\nu^{k, s}$,
where $\sigma_\nu^{k, s}$ is the $1\mbox{--}\sigma$ error of the least-square
polynomial fit. The focus surface measurements per array are obtained
as weighted averages of the five \bm\ sequences as in the
following:
\begin{equation}
\label{eq:mv_focus_surf}
z_\nu^{(k)} = \left( \sigma_\nu^{(k)} \right)^2 \,  \sum_s \frac{z_\nu^{k,s}}{\left(\sigma_\nu^{k,s}\right)^2}\, \,  ,
\end{equation}
with uncertainties 
\begin{equation}
\label{eq:error_mv_focus_surf}
\sigma_\nu^{(k)} = \left[ \sum_s \frac{1}{\left(\sigma_\nu^{k,s}\right)^2}\right]^{-1/2}\, .
\end{equation}

We present NIKA2 focus surfaces per array obtained as in
Eq.~\ref{eq:mv_focus_surf} for the method of flux density maximization  
in Fig.~\ref{fig:focus-surfaces}.
We further check that the three flavours of focus-estimators provide
us with focus surfaces per array that are in good agreement with each
others. Furthermore, they have a non-axisymmetrical flatten bowl
shape, which is well consistent with expectations from optical
simulation using ZEMAX but with slightly higher curvature amplitude.
Namely, the median defocus (that is the relative focus w.r.t. the
central focus) across the detectors is about
$-0.1~\rm{mm}$ for the three arrays. Maximum defocus values of about
$-0.6~\rm{mm}$ are found for detectors corresponding to the outer top and
left regions of the FOV.

We estimate the uncertainty of the focus
surface measurements using the standard deviation between the three
estimators $z_\nu^{(k)}|_{\rm{fwhm}}$, $z_\nu^{(k)}|_{\rm{peak}}$ and
$z_\nu^{(k)}|_{\rm{flux}}$. We found approximatively homogeneous
standard deviation across the FOV with median values of about
$0.03~\rm{mm}$. We also verified the stability of the focus surfaces by comparing
results from a series of \bm\ sequences acquired at various dates and
under various atmospheric conditions.

{\lp In addition, using optical simulation, we have found that the
variations of beam aberrations are much smaller than the
diffraction pattern, resulting in a quasi invariant beam for a
wide range of defocus up to about $0.3\, \rm{mm}$ around the optimal
focus. For larger defocus, however, the beam starts to deteriorate in
some region of the image plane.}
{\lp Using the results from the measurements and the optical
simulation, we optimize the focus of all detectors by setting the
focus at the value estimated at the center of the arrays shifted of
$-0.2\,\rm{mm}$.}

%
%
  \section{Photometric correction}
  \label{se:photometric_correction}

The \emph{baseline} calibration (Sect.~\ref{se:baseline_calibration}) relies on
the \emph{baseline} scan selection, as defined in
Sect.~\ref{se:data_selection}, to mitigate the impact of
the \afternoon\ variation effect, as evidenced in Sect.~\ref{se:beam_variation}.
By contrast, in this section, we
address the issue of calibrating even during the observing periods
impacted by the \afternoon\ beam variation effect. We discuss an
alternative calibration method that
relies on a photometric correction depending on the beam size.
{\lp The key idea is to perform a joint monitoring of flux estimates
  using the fixed-width Gaussian amplitude and of the beam size.}
The objective is both to cross-check the baseline calibration results
using more observation scans and to anticipate on future developments
that could be deployed if an accurate beam monitoring is performed.

\subsection{Photometric correction method}
\label{se:photometric_correction_method}

When the beam size broadens due to e.g. \afternoon\ beam effect, the flux
density is smeared in a larger solid angle and the flux density estimator, which
is based on the amplitude fit of a Gaussian beam of fixed FWHM (see
Sect.~\ref{se:photometric_system}) is biased towards low flux
densities.

Indeed, up to a normalization factor, the flux density
estimator in the reference photometric system (Sect.~\ref{se:photometric_system}) applied to a map
$M(\theta,\phi)$ reads

\begin{equation}
  \hat{S}  = \int \int M(\theta, \phi)\, e^{-\frac{d(\theta,\phi)^{2}}{2\sigma_{0}^{2}}} \sin \theta d\theta d\phi\,,
  \label{eq:flux_density_estimator}
\end{equation}

where $d(\theta,\phi)$ is the angular distance and $\sigma_0$
corresponds to FWHM$_0$
, as defined in Table~\ref{tab:definitions}.

Modelling the map of a point-like source with a single Gaussian of
width $\sigma$ and of amplitude $\mathcal{A}$ as

\begin{equation}
  M(\theta, \phi) = \mathcal{A} e^{-\frac{\theta^{2}}{2\sigma^2}}\,,
  \label{eq:pointsource_map}
\end{equation}

leads to

\begin{equation}
  \hat{S}  = 2\pi \sigma^2 \mathcal{A} \,  \frac{\sigma_0^2}{(\sigma^2 + \sigma_0^2)}.
  \label{eq:gaussian_star}
\end{equation}
As the map is calibrated using the reference Gaussian beam
(Sect.~\ref{se:photometric_system}), the absolute calibration factor 
has a beam dependency that compensates the
$\sigma_0^2/(\sigma^2 + \sigma_0^2)$ factor in Eq.~\ref{eq:gaussian_star}.

Assume that we observe the source under stable conditions and
denote with a $\star$, the associated amplitude $\mathcal{A}_\star$, beam width
$\sigma_\star$ and flux estimate $\hat{S}_\star$. Using aperture
photometry, the energy conservation ensures that one has
$2\pi\sigma^2 \, \mathcal{A} = 2\pi\sigma_\star^2 \, \mathcal{A_\star}$.

To retrieve the flux estimate $\hat{S}_\star$ from the flux density
estimate $\hat{S}$, as obtained for any observations, a
corrected flux density $\hat{S}_{\rm{pc}}$ can be obtained using 
\begin{equation}
  \hat{S}_{\rm{pc}} = f(\sigma) \, \hat{S},
\end{equation} 
where the photometric correction is 
\begin{equation}
  f(\sigma) = \frac{(\sigma^2 + \sigma_0^2)}{(\sigma_\star^2+\sigma_0^2)}, 
\end{equation} 
so that $f(\sigma) = 1$ if $\sigma=\sigma_\star$.

\begin{figure}[ht!]
  \begin{center}
    \includegraphics[clip=true, trim={0.9cm, 0.1cm, 0.5cm, 0.5cm}, width=0.4725\textwidth]{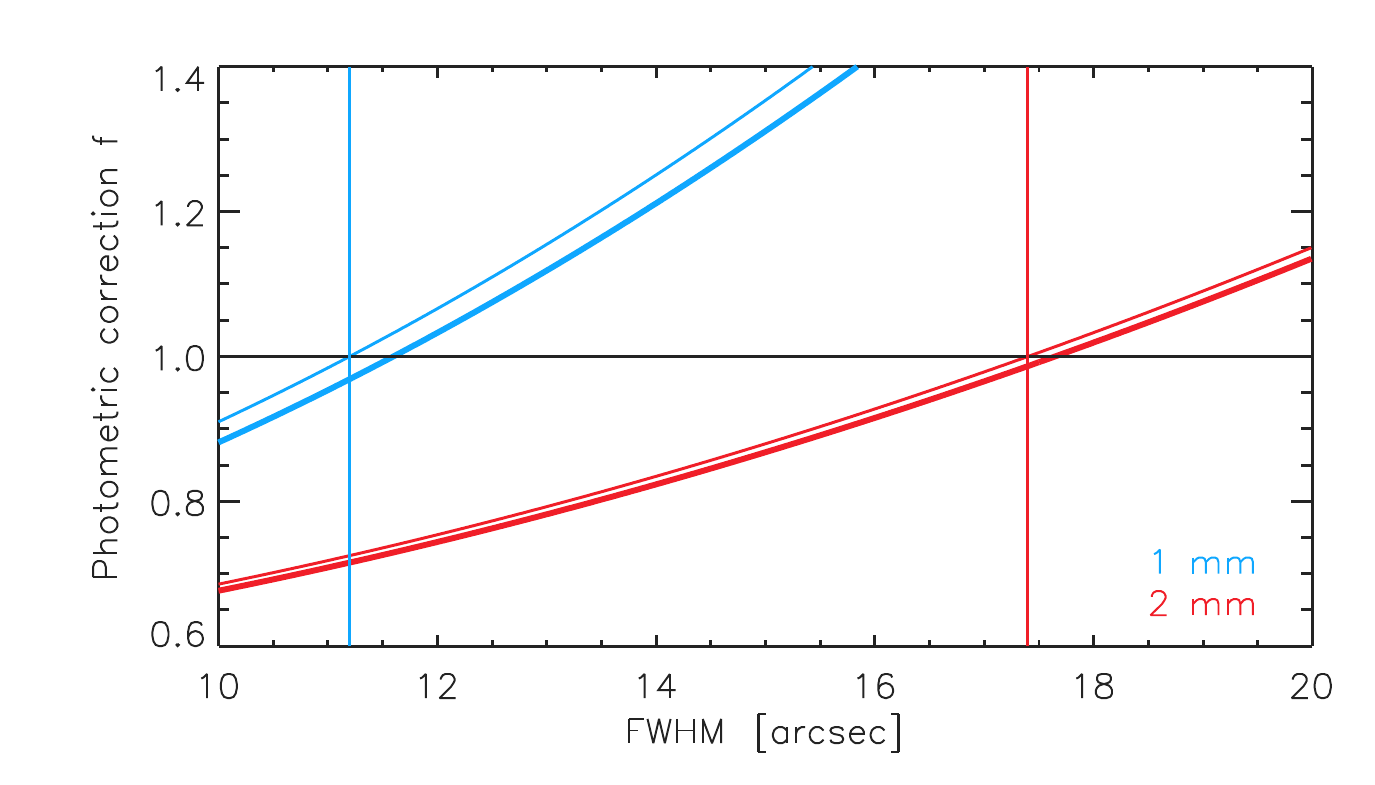}
    \caption[Photometric correction]{Magnitude of the beam photometric
      correction $f(\sigma)$ as a function of the actual FWHM, at 1\,mm (blue)
      and 2\,mm (red). Thick lines correspond to a choice of $\sigma_\star$
      derived on a very bright source like Uranus. Thin lines are for sources of
      1\,Jy at 1 and 2\,mm.}
\label{fig:f_sigma}
\end{center}
\end{figure}

Fig.~\ref{fig:f_sigma} shows how $f(\sigma)$ varies with the actual beam width
and for two choices of $\sigma_\star$ depending on the flux of the observed source.

The beam size in stable atmospheric conditions $\sigma_\star$ is
determined by fitting the 2D Gaussian beam on the series of scans of
source with varying flux densities that have been used for the beam
characterization in Sect.~\ref{se:mainbeam}. However, $\sigma_\star$
is not equivalent to the main beam Gaussian size since the side lobes
and first error beams are not taken into account for this beam size
monitoring. The $\sigma_\star$ estimates are slightly
larger for bright sources due to the contribution of the first side
lobes and error beams, which are above the noise level, in the Gaussian fit.

\subsection{Monitoring of the temperature-induced beam size variation}
\label{ap:beam_monitoring}

The photometric correction thus relies on the measure of the current beam size
$\sigma$. The induced uncertainties on the flux density measurements depend on
the precision of the beam size determination. Here we
consider two methods to monitor the beam size.

The temperature-induced beam size variation is primarily monitored
using 2D Gaussian fit on all the available bright source observations,
as presented in Sect.~\ref{se:beam_variation}. 

For a finer sampling of the observation time, we also consider using
the pointing scans for beam monitoring. As discussed in
Sect.~\ref{se:pointing}, the telescope pointing is
monitored on a hourly basis during observation using {\tt pointing}
scans. As these scans consist of two sub-scans in azimuth and
elevation of about 10 seconds of integration time each, they
can also be used to make a map of the pointing source. For each campaign,
we thus have on hands hundreds of maps of mostly point-like bright
sources, which can also be used to monitor the beam size. 
For this purpose, {\tt pointing} scans are reduced 
and projected onto maps of $2''$ resolution using
the data analysis pipeline described in Sect.~\ref{se:dataproc}.
Fitting an elliptical 2D Gaussian to this map, we compute the geometrical FWHM.
{\tt Pointing} scans on {\lp slightly extended} sources, such as NGC7027,
are discarded from the analysis.

For each {\tt pointing}, we also seek for signs of atmospheric
anomalous refraction. There are
enough KIDs per observation band to make an independent map using only
one subscan, i.e. 10 seconds of integration time.
For each of the four cross subscans, we thus estimate the position of the best
2D Gaussian that fits the map. We compute the deviation between each
subscan-derived position and the best-fit position using the entire
scan. An anomalous refraction event is detected when the difference is
above $2.5''$ for at least one subscan. We find that the \afternoon\
beam size variation effect is due to anomalous refraction for
between one third and one half of the scans, as reported in
Sect.~\ref{se:beam_variation}.

The pointing-based FWHM estimates constitute a time-sampling of the
beam size during the whole observation campaign. They can serve to
estimate the beam size of any observation scans, in particular
towards sources too faint for a direct FWHM 
fit to be made on the map. However, we
expect less accuracy than using standard OTF raster
scans of bright sources due to the shorter integration time and the
fact that only the innermost KIDs in the array see the source.
To mitigate the dispersion, the time-stamped
pointing-based FWHM
is filtered with a running median on a 70-minute
width time window. Then, the FWHM 
at the time of the considered scans is
interpolated from the median-filtered pointing-based FWHM.
\begin{figure}[ht!]
  \begin{center}
    \includegraphics[clip=true, trim={0.9cm, 0.5cm, 0.5cm, 0.5cm}, width=0.4725\textwidth]{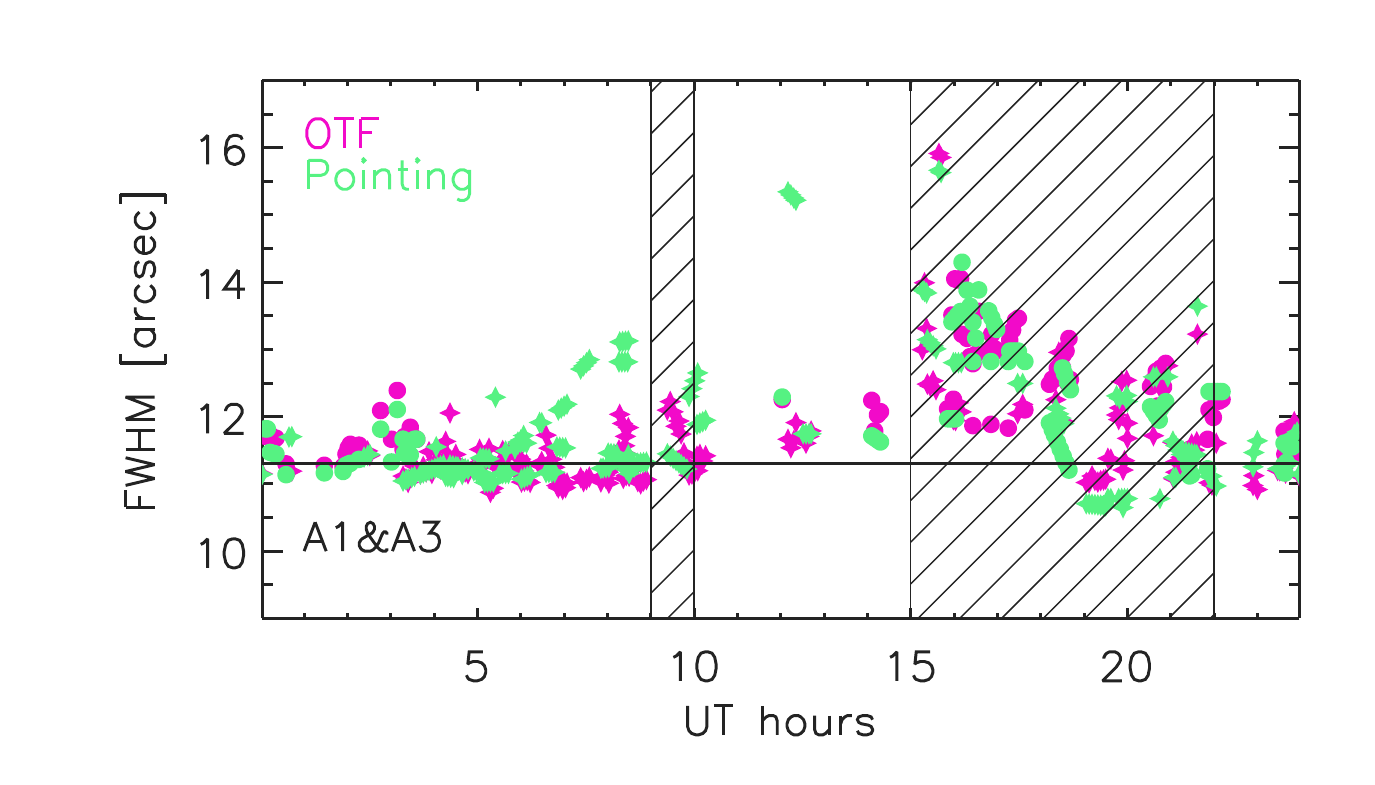}
    \includegraphics[clip=true, trim={0.5cm, 0.5cm, 0.5cm, 0.5cm}, width=0.4875\textwidth]{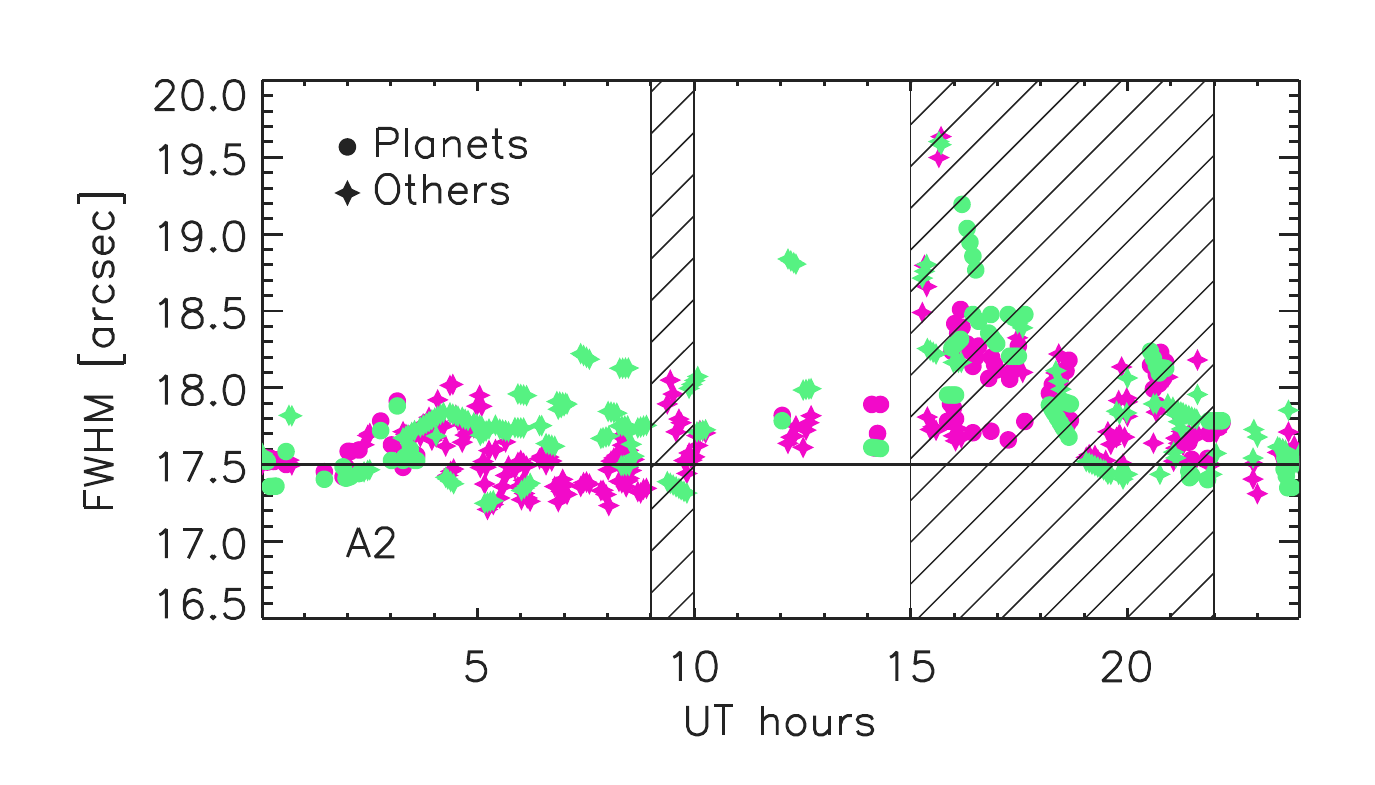}
    \caption[Beam size monitoring comparison]{Beam size monitoring.
     'OTF'-labelled pink data points show the FWHM estimates from a 2D
    Gaussian fit on the maps of OTF raster scans towards bright
    sources, whereas the 'Pointing'-labelled light green data points
    are FWHM estimates obtained by interpolating the {\lp
    median-filtered} pointing-based FWHM at the time of the
    scans. {\lp The
    pointing-based FWHM estimates are obtained by fitting a 2D Gaussian on the
    maps of {\tt pointing} scans.}}
\label{fig:beam_monitoring_compare}
\end{center}
\end{figure}
Figure~\ref{fig:beam_monitoring_compare} shows two different FWHM 
estimates for the same data set: on one hand the best-fit FWHM 
estimates on the OTF-scan map and on the other hand the interpolation from
the pointing-based FWHM 
monitoring. The two estimates show the same global variations as a
function of the UT hours. They are well in agreement with each
other, although the pointing-based estimates have more dispersion and
a few outliers as expected.

\subsection{The two case studies}

We perform two case studies that correspond to the two beam monitoring
methods discussed above.\\

\noindent \emph{Demonstration case} This method named {\tt PC-demo} hereafter, uses a
photometric correction based on the beam monitoring with bright source
scans. Both the 2D Gaussian FWHM fit and the FWHM$_0$ photometry are performed
on the map of the source. This method thus applies only to point-like sources
that are bright enough for an accurate fit of the beam on a single scan.

To capture only the beam size variations driven by the
observing conditions (primary mirror deformations, anomalous
refraction, elevation), a small correction $\delta_{\rm{FWHM}}$ has to be made to
the 2D Gaussian beam FWHM estimate for bright sources. The estimate of the
actual Gaussian size $\sigma$ is
\begin{equation}
  \hat{\sigma} = \frac{(\rm{FWHM} - \delta_{\rm{FWHM}})}{2\sqrt{2 \ln{2}}}, 
\end{equation} 
where the offset $\delta_{\rm{FWHM}}$ is null for faint or moderately
bright point sources, and non-zero for bright sources.
As for $\sigma_\star$, the 2D Gaussian fit yields slightly broader
FWHM for bright sources (e.g. planets) to accommodate
for the side lobes and error beams that are measured with high signal-to-noise.
For Uranus, $\delta_{\rm{FWHM}}$ includes also the beam widening due
to Uranus disc, as discussed in Sect.~\ref{se:mainbeam_results}.
We measure Uranus $\delta_{\rm{FWHM}}$
by comparing the average 
FWHM estimates using Uranus
scans and using MWC349 scans, we found $\delta_{\rm{FWHM}} = 0.4''$ at
1\,mm and $\delta_{\rm{FWHM}} = 0.25''$ at 2\,mm, which basically
distributes as one half being due to Uranus finite extension and the
other half stemming from the side lobes.\\

\noindent \emph{Practical case using pointing scans} This method,
named {\tt PC-point} hereafter, performs a photometric correction based on the beam monitoring with
pointing scans. 
Unlike {\tt PC-demo}, this method is usable even for sources fainter than
about one Jy \nico{but relies on interpolations between beam width estimates on
  other scans}. For Uranus, this value is corrected for the diameter size, as in
Sect.~\ref{se:mainbeam_results}. No other FWHM offset correction is needed since
pointing scan maps have a low signal-to-noise ratio that prevents the
geometrical FWHM from being significantly affected by the side lobes.

\subsection{Absolute calibration with a photometric correction}

We perform the absolute calibration by i) implementing the reference
method described in Sect.~\ref{se:practical_calib}, ii) correcting the
atmospheric attenuation using the {\tt corrected skydip} opacity
estimates, and iii) using the photometric correction of
Appendix~\ref{se:photometric_correction_method}.

\begin{figure}[!htbp]
  \begin{center}
    \begin{overpic}[clip=true, trim={0, -0.3cm, -0.3cm, 0},width=0.525\linewidth]{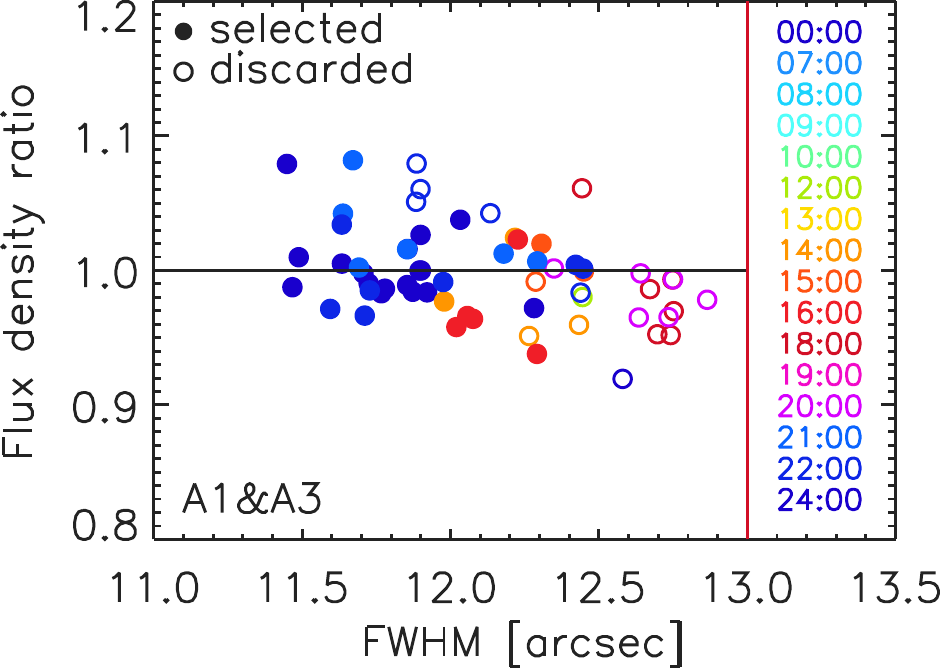}
       \put(18,25){\footnotesize PC-demo}
    \end{overpic}
    \includegraphics[clip=true, trim={0.7cm, -0.3cm, -0.25cm, 0}, width=0.465\linewidth]{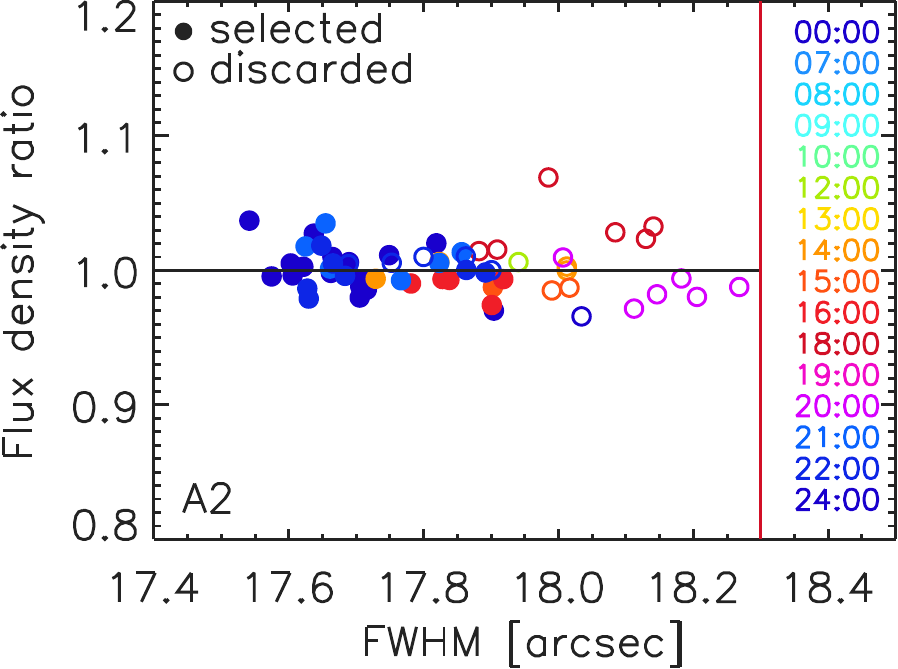}
    \begin{overpic}[clip=true, trim={0, -0.3cm, -0.3cm, 0},width=0.525\linewidth]{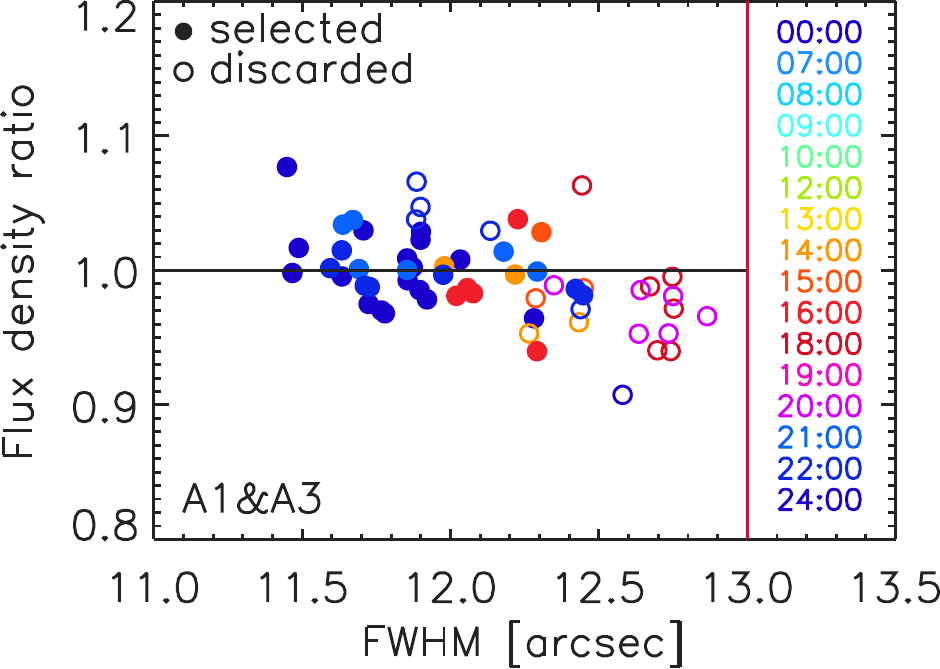}
      \put(18,25){\footnotesize PC-point}
    \end{overpic}
    \includegraphics[clip=true, trim={0.7cm, -0.3cm, -0.25cm, 0}, width=0.465\linewidth]{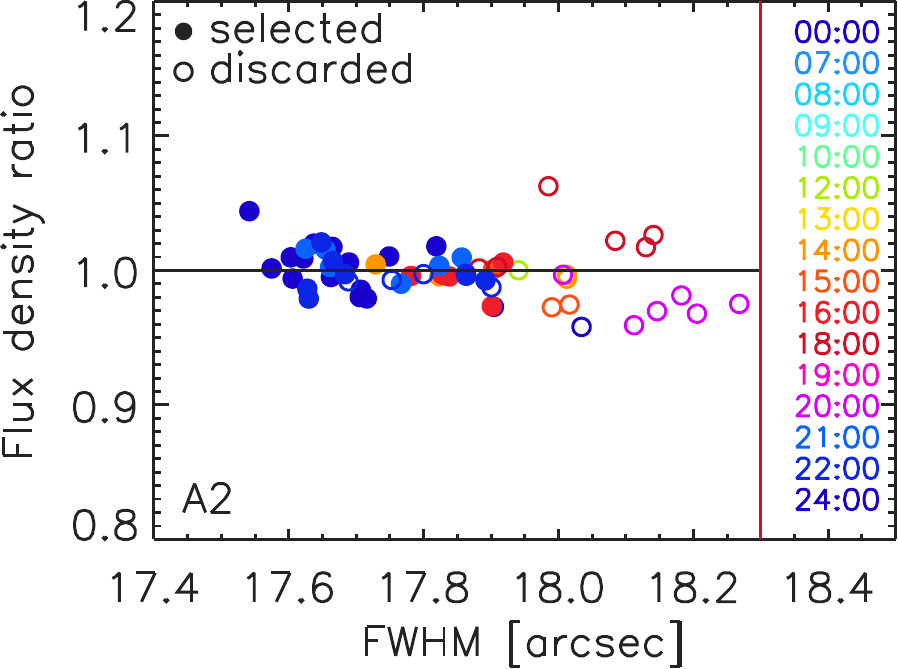}
    \vspace{-0.5cm}
    \caption[Uranus flux density stability against FWHM]{
      \small{Uranus flux density ratio vs beam size for calibration
  with photometric correction. The ratio of 
      Uranus measured flux densities to expectations as a function of the
      measured 2D Gaussian beam FWHM is shown for the $1$-mm array
      combination (left column) and for array 2 (right column) after absolute
      calibration using (\emph{first row}) the {\tt PC-demo} and (\emph{second
        row}) the {\tt PC-point} photometric corrections. These plots
      include all Uranus scans acquired during N2R9, N2R12 and N2R14
      campaigns and whose beam FWHMs are below the threshold indicated
      by the vertical red lines, (open circles), as
      well as the scans that met the baseline scan selection criteria (filled
      circles). 38 scans now pass the \emph{baseline} selection criteria vs 26
      only without photometric correction (see also Fig.~\ref{fig:calib_uranus_vs_fwhm_all}).}}
\label{fig:calib_uranus_vs_fwhm_photocorr}
\end{center}
\end{figure}

Using the photometric correction alleviates the need of
performing a scan selection based on the observation time. However,
the scans from which the absolute calibration is derived, are selected
on the FWHM estimate using the same criteria as for the baseline
calibration, that are FWHM thresholds of $12.5''$ at $1\, \rm{mm}$ and $18''$ at
$2\, \rm{mm}$. Thus, only the scans that are moderately affected by the beam
effect are included in the absolute calibration in order not to
include twice the photometric correction uncertainties in the error
budget (once for the absolute calibration and once for the photometry).

Figure~\ref{fig:calib_uranus_vs_fwhm_photocorr} presents the Uranus
measured-to-predicted flux density ratio as a function of the beam FWHM
after the photometric correction with the {\tt PC-demo} and
{\tt PC-point} methods. The flux
density is stable against the beam FWHM within uncertainties for both
wavelengths.

\begin{table}[!htbp]
\caption[Comparison of calibration results using five methods]{Comparison of absolute calibration results using five methods}
\label{tab:Abs_calibration_results_all}
\centering
\begin{tabular}{clrrrrr}
  \hline\hline
  \noalign{\smallskip}
  \multicolumn{2}{c}{}  &  baseline  & TM\tablefootmark{a}  &  SD\tablefootmark{b} & PC-d\tablefootmark{c} & PC-p\tablefootmark{d}  \\
  \noalign{\smallskip}
  \hline\hline
   \multicolumn{2}{c}{$\#$ scans} & 26    &       26  &    26    &    38           &    38 \\ 
  \hline
  \noalign{\smallskip}
   Ratio  &  1mm         &   1.00  &  0.95   &  1.06    &   1.01    &   1.01  \\
          &  2mm         &   1.00  &  0.94   &  0.99    &   1.01    &   1.01  \\
  \hline
  \noalign{\smallskip}
   RMS    &  1mm           &  3.3    &   4.5   &   3.3    &    3.1    &   2.6 \\
   $[\%]$ &  2mm           &  1.6    &   2.6   &   1.5    &    1.5    &   1.5 \\
\hline
\end{tabular}
\tablefoot{ Results based on 
    \tablefoottext{a}{the {\tt Taumeter} opacity correction}
    \tablefoottext{b}{the {\tt Skydip} opacity correction}
    \tablefoottext{c}{the {\tt PC-demo} photometry correction}
    \tablefoottext{d}{the {\tt PC-point} photometry correction}
    }
\end{table}

We further quantify the
difference between all the calibration methods that have been tested
in evaluating i) the average absolute calibration factor
with respect to the baseline calibration factor and
ii) the rms dispersion of the measured-to-modelled flux ratios. These
quantities are gathered in Table~\ref{tab:Abs_calibration_results_all}
in the rows labelled 'Ratio' and 'RMS' respectively. 

We find that, resorting to a photometric correction i) allows us to use $45\%$ more
scans for the absolute calibration, ii) has a negligible impact on
the absolute calibration factor and iii) yields a small reduction of
the flux density ratio dispersion. For the absolute calibration, the
{\tt PC-point} method performs as well as the {\tt PC-demo} one.
Photometry capability and stability when using a photometric
correction are further tested in Sect.~\ref{se:photometry}.\\
  
\end{appendix}

\end{document}